\documentclass[a4paper,11pt]{article}
\pdfoutput=1
\usepackage{jheppub}

\usepackage{graphicx}
\usepackage[figuresright]{rotating}
\usepackage{bm,amsmath,amssymb}
\usepackage[mathscr]{eucal}
\usepackage{multirow}
\usepackage{xcolor}
\usepackage{mathrsfs}
\usepackage{mathtools}
\usepackage{slashed}
\usepackage{graphics}
\usepackage{graphicx}
\usepackage{subfigure}
\usepackage{dsfont}
\usepackage{longtable}
\usepackage{bbm} 
\usepackage{float}
\usepackage{tcolorbox}
\usepackage{lipsum}
\usepackage{cancel}
\usepackage{enumerate}
\definecolor{lcolor}{rgb}{0.,0.0,0.}
\definecolor{citcolor}{rgb}{0,0.,0.5}
\usepackage{soul}

\newcommand{\rmL}{\textrm{L}}
\newcommand{\Hcal}{\mathcal{H}}

\newcommand{\Rcal}{\mathcal{R}}

\newcommand{\Jcal}{\mathcal{J}}
\newcommand{\vect}[1]{\boldsymbol{#1}_{\perp}}

\newcommand{\kgt}{\boldsymbol{k_{g\perp}}}

\newcommand{\Pt}{\vect{P}}

\newcommand{\qt}{\vect{q}}
\newcommand{\lt}{\vect{l}}

\newcommand{\at}{\vect{a}}
\newcommand{\bt}{\vect{b}}
\newcommand{\Kt}{\vect{K}}

\newcommand{\ruupt}{\boldsymbol{r}_{uu'}}
\newcommand{\ktone}{\boldsymbol{k_{1\perp}}}
\newcommand{\kttwo}{\boldsymbol{k_{2\perp}}}
\newcommand{\ltone}{\boldsymbol{l_{1\perp}}}
\newcommand{\lttwo}{\boldsymbol{l_{2\perp}}}
\newcommand{\ltthre}{\boldsymbol{l_{3\perp}}}

\newcommand{\xt}{\vect{x}}
\newcommand{\yt}{\vect{y}}
\newcommand{\zt}{\vect{z}}
\newcommand{\ut}{\vect{u}}

\newcommand{\Xt}{\vect{X}}
\newcommand{\Xttilde}{\boldsymbol{\tilde{X}_\perp}}

\newcommand{\rt}{\vect{r}}

\newcommand{\rxyt}{\boldsymbol{r}_{xy}}
\newcommand{\rbbpt}{\boldsymbol{r}_{bb'}}
\newcommand{\etx}{\boldsymbol{\epsilon_{x\perp}}}
\newcommand{\ety}{\boldsymbol{\epsilon_{y\perp}}}
\newcommand{\etz}{\boldsymbol{\epsilon_{z\perp}}}

\newcommand{\rzxt}{\boldsymbol{r}_{zx}}
\newcommand{\rzyt}{\boldsymbol{r}_{zy}}
\newcommand{\rzbt}{\boldsymbol{r}_{zb}}
\newcommand{\rzbpt}{\boldsymbol{r}_{zb'}}

\newcommand{\rxytp}{\boldsymbol{r}_{x'y'}}
\newcommand{\rxxtp}{\boldsymbol{r}_{xx'}}
\newcommand{\ryytp}{\boldsymbol{r}_{yy'}}

\newcommand{\rzxpt}{\boldsymbol{r}_{zx'}}

\newcommand{\der}{\mathrm{d}}

\newcommand{\Tr}{\mathrm{Tr}}
\newcommand{\deltatwo}{\delta^{\rm(2)}_z}

\title{Back-to-back inclusive dijets in DIS at small $x$: Gluon Weizs\"{a}cker-Williams distribution at NLO }
\author[a]{Paul Caucal,}
\emailAdd{caucal@subatech.in2p3.fr}
\author[b,c,d,e]{Farid Salazar,}
\emailAdd{salazar@physics.ucla.edu}
\author[f]{Bj\"{o}rn Schenke,}
\emailAdd{bschenke@bnl.gov}
\author[g]{Tomasz Stebel,}
\emailAdd{tomasz.stebel@uj.edu.pl}
\author[f]{Raju Venugopalan}
\emailAdd{raju@bnl.gov}
\affiliation[a]{SUBATECH UMR 6457 (IMT Atlantique, Université de Nantes, IN2P3/CNRS), 4 rue Alfred Kastler,44307 Nantes, France}

\affiliation[b]{Department of Physics and Astronomy, University of California, Los Angeles, California 90095, USA}
\affiliation[c]{Mani L. Bhaumik Institute for Theoretical Physics, University of California, Los Angeles, California 90095, USA}
\affiliation[d]{Nuclear Science Division, Lawrence Berkeley National Laboratory, Berkeley, California 94720, USA}
\affiliation[e]{Physics Department, University of California, Berkeley, California 94720, USA}
\affiliation[f]{Physics Department, Brookhaven National Laboratory, Upton, NY 11973, USA}
\affiliation[g]{Institute of Theoretical Physics, Jagiellonian University, ul. Lojasiewicza 11, 30-348 Krak\'{o}w, Poland}

\abstract{
In \cite{Caucal:2022ulg}, we performed the first complete computation of the back-to-back inclusive dijet cross-section in Deeply Inelastic Scattering (DIS) at small $x_{\rm Bj}$ to next-to-leading order (NLO) in the Color Glass Condensate effective field theory (CGC EFT). We demonstrate here that for dijets with relative transverse momentum 
$P_\perp$ and transverse momentum imbalance $q_\perp$, to leading power in $q_\perp/P_\perp$, the cross-section for longitudinally polarized photons can be fully factorized into the product of a perturbative impact  factor and the nonperturbative Weizs\"{a}cker-Williams (WW) transverse momentum dependent (TMD) gluon distribution to NLO accuracy.  The impact factor can further be expressed as the product of a universal soft factor which resums Sudakov double and single logs in $P_\perp/q_\perp$ and a coefficient function given by a remarkably compact analytic expression. We show that in the CGC EFT the WW TMD satisfies a kinematically constrained JIMWLK renormalization group evolution in rapidity. This factorization formula is valid  to all orders in $Q_s/q_\perp$ for $q_\perp, Q_s \ll P_\perp$, where $Q_s$ is the semi-hard saturation scale that grows with rapidity.
}

\begin{document}
\maketitle
\newpage 

\section{Introduction}

The Electron-Ion Collider (EIC)~\cite{AbdulKhalek:2021gbh} will provide fundamental insight into the spatial and momentum distributions of partons in protons and nuclei. A particular promise of this high energy collider is the likelihood of robust extraction of transverse momentum dependent distributions (TMDs) and generalized parton distributions (GPDs) of gluons, with the latter being sensitive to  spatial impact parameter distributions inside protons and nuclei~\cite{Accardi:2012qut,Aschenauer:2017jsk}. At small $x_{\rm Bj}$, these distributions also provide a window into the onset of gluon saturation, characterized by an energy-dependent emergent saturation scale $Q_s$, larger than intrinsically nonperturbative QCD scales~\cite{Gribov:1984tu,Mueller:1985wy}. 

A particularly intriguing prospect is that of extracting information on the gluon Weizs\"{a}cker-Williams (WW) TMD distribution at small $x_{\rm Bj}$. In light cone gauge, it represents the light cone quantized number distribution of gluons  as a function of their transverse momenta in a proton or nucleus~\cite{McLerran:1993ka,McLerran:1993ni,McLerran:1998nk}. 
It was shown some time ago~\cite{Dominguez:2011wm,Dominguez:2011br} that the leading order (LO) DIS back-to-back dijet cross-section, for dijets with relative transverse momentum 
$P_\perp$ and transverse momentum imbalance $q_\perp$ satisfying $P_\perp \gg q_\perp$, can be expressed as the 
factorized product of a perturbative ``impact factor" and the nonperturbative WW distribution. The WW distribution can be decomposed into two Lorentz structures: a trace component corresponding to the unpolarized WW distribution and a traceless component representing a linearly polarized distribution~\cite{Mulders:2000sh}. A first study of the behavior of the two distributions at LO in the CGC EFT \cite{Iancu:2003xm,Gelis:2010nm,Kovchegov:2012mbw,Albacete:2014fwa,Blaizot:2016qgz,Morreale:2021pnn} within the framework of the McLerran-Venugopalan model suggested that while these distributions behave similarly for $q_\perp > Q_s$, they have qualitatively different behaviour in the saturation regime of $q_\perp \leq Q_s$~\cite{Metz:2011wb,Marquet:2016cgx}. A numerical study of azimuthal moments of 
the LO back-to-back distribution shows that the linearly polarized WW distribution is strongly suppressed for large nuclei in EIC kinematics for low $q_\perp$ but grows rapidly with increasing $q_\perp$~\cite{Dumitru:2015gaa,Dumitru:2016jku,Dumitru:2018kuw,Mantysaari:2019hkq,Boussarie:2021ybe}.

For robust predictions of high energy cross-sections in perturbation theory, it is essential that the computations be performed  to at least next-to-leading order (NLO) accuracy that include corrections of $\mathcal O(\alpha_s)$ to the LO cross-section \cite{Chirilli:2011km,Altinoluk:2014eka,Boussarie:2016bkq,Boussarie:2016ogo,Beuf:2017bpd,Hanninen:2017ddy,Roy:2019cux,Roy:2019hwr,Beuf:2020dxl,Liu:2020mpy,Shi:2021hwx,Caucal:2021ent,Taels:2022tza,Beuf:2021srj,Mantysaari:2022kdm,Liu:2022ijp,Wang:2022zdu,Bergabo:2022tcu,Bergabo:2022zhe,Fucilla:2022wcg,Beuf:2022kyp,vanHameren:2022mtk,Hentschinski:2020tbi,Hentschinski:2021lsh,Ivanov:2012iv,Celiberto:2022fgx}. For back-to-back dijets, such NLO computations are essential to confirm that the aforementioned LO factorization into the impact factor and the WW TMD distribution holds to this accuracy. An additional essential feature of such NLO corrections is that they can be significantly enhanced in certain kinematic regions of the phase space of interest. 

Of particular interest to us are ``slow" NLO real and virtual gluon emissions, with light cone longitudinal momenta $k^+ \ll P^+$, where $P^+$ is the (large) momentum of the target nucleus. Their  contributions are parametrically of magnitude $\alpha_s \ln(1/x)$ ($x= k^+/P^+$) and are therefore $\mathcal{O}(1)$ contributions of the same magnitude as the LO contribution at small momentum fractions $x\ll 1$. Further, there are contributions $\alpha_s^N \ln^N(1/x)\sim 1$ that appear at each $N$th loop order in perturbation theory. Fortunately, the all-order resummation of such leading logarithmic corrections in $x$ (LL$x$) can be performed, ensuring the validity of perturbation theory which would otherwise break down. The renormalization group (RG) equation through which this is achieved is the Balitsky-Fadin-Kuraev-Lipatov (BFKL) equation~\cite{Kuraev:1977fs,Balitsky:1978ic}. In the gluon saturation regime, due to the large power corrections from coherent multiple scattering that are enhanced by the saturation scale, the corresponding LL$x$ equations generalize to the Balitsky-Kovchegov (BK)~\cite{Balitsky:1995ub,Kovchegov:1999yj} and Balitsky-JIMWLK equations~\cite{Balitsky:1995ub,JalilianMarian:1997dw,JalilianMarian:1997jx,Iancu:2000hn,Iancu:2001md}. The state-of-the-art are RG equations that resum next-to-leading-log $\alpha_s^{N+1}\ln^N(1/x)$ contributions (NLL$x$); these are respectively the NLL$x$ BFKL equation~\cite{Fadin:1998py,Ciafaloni:1998gs} and the NLL$x$ JIMWLK equation~\cite{Balitsky:2013fea,Kovner:2014lca,Dai:2022imf}.

These small $x$ resummations are crucial to understand the energy (or rapidity) evolution of expectation values of correlators of Wilson lines that are the fundamental nonperturbative quantities in the CGC EFT; the WW distribution corresponds to one such correlator. In isolating such slow gluon contributions that first appear at NLO, one has to in practice introduce a cutoff $x_f$ to separate the large logs $\alpha_s \ln(x_0/x_f)\sim 1$ (with $x_0$ denoting fast target modes $k^+\sim P^+$) from logarithms 
$\alpha_s \ln(x_f/x_{\rm Bj})$ that are parametrically of order $\mathcal O (\alpha_s)$ that are then absorbed into the impact factor. Indeed it is the required independence of cross-sections from this cutoff that leads to the small $x$ RG equations which resum the small $x$ logs to all orders in perturbation theory. 

Since the small $x$ RG equations are well-known, much of our focus in our previous papers \cite{Caucal:2021ent,Caucal:2022ulg} was on the computation of the NLO impact factor for dijets. For back-to-back dijets, as  discussed in \cite{Caucal:2022ulg}, this computation of the impact factor is complicated by the appearance of so-called Sudakov logarithms that suppress the cross-section~\cite{Catani:1989ne,Catani:1990rp,Catani:1994sq}. At NLO, these are the double logarithms $\alpha_s \ln^2 (P_\perp^2/q_\perp^2)$ and single logarithms $\alpha_s \ln(P_\perp^2/q_\perp^2)$ which can be of $\mathcal O (1)$ when the back-to-back dijet relative transverse momentum $P_\perp \gg q_\perp$. The resummation of such large logarithms in perturbative QCD  is achieved through the Collins-Soper-Sterman (CSS) formalism~\cite{Collins:1981uk,Collins:1981uw,Collins:1984kg} (see also \cite{delCastillo:2020omr,Kang:2020xez}). Numerical estimates of the impact of the resulting Sudakov suppression on semi-inclusive dihadron  and dijet final states in small $x$ kinematics accessible at the EIC  suggest that such contributions are likely important and need to be quantified~\cite{Zheng:2014vka,vanHameren:2021sqc,Zhao:2021kae}.

The first NLO study of the combined contributions from  small $x$ logs (from  modes $x_0 > x\geq x_f$) and the Sudakov logs from the impact factor (for $x_{\rm Bj} < x\leq x_f$)  was performed in \cite{Mueller:2012uf,Mueller:2013wwa} where the authors considered a number of final states in proton-nucleus and electron-nucleus collisions, including back-to-back dijets. In particular, the factorization of double Sudakov logs from small $x$ evolution was discussed in this work. Single Sudakov logs in this context were considered in \cite{Xiao:2017yya,Hatta:2020bgy,Hatta:2021jcd}. In \cite{Caucal:2022ulg}, we exploited the formalism for high order computations in DIS at small $x$ developed in \cite{Roy:2018jxq,Roy:2019hwr,Roy:2019cux}, and applied to inclusive dijets in \cite{Caucal:2021ent}, to compute all real and virtual contributions to inclusive back-to-back  dijets. This yielded all  Sudakov double and single logarithmic contributions (including terms not previously accounted for in the small $x$ literature) as well as all finite $\mathcal O (\alpha_s)$ contributions to the NLO impact factor computed for the first time. 

A key finding of our work in \cite{Caucal:2022ulg} (also deduced simultaneously in \cite{Taels:2022tza} for photo-produced dijets) is that it is essential to impose lifetime ordering in addition to rapidity ordering on small $x$ evolution in order to recover the right sign of the contribution from the Sudakov logarithms. Lifetime ordering was understood to be important previously but only in the context of modifying the kernel of the BFKL renormalization group (RG) equation~\cite{Kuraev:1977fs,Balitsky:1978ic} for fully inclusive cross-sections at next-to-leading logarithmic accuracy as observed for NLL$x$ BFKL~\cite{Fadin:1998py,Ciafaloni:1998gs} in~\cite{Salam:1998tj,Ciafaloni:1999yw,Ciafaloni:2003rd} and for the non-linear JIMWLK RG equation~\cite{JalilianMarian:1997dw,JalilianMarian:1997jx,Iancu:2000hn,Iancu:2001md} in~\cite{Iancu:2015vea,Lappi:2016fmu,Ducloue:2019ezk}.

Significant insight is gained from the azimuthal angle moments $\cos(n\phi)$ where $\phi$ is the angle between $\Pt$ and $\qt$ of the back-to-back cross-section. At leading order, the zeroth moment of the cross-section is proportional to the unpolarized WW TMD distribution while the second moment is proportional to the linearly polarized TMD distribution. (All odd moments vanish due to quark-antiquark symmetry in the cross-section.) At NLO, due to their interplay with soft gluon radiation, both TMDs contribute to all even moments. In \cite{Caucal:2022ulg}, we classified the NLO coefficient function - defined as all the NLO terms in the impact factor not containing Sudakov logarithms -  as arising from two classes of diagrams. One  class of such diagrams contain the contributions of  virtual graphs in the back-to-back cross-section that manifestly satisfied TMD factorization for the correlation limit corresponding to $P_\perp \gg q_\perp$. The other class of diagrams, corresponding to real and virtual diagrams where the gluon crosses the shockwave, appeared to have a complicated color structure different from the WW gluon TMD. When both the correlation limit and the dilute limit ($P_\perp \gg q_\perp \gg Q_s$) are satisfied, TMD factorization is manifestly recovered. However, when 
$q_\perp \sim Q_s$, the color correlators did not appear to reduce to the WW TMDs in the correlation limit and therefore appeared to violate TMD factorization \cite{Caucal:2022ulg,Taels:2022tza}. 

In this paper, we will revisit the computation in  \cite{Caucal:2022ulg} of the NLO coefficient function contributing to the NLO impact factor\footnote{The term containing the exponentiated large Sudakov logs of $P_\perp/q_\perp$, and multiplying the coefficient function in the impact factor, is often called the ``soft factor" interchangeably with ``Sudakov factor".} 
%that do not contain Sudakov logarithms
and demonstrate that TMD factorization can be proven to NLO when $q_\perp, Q_s \ll P_\perp$. 
The terms that previously appeared to violate TMD factorization in the dense saturated regime of $q_\perp \sim Q_s$ (real and virtual diagrams corresponding to gluons crossing the shockwave) are shown explicitly to satisfy TMD factorization. We compute the coefficient functions, including so-called ``hard factors" and other finite $\mathcal{O}(\alpha_s)$ terms, for all the factorized contributions. For the longitudinally polarized cross-section that we focus on in this paper, we obtain a remarkably compact and numerically tractable expressions for the coefficient functions in the factorized NLO result for angular moments $\cos(n\phi)$ of the back-to-back cross-section. 
 
We will present numerical results that demonstrate the relative magnitude of the contributions of the soft factor and those of the coefficient function on the NLO impact factor and their overall magnitude compared to the LO results. 
Detailed numerical studies including their implications for the extraction of WW distributions, and their interplay with the effects arising from gluon saturation,  for {\it both} longitudinally and transversely polarized virtual photon-nucleus back-to-back dijet cross-sections, will be presented separately.

The paper is organized as follows. In Section~\ref{sec:review}, we give a brief summary of the main results of \cite{Caucal:2022ulg} and fix the notations and conventions used throughout the paper. The third and fourth sections are dedicated to the analytic calculation of the back-to-back limit of virtual and real NLO corrections. In Section~\ref{sec:WW-TMD}, we put everything together to  obtain the final result which is given in  Section~\ref{sec:summary1} for the TMD factorized  form of the back-to-back NLO dijet cross-section, with analytic expressions provided for the coefficient function and the soft factor. We also present in Section~\ref{sec:numerics} a preliminary numerical study of the NLO results that demonstrates the relative importance of the different contributions. We end in Section~\ref{sec:Conclusions} with a summary and a discussion of future work. 

The paper is supplemented by four appendices. Appendix\,\ref{app:virtual-details} provides details of the computation of the ``master integrals" that appear in the NLO hard factor from the virtual diagrams where the gluon crosses the shockwave. Appendix\,\ref{app:power-suppressed} provides a mathematical justification of the correlation expansion that shows that higher order terms are suppressed in the dijet back-to-back limit. Appendix\,\ref{app:sudakov} is dedicated to the analytic computation of the Sudakov form factor using the running of the strong coupling at two loops. Finally, appendix\,\ref{app:useful-int} collects some useful integrals.

\section{Review of computation in \cite{Caucal:2022ulg} of inclusive back-to-back dijet cross-section}
\label{sec:review}

In this section, we will summarize the main results of \cite{Caucal:2022ulg} for the  inclusive back-to-back dijet cross-section in DIS at small Bjorken $x_{\rm Bj}$.
This section is structured as follows. After kinematic preliminaries, we will begin with the result for the cross-section in the CGC EFT at LO, which is factorized as the product of the hard LO impact factor and the Weizs\"{a}cker-Williams gluon transverse momentum distribution. We will then discuss our NLO results. We first outline all the real and virtual contributions that appear at NLO and state the results for the Sudakov form factor that are explicitly derived in \cite{Caucal:2022ulg}. A key result is that recovering the right Sudakov factor requires imposing a kinematic lifetime ordering constraint on small $x$ evolution. In addition to the Sudakov terms, there are additional contributions whose structure we will 
outline. In particular, we show that the lifetime ordering constraint is also required for the NLO coefficient function. %that are not accompanied by Sudakov logarithms to NLO accuracy.
The explicit computation of these contributions will be the primary focus of this paper. 

For the DIS process $\gamma_{\lambda}^*+A\to q\bar q+X$,  $\gamma_\lambda^*$ is the virtual photon with polarization $\lambda=\rm L, T$ (longitudinal or transverse) and $q$ and $\bar q$ label the two produced jets (or quark-antiquark pair at the parton level). The four momenta of the two produced jets are labeled $k_1^\mu$ and $k_2^\mu$, and $q^\mu$ is the four-momentum of the virtual photon. We will work in the dipole frame where  the virtual photon propagates along the minus light cone direction while the proton or nucleus $A$, with 4-momentum $P_A^\mu$ propagates along the plus light-cone direction:
\begin{align}
   q^\mu=(-Q^2/(2q^-),q^-,\vect{0}) \,, \quad \quad \quad
   P_A^\mu=(P_A^+,M^2/(2P_A^+),\vect{0}) \,.
\end{align}
Table~\ref{tab:kinematics_dijets} summarizes all the relevant kinematic variables for this process. Throughout the paper, we will work in the high-energy limit  $W^2\gg Q^2, P_\perp^2$.

\begin{table}[h]
\centering
\caption{ Kinematic variables} 
\label{tab:kinematics_dijets}
\vspace{0.2cm}
\begin{tabular}{ll}
\hline \hline
$P_A $ & nucleus four-momentum \\
$P_n $ & nucleon four-momentum \\
$M$, $m_n$ & nucleus and nucleon mass \\
$k_e \ (k_e')$ & incoming (outgoing) electron four-momentum \\ 
$q=k_e-k_e'$ & virtual photon four-momentum \\
$Q^2=-q^2$ & virtuality squared of the exchanged photon\\
$s=(P_n+k_e)^2$ & nucleon-electron center of momentum energy squared  \\
$W^2=(P_n+q)^2$ & nucleon-virtual photon center of momentum energy squared\\
$x_{\rm Bj}=\frac{Q^2}{2P_n\cdot q}=\frac{Q^2}{W^2+Q^2-m_n^2}$ & Bjorken $x$ variable\\
$y=\frac{W^2+Q^2-m_n^2}{s-m_n^2}$ & inelasticity\\
$k_1,k_2$ & quark (antiquark) four-momentum\\ 
$z_{i=1,2}=k_i^-/q^-$ & quark (antiquark) longitudinal momentum fraction w.r.t.\ $q^-$ \\
$\ktone, \kttwo$ &  quark (antiquark) transverse momentum \\
$\eta_{i=1,2}=\ln\left(\frac{\sqrt{2}z_iq^-}{|\boldsymbol{k}_{i\perp}|}\right)$ & quark (antiquark) rapidity \\
$M_{q\bar q}^2=(k_1+k_2)^2$ & invariant mass squared of the dijet system \\
$\Pt=z_2\ktone-z_1\kttwo$ & dijet relative transverse momentum \\
$\qt=\ktone+\kttwo$ & dijet transverse momentum imbalance\\
\hline \hline
\end{tabular}
\end{table}

The full DIS cross-section is determined by convoluting the cross-section for $\gamma^*+A\to q\bar q +X$ with the longitudinal and transverse photon fluxes defined as
\begin{align}
    f_{\lambda=\mathrm{L}}(x_{\rm Bj}, Q^2)&=\frac{\alpha_{\mathrm{em}}}{\pi Q^2 x_{\rm Bj}}(1-y)\,,\\
    f_{\lambda=\mathrm{T}}(x_{\rm Bj}, Q^2)&=\frac{\alpha_{\mathrm{em}}}{2\pi Q^2 x_{\rm Bj}}[1+(1-y)^2]\,,
\end{align}
where $s$ is the center of mass energy squared per nucleon in the $e+A$ collision, $y=Q^2/(s\, x_{\rm Bj})$ denotes the inelasticity (in the limit $s\gg m_n^2$ where $m_n$ is the nucleon mass), and $\alpha_{\rm em}$ is the electromagnetic fine structure constant. For fixed $s$, the $e+A\to e'+q\bar q+X$ cross-section is given by
\begin{equation}
    \frac{\der\sigma^{e+A\to e'+q\bar q+X}}{\der x_{\rm Bj} \der Q^2\der^2\ktone\der^2\kttwo\der\eta_1\der\eta_{2}}=\sum_{\lambda=\mathrm{L,T}}f_{\lambda}(x_{\rm Bj}, Q^2) \ \frac{\der \sigma^{\gamma_{\lambda}^*+A\to q\bar{q}+X}}{ \der^2 \ktone \der^2 \kttwo \der \eta_1 \der \eta_{2}} \,.
\end{equation}
This expression is valid to all orders in the strong coupling constant. It is therefore sufficient to consider only the perturbative expansion in $\alpha_s$ of the $\gamma^*+A\to q\bar q +X$ cross-section. In the NLO discussions in this paper, we will focus on the longitudinally polarized $\gamma^*_{\rm L}$ dijet cross-section alone, the expressions for which are more compact and easier to derive than that of the transversely polarized $\gamma^*_{\rm T}$ dijet cross-section. The results for the transversely polarized photon case, in combination with the results here, will be summarized in a forthcoming paper. 

\subsection{Dijet back-to-back limit at LO}

After a brief overview of the inclusive dijet cross-section in the CGC EFT at leading order, we define the back-to-back kinematics and provide the TMD factorized expression of the CGC formula in this limit.
We refer the reader to Section 2 in \cite{Caucal:2022ulg} for details.

\subsubsection{Leading order cross-section in the CGC EFT}

\begin{center}
    \begin{figure}[H]
    \centering
    \includegraphics[width=0.55\textwidth]{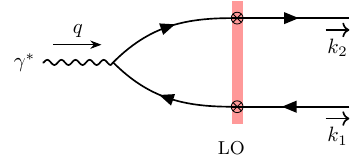}
    \caption{Leading order contribution to the amplitude for dijet production. The cross symbol on the quark and antiquark legs refers to the CGC quark effective vertex. The four-momenta of the quark and antiquark read $k_1 = \left( \ktone^2 / (2 z_1 q^-) , z_1 q^-, \ktone \right) $ and $k_2 = \left( \kttwo^2 / (2 z_2 q^-) , z_2 q^-, \kttwo \right) $ respectively.}
    \label{fig:feyLO}
    \end{figure}
\end{center}

As discussed at length in previous work~\cite{Benic:2016uku,Roy:2018jxq,Roy:2019cux,Caucal:2021ent}, weak coupling computations in the CGC EFT can be performed using standard Feynman diagrams techniques in the light cone gauge $A^-=0$, supplemented by dressed vertices corresponding to the coherent multiple scattering of a quark or a gluon inside the classical (over occupied) gauge fields of the target. These vertices are denoted, respectively, by cross and bullet symbols in Fig.\,\ref{fig:feyLO} and Fig.~\ref{fig:NLO-dijet-all-diagrams}. 

\label{sub:LO}

For the leading order production of a quark-antiquark pair in DIS at small $x_{\rm Bj}$, the only relevant Feynman diagram is the one shown in Fig.\,\ref{fig:feyLO}. After evaluating and squaring the Feynman amplitude associated with diagram~\ref{fig:feyLO}, the fully differential leading order cross-section for inclusive production of two jets in the CGC can be written in the compact form\footnote{For a detailed derivation see e.g. Sec.\,2 in \cite{Caucal:2021ent}.}
\begin{align}
    \left.\frac{\der \sigma^{\gamma_{\lambda}^{*}+A\to q\bar{q}+X}}{ \der^2 \ktone \der^2 \kttwo \der \eta_1 \der \eta_{2}}\right|_{\rm LO}  \!\!\!\!\!\!\! =  \frac{\alpha_{\mathrm{em}} e_f^2 N_c\deltatwo}{(2\pi)^6} \int &\der^8 \Xt e^{-i\ktone\cdot\rxxtp}e^{-i\kttwo\cdot\ryytp}\nonumber\\
    &\times\Xi_{\mathrm{LO}}(\xt,\yt;\xt', \yt')  \Rcal_{\mathrm{LO}}^{\lambda}(\rxyt,\rxytp)\label{eq:dijet-LO-cross-section} \,.
\end{align}
In this expression, $e_f^2$ is the sum of the squares of the light quark fractional charges, and $\deltatwo=\delta(1-z_1-z_2)$ is an overall longitudinal momentum conserving delta function. The differential measure $\der^8 \Xt$ comes from the transverse coordinate integration contained in the CGC vertices and reads
\begin{equation}
    \der^8\Xt=\der^2\xt\der^2\xt'\der^2\yt\der^2\yt' \,,
\end{equation}
with $\xt$ ($\yt$) the transverse coordinate at which the quark (antiquark) crosses the shockwave in the amplitude (and similarly with prime coordinates for the complex conjugate amplitude). Differences of generic transverse coordinates $\at$ and $\bt$ will be labeled $\boldsymbol{r}_{ab}$ with
\begin{equation}
    \boldsymbol{r}_{ab}\equiv\at-\bt\,.
\end{equation}

 The integrand in Eq.\,\eqref{eq:dijet-LO-cross-section} is expressed as the convolution of a perturbative ``impact factor" $\mathcal{R}_{\rm LO}^\lambda$ corresponding to the QED splitting of the virtual photon into the quark-antiquark pair with the color correlator $\Xi_{\rm LO}$ describing the interaction of the pair with the small $x$ gluons of the target. At LO, the explicit expressions for these impact factors are 
\begin{align}
    \Rcal_{\mathrm{LO}}^{\mathrm{L}}(\rxyt,\rxytp) &=  8 z_1^3 z_{2}^3  Q^2 K_0(\bar{Q} r_{xy}) K_0(\bar{Q} r_{x'y'}) \,, \label{eq:dijet-NLO-LLO} \\
    \Rcal_{\mathrm{LO}}^{\mathrm{T}}(\rxyt,\rxytp) &=  2 z_1 z_{2} \left[z_1^2 + z_{2}^2 \right]  \frac{\rxyt \cdot \rxytp}{r_{xy} r_{x'y'}}  \bar{Q}^2K_1(\bar{Q} r_{xy}) K_1(\bar{Q}r_{x'y'})\label{eq:dijet-NLO-TLO} \,,
\end{align}
respectively, for longitudinally and transversely polarized virtual photons. The effective virtuality $\bar Q$ is defined as $\bar Q^2=z_1z_2 Q^2$ and $K_n(x)$ are the modified Bessel functions of the second kind and order $n$.
 
The color correlator $\Xi_{\rm LO}$ is a CGC average of the product of Wilson lines at the  projectile rapidity scale $Y$. Indeed, any observable $\mathcal{O}$ computed in the CGC EFT must be averaged over the large $x$ color charge configurations in the target nucleus as
\begin{equation}
    \langle \mathcal{O}\rangle_{Y}=\int\mathcal{D}[\rho_A]W_{Y}[\rho_A]\mathcal{O}[\rho_A]\,,
\end{equation}
where the color charge configuration $\rho_A$ is drawn from a stochastic gauge invariant weight functional $W_{Y}[\rho_A]$  with $Y=\ln(z)$, for a given typical fraction  $z$ of the projectile $q^-$ momentum corresponding to the kinematics of the observable $\mathcal{O}$.

For inclusive dijet production in DIS, this correlator is given by 
\begin{align}
    \Xi_{\mathrm{LO}}(\xt,\yt;\xt', \yt')&=\frac{1}{N_c}\left\langle\Tr\left[\left(V(\xt)V^\dagger(\yt)-\mathbbm{1}\right)\left(V(\yt')V^\dagger(\xt')-\mathbbm{1}\right)\right]\right\rangle_{Y}\label{eq:xiLO-def}\\
    &=\left \langle Q_{xy,y'x'} - D_{xy} -  D_{y'x'} + 1 \right \rangle_{Y} \,,
\end{align}
where $V(\xt)$, in the fundamental representation of $\mathfrak{su}(3)$ whose generators $t^a$ are the Gell-Mann matrices, is the lightlike path ordered Wilson line
\begin{equation}
    V_{ij}(\xt)=\mathcal{P}\exp\left(ig\int_{-\infty}^{\infty}\der z^-A^{+,a}_{\rm cl}(z^-,\xt)t^a_{ij}\right)\,.
\end{equation}
This Wilson line factor resums to all orders the coherent multiple scatterings between the quark and the small $x$ gluons in the target. The gauge invariant observables  corresponding to these Wilson lines factors are the aforementioned expectation values of the dipole $D$ and quadrupole $Q$ operators:
\begin{align}
 D_{xy}=\frac{1}{N_c}\Tr\left( V(\xt)V^\dagger(\yt)\right) \,,\quad Q_{xy,y'x'}=\frac{1}{N_c}\Tr \left(V(\xt)V^\dagger(\yt)V(\yt')V^\dagger(\xt')\right) \,.
\end{align}

\subsubsection{Back-to-back kinematics}

The back-to-back limit of the LO inclusive DIS dijet cross-section is defined by the hierarchy of scales,
\begin{equation}
    \Pt^2\gg\qt^2\,,
\end{equation}
where $\Pt$ and $\qt$ denote respectively the relative transverse momentum of the dijet system and the dijet transverse momentum imbalance (see Table~\ref{tab:kinematics_dijets}).

The tree-level kinematics for the process $\gamma^*+g\to q\bar q$, where $g$ denotes a gluon in the nucleus wave-function, imposes the following constraint for the plus longitudinal momentum fraction of the gluon relative to $P_n^+$:
\begin{equation}
    x_g=\frac{Q^2+M_{q\bar q}^2+\qt^2}{W^2+Q^2-m_n^2}= x_{\rm Bj}\left(1+\frac{M^2_{q\bar q}}{Q^2}\right)+\mathcal{O}\left(\frac{q_\perp^2}{W^2}\right)\,, \label{eq:xg-dijet-b2b}
\end{equation}
where the second equality is valid in the Regge $W^2\gg Q^2$ . We have also introduced the dijet invariant mass $M_{q\bar q }$ defined as 
\begin{equation}
    M_{q\bar q}^2\equiv(k_1+k_2)^2=\frac{\Pt^2}{z_1z_2} \,.
\end{equation}
At small $x$, as noted previously, the saturation scale $Q_s(x)$ which grows with decreasing $x$ 
emerges as a semi-hard scale that plays an important role in 
the many-body dynamics. A priori, there is no constraint on the magnitude of $Q_s$ relative to $q_\perp$ or $P_\perp$. In this work, we will assume that 
\begin{equation}
    P_\perp \gg Q_s\,,
\end{equation}
 without any assumption on the magnitude of the ratio $Q_s/q_\perp$; saturation effects are large\footnote{In the language of \cite{Altinoluk:2019wyu}, the limit $ P_\perp \gg Q_s$ is referred to as
 neglecting ``genuine" saturation (higher twist) corrections relative to the hard scale. While such contributions are indeed genuine saturation contributions, this terminology can be confusing because saturation plays an equally important role when $q_\perp\lesssim Q_s$. It is responsible for the qualitative change in the behavior of the WW distribution below the saturation scale.} when $q_\perp\lesssim Q_s$. When $q_\perp\gg Q_s$, 
 back-to-back dijets probe the dilute or universal TMD regime which is insensitive to gluon saturation.

\subsubsection{LO TMD factorized expression}

The overlap between the CGC and TMD formalism for back-to-back inclusive dijets was discussed previously in \cite{Dominguez:2010xd,Dominguez:2011wm}. This overlap becomes manifest when one performs an expansion in both $q_\perp/P_\perp$ and $Q_s/P_\perp$ and keeps the leading terms in this double expansion. 
Formally, one initiates this power counting with the change of variables,
\begin{align}
&\ut = \xt-\yt\,,\quad\bt = z_1 \xt+z_{2}\yt \label{eq:LO-b2b-variable}\,,\\
&\der^8\Xt\to \der^8\Xttilde=\der^2\ut\der^2\ut'\der^2\bt\der^2\bt'\,,
\end{align}
in the leading order CGC cross-section:
\begin{align}
    \left.\frac{\der \sigma^{\gamma_{\lambda}^{*}+A\to q\bar{q}+X}}{ \der^2 \Pt \der^2 \qt \der \eta_1 \der \eta_{2}}\right|_{\rm LO}  \!\!\!\!\!\!\! = & \frac{\alpha_{\mathrm{em}} e_f^2 N_c\deltatwo}{(2\pi)^6}    \int \der^8 \Xttilde e^{-i\Pt\cdot\ruupt}e^{-i\qt\cdot\rbbpt}\Rcal_{\mathrm{LO}}^{\lambda}(\ut,\ut')\nonumber\\
    &\times\Xi_{\mathrm{LO}}(\bt+z_{2}\ut,\bt-z_1\ut;\bt'+z_{2}\ut',\bt'-z_1\ut') \,, \label{eq:dijet-LO-rtbt}
\end{align}
followed by an expansion of the color correlator $\Xi_{\rm LO}$ for $u_\perp\ll b_\perp$ and $u_\perp'\ll b_\perp'$:
\begin{equation}
    \Xi_{\mathrm{LO}}=\ut^i\ut'^j\times\frac{1}{N_c}\left\langle\Tr\left[V(\bt)\left(\partial^iV^\dagger(\bt) \right) \left(\partial^jV(\bt')\right)V^\dagger(\bt')\right]\right\rangle_{Y}+\mathcal{O}\left(u_\perp^2u_\perp',u_\perp u_\perp'^2\right) \,.
    \label{eq:correlation_limit_expansion}
\end{equation}
 After  the necessary Fourier transform is performed, each additional power of $\ut$ or $\ut'$ brings an extra power of $1/P_\perp$ times a transverse scale of order $q_\perp$ or $Q_s$. The latter appears via the additional transverse coordinate derivative of a Wilson line in the Taylor expansion of $\Xi_{\rm LO}$. Therefore the truncation in Eq.\eqref{eq:correlation_limit_expansion} of the expansion of the color correlator captures the leading term in the limit $P_\perp\gg q_\perp, Q_s$ and the order of magnitude of the neglected terms is $\mathcal{O}(\textrm{max}(q_\perp,Q_s)/P_\perp)$. 

The operator in the leading term in this power counting is nothing other than the small $x$ Weizs\"{a}cker-Williams distribution $\hat G^{ij}_{Y}(\bt,\bt')$ \cite{Boussarie:2021ybe}: 
\begin{align}
    \hat G^{ij}_{Y}(\bt,\bt')&\equiv\frac{-2}{\alpha_s}\left\langle\Tr\left[V(\bt) \left(\partial^iV^\dagger(\bt) \right) V(\bt') \left(\partial^jV^\dagger(\bt') \right)\right]\right\rangle_{Y}\,,
    \label{eq:WWTMD}
\end{align}
whose Fourier transform is the WW TMD distribution:
\begin{align}
    G^{ij}_{Y}(\qt)&=\int\frac{\der^2\bt\der^2\bt'}{(2\pi)^4}e^{-i\qt\cdot(\bt-\bt')}\hat G^{ij}_{Y}(\bt,\bt')\,. 
\end{align}

One can rewrite the LO inclusive dijet DIS cross-section in back-to-back kinematics as the factorized expression\footnote{Henceforth, we
will write  $\left.\frac{\der \sigma^{\gamma_{\lambda}^{*}+A\to q\bar{q}+X}}{ \der^2 \Pt \der^2 \qt \der \eta_1 \der \eta_{2}}\right|\rightarrow \der \sigma^{\gamma_{\lambda}^{*}+A\to q\bar{q}+X}$, to avoid carrying around the differential phase space factor.}
\begin{align}
    \der \sigma^{\gamma_{\lambda}^{*}+A\to q\bar{q}+X}_{\rm LO} &=  \alpha_{\rm em}e_f^2\alpha_s\deltatwo\mathcal{H}_{\rm LO}^{\lambda,ij}(\Pt) \int\frac{\der^2\rbbpt}{(2\pi)^4}e^{-i\qt\cdot\rbbpt}\hat G^{ij}_{Y}(\bt,\bt')+\mathcal{O}\left(\frac{q_\perp}{P_\perp},\frac{Q_s}{P_\perp}\right) \,.
    \label{eq:diff_xsec_TMD_LO}
\end{align}
Here the LO coefficient function in this case can be expressed in terms of a ``hard factor" $\Hcal_{\rm LO}^{\lambda,ij}$, defined to be
\begin{equation}
    \mathcal{H}_{\rm LO}^{\lambda,ij}(\Pt)=\frac{1}{2}\int\frac{\der^2\ut}{(2\pi)}\frac{\der^2\ut'}{(2\pi)}e^{-i\Pt\cdot\ruupt}\ut^i\ut'^j\Rcal_{\rm LO}^\lambda(\ut,\ut')\label{eq:hard-factor-def} \,.
\end{equation}
A straightforward calculation using Bessel identities gives 
\begin{align}
\Hcal_{\rm LO}^{\lambda=\textrm{L},ij}(\Pt)&=16z_1^3z_2^3Q^2\frac{\Pt^i\Pt^j}{(\Pt^2+\bar Q^2)^4}\,,\\
\Hcal_{\rm LO}^{\lambda=\textrm{T},ij}(\Pt)&=z_1z_2(z_1^2+z_2^2)\left\{\frac{\delta^{ij}}{(\Pt^2+\bar Q^2)^2}-\frac{4\bar Q^2\Pt^i\Pt^j}{(\Pt^2+\bar Q^2)^4}\right\}\,,
\end{align}
for the leading order hard factors, respectively for the longitudinally and transversely polarized virtual photon cross-sections.

A further simplification occurs when one decomposes this distribution further into a trace part (the unpolarized WW gluon TMD $G^0_Y(\qt)$) and a traceless piece (the linearly polarized WW TMD $ h^0_Y(\qt)$) as 
\begin{equation}
    G^{ij}_Y(\qt)=\frac{1}{2}\delta^{ij}G^0_Y(\qt)+\frac{1}{2}\left[\frac{2\qt^i\qt^j}{\qt^2}-\delta^{ij}\right]h^0_Y(\qt) \,.
\end{equation}
Substituting the above expressions back into Eq.~\eqref{eq:dijet-LO-rtbt} 
then gives the very simple factorized expressions for the LO longitudinally and transversely polarized back-to-back dijet cross-sections, respectively: 
\begin{align}
     \left.\frac{\der \sigma^{\gamma_{\rm L}^{*}+A\to q\bar{q}+X}}{ \der^2 \Pt \der^2 \qt \der \eta_1 \der \eta_{2}}\right|_{\rm LO}  \!\!\!\!\!\!\! &=  \alpha_{\rm em}e_f^2\alpha_s\deltatwo\times\frac{8(z_1z_2)^3Q^2\Pt^2}{(\Pt^2+\bar Q^2)^4}\nonumber\\
     &\times\left[G^0_{Y}(\qt)+\cos(2\phi)h^0_{Y}(\qt)\right]\,,\label{eq:LOb2b-L}\\
     \left.\frac{\der \sigma^{\gamma_{\rm T}^{*}+A\to q\bar{q}+X}}{ \der^2 \Pt \der^2 \qt \der \eta_1 \der \eta_{2}}\right|_{\rm LO}  \!\!\!\!\!\!\! &=  \alpha_{\rm em}e_f^2\alpha_s\deltatwo \times z_1 z_2 (z_1^2+z_2^2)\frac{\Pt^4+\bar Q^4}{(\Pt^2+\bar Q^2)^4}\nonumber\\     &\times\left[G^0_{Y}(\qt)-\frac{2\bar Q^2\Pt^2}{\Pt^4+\bar Q^4}\cos(2\phi)h^0_{Y}(\qt)\right]\,.\label{eq:LOb2b-T}
\end{align}
By inspection, the azimuthal angle average of Eqs.\,\eqref{eq:LOb2b-L} and \eqref{eq:LOb2b-T} is only proportional to the unpolarized WW TMD. Conversely, the $\langle \cos(2\phi)\rangle$ moment of these cross-sections (the elliptic anisotropy) is proportional to the ratio of linearly polarized to the unpolarized distribution.  As we will discuss, soft gluon radiation at NLO  quantitatively modifies these LO results. 

For future reference, the LO hard factor will appear in the NLO cross-section in the form of the trace 
\begin{align}
    \Hcal_{\rm LO}^{0,\lambda}(\Pt)\equiv\frac{1}{2}\Hcal_{\rm LO}^{\lambda,ii}(\Pt)=\left\{
    \begin{array}{ll}
        \frac{8(z_1z_2)^3Q^2\Pt^2}{(\Pt^2+\bar Q^2)^4} & \mbox{ for }\lambda=\rm L\\
       z_1 z_2 (z_1^2+z_2^2)\frac{\Pt^4+\bar Q^4}{(\Pt^2+\bar Q^2)^4} & \mbox{ for }\lambda=\rm T \,. 
    \end{array}
\right. 
\end{align}
Due to the large gluon occupancy at small $x$, the WW gluon distribution $G^{ij}_Y(\qt)$ is parametrically of order $\mathcal{O}(1/\alpha_s)$ \cite{Iancu:2003xm}. Thus the LO cross-section is parametrically $\mathcal{O}(1)$, and the NLO corrections are of order $\mathcal{O}(\alpha_s)$.

\subsection{Back-to-back limit at NLO: Sudakov suppression and kinematically constrained small $x$ evolution}
\label{sub:b2b-NLO}

\begin{figure}[tbh]
    \centering
    \includegraphics[width=1\textwidth]{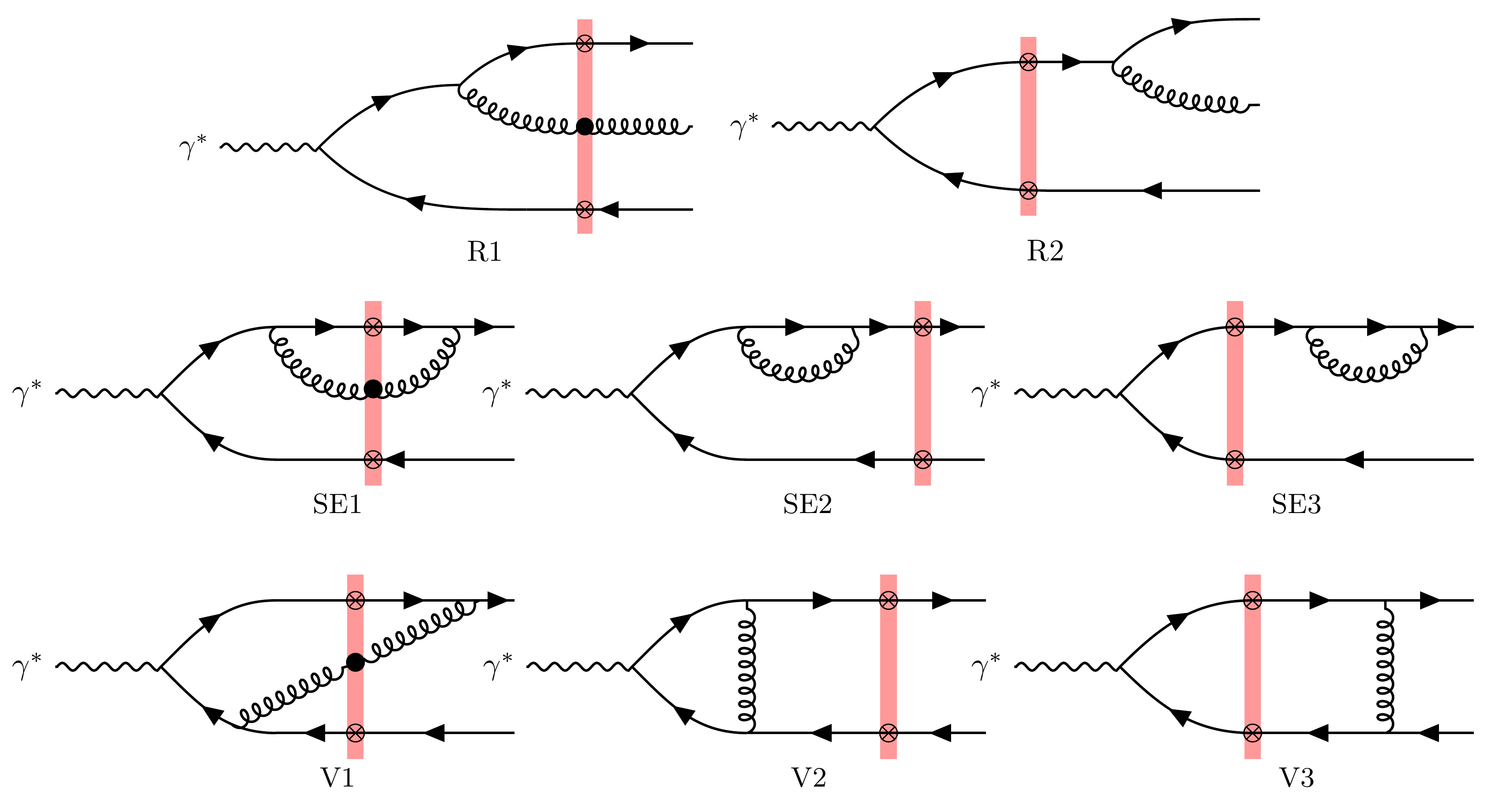}
    \caption{Feynman diagrams that contribute to the inclusive dijet cross-section at NLO. Top: Real gluon emission diagrams. Middle: Self energy  diagrams. Bottom: Vertex correction diagrams. Diagrams obtained from $q \leftrightarrow \bar{q}$ interchange are not shown, and will be labeled with an additional prime index. (For example, $\rm R_2\rightarrow \rm R_2^\prime$.) Only diagrams in which the gluon does not scatter off the shockwave (represented by the red band) contribute to the Sudakov double and single logarithms.}
    \label{fig:NLO-dijet-all-diagrams}
\end{figure}

The NLO inclusive dijet cross-section at NLO in general kinematics was first computed in the CGC EFT in \cite{Caucal:2021ent}. 
All the Feynman diagrams that contribute at this order are shown in Fig.~\ref{fig:NLO-dijet-all-diagrams}
At NLO, there is an additional CGC vertex (represented by the bullet symbol)  associated with the coherent multiple scattering scattering of the radiated gluon off the shockwave~\cite{McLerran:1994vd,Ayala:1995hx,Belitsky:2002sm,Roy:2018jxq}. In order to obtain an infrared and collinear finite result, one has to introduce jets. The jet definitions that we consider throughout this paper are generalized $k_t$ algorithms, which all give the same NLO impact factor up to powers of $R^2$ corrections, with $R$ the jet radius~\cite{Salam:2010nqg}.

In \cite{Caucal:2022ulg}, we computed the back-to-back limit of our NLO results in \cite{Caucal:2021ent} and expressed the results in terms of the
moments of the azimuthal angle $\phi$ between between $\Pt$ and $\qt$. 
For the zeroth moment of the distribution 
(the angle averaged cross-section), we obtained the result (up to terms of order $\mathcal{O}(R^2)$),
\begin{align}
    \der \sigma^{(0),\lambda=\rm L}&=\alpha_{\rm em}\alpha_s e_f^2\deltatwo\Hcal_{\rm LO}^{0,\lambda=\rm L}(\Pt)\int\frac{\der^2\rbbpt}{(2\pi)^4}e^{-i\qt\cdot\rbbpt}\hat G^0_{Y_f}(\rbbpt)\mathcal{S}(\Pt^2,\rbbpt^2) \nonumber\\
    \times&\left\{1+\frac{\alpha_sC_F}{\pi}\left[\frac{3}{2}\ln(c_0^2)-3\ln(R)+\frac{1}{2}\ln^2\left(\frac{z_1}{z_2}\right)+\frac{17}{2}-\frac{5\pi^2}{6}\right]\right.\nonumber\\
    &+\frac{\alpha_sN_c}{2\pi}\left[\ln\left(\frac{z_f^ 2}{z_1z_2}\right)\ln(c_0^2)-\ln^2\left(\frac{Q_f^2c_0^2}{\Pt^2}\right)\right]\,\nonumber\\
    &\left.+\frac{\alpha_sN_c}{2\pi}\left[\frac{1}{2}\ln\left(\frac{z_1z_2}{z_f^2}\right)-\frac{3C_F}{2N_c}\right]\frac{\Hcal_{\rm NLO,1}^{\lambda=\textrm {L},ii}(\Pt)}{2\Hcal_{\rm LO}^{0,\lambda=\rm L}(\Pt)}-\frac{\alpha_s}{2\pi N_c}\frac{\Hcal_{\rm NLO,2}^{\lambda=\textrm {L},ii}(\Pt)}{2\Hcal_{\rm LO}^{0,\lambda=\rm L}(\Pt)}\right\}\nonumber\\
    &+\alpha_{\rm em}\alpha_s e_f^2\deltatwo\Hcal_{\rm LO}^{0,\lambda=\rm L}(\Pt)\int\frac{\der^2\rbbpt}{(2\pi)^4}e^{-i\qt\cdot\rbbpt}\hat h^0_{Y_f}(\rbbpt)\mathcal{S}(\Pt^2,\rbbpt^2)\nonumber\\
    &\times\frac{\alpha_sN_c}{2\pi}\left\{1+\frac{2C_F}{N_c}\ln(R^2)-\frac{1}{N_c^2}\ln(z_1z_2)\right\}\nonumber\\
    &+\der \sigma^{(0),\lambda=\rm L}_{\rm other}\,.
    \label{eq:zeroth-moment-final}
\end{align}

For the second moment of the distribution\footnote{As noted previously, odd moments vanish.}, we obtained 
\begin{align}
    \der \sigma^{(2),\lambda=\rm L}&=\alpha_{\rm em}\alpha_s e_f^2\deltatwo\Hcal_{\rm LO}^{0,\lambda=\rm L}(\Pt)\int\frac{\der^2\rbbpt}{(2\pi)^4}e^{-i\qt\cdot\rbbpt}\frac{\cos(2\theta)}{2}\hat h^0_{Y_f}(\rbbpt)\mathcal{S}(\Pt^2,\rbbpt^2) \nonumber\\
    \times&\left\{1+\frac{\alpha_sC_F}{\pi}\left[\frac{3}{2}\ln(c_0^2)-4\ln(R)+\frac{1}{2}\ln^2\left(\frac{z_1}{z_2}\right)+\frac{17}{2}-\frac{5\pi^2}{6}\right]\right.\nonumber\\
    &+\frac{\alpha_sN_c}{2\pi}\left[-\frac{5}{4}+\ln\left(\frac{z_f^ 2}{z_1z_2}\right)\ln(c_0^2)-\ln^2\left(\frac{Q_f^2c_0^2}{\Pt^2}\right)\right]+\frac{\alpha_s}{4\pi N_c}\ln(z_1 z_2)\,\nonumber\\
    &\left.+\frac{\alpha_sN_c}{2\pi}\left[\frac{1}{2}\ln\left(\frac{z_1z_2}{z_f^2}\right)-\frac{3C_F}{2N_c}\right]\frac{\Hcal_{\rm NLO,1}^{\lambda=\textrm {L},ii}(\Pt)}{2\Hcal_{\rm LO}^{0,\lambda=\rm L}(\Pt)}-\frac{\alpha_s}{2\pi N_c}\frac{\Hcal_{\rm NLO,2}^{\lambda=\textrm {L},ii}(\Pt)}{2\Hcal_{\rm LO}^{0,\lambda=\rm L}(\Pt)}\right\}\nonumber\\
    &+\alpha_{\rm em}\alpha_s e_f^2\deltatwo\Hcal_{\rm LO}^{0,\lambda=\rm L}(\Pt)\int\frac{\der^2\rbbpt}{(2\pi)^4}e^{-i\qt\cdot\rbbpt}\frac{\cos(2\theta)}{2}\hat G^0_{Y_f}(\rbbpt)\mathcal{S}(\Pt^2,\rbbpt^2)\nonumber\\
    &\times\frac{\alpha_sN_c}{\pi}\left\{1+\frac{2C_F}{N_c}\ln(R^2)-\frac{1}{N_c^2}\ln(z_1z_2)\right\}\nonumber\\
    &+\der \sigma^{(2),\lambda=\rm L}_{\rm other}\,.
    \label{eq:second-moment-final}
\end{align}
Results for the higher moments, and for transversely polarized photons were also provided in \cite{Caucal:2022ulg} in the appendices. Final results for transversely polarized photons follow analogously to those that we will obtain here; as noted earlier, the combined results will be discussed in a subsequent paper. 

We will show in this paper that the expressions provided above in Eqs.~\eqref{eq:zeroth-moment-final} and \eqref{eq:second-moment-final} 
can be greatly simplified enabling one to numerical evaluate with relative ease. However before we proceed to do so in the following sections, it will be useful to first explain key features of the above results.

\paragraph{Sudakov form factor.} 
The function $\mathcal{S}(\Pt^2,\rbbpt^2)$ is the Sudakov or ``soft" form factor. At LO in the coupling, this quantity is unity 
and one recovers in the above expressions the result for the zeroth and second azimuthal moment of  Eq.~\eqref{eq:LOb2b-L}, respectively, in terms of the unpolarized and linearly polarized WW TMD distributions. 

At NLO, the function contains the Sudakov double and single logarithms derived explicitly in \cite{Caucal:2022ulg}, which give $\mathcal O(1)$ contributions when $P_\perp \gg q_\perp$. Assuming  CSS evolution 
holds at small $x$, the all order contribution of the resummation of these  Sudakov logarithms 
is given by 
\begin{equation}
   \mathcal{S}(\Pt^2,\rbbpt^2)=\exp\left(-\int_{c_0^2/\rbbpt^2}^{\Pt^2}\frac{\der\mu^2}{\mu^2}\frac{\alpha_s(\mu^2)N_c}{\pi}\left[\frac{1}{2}\ln\left(\frac{\Pt^2}{\mu^2}\right)+\frac{C_F}{N_c}s_0-s_f\right]\right) \,.
   \label{eq:CSS}
\end{equation}
The coefficients of the single Sudakov $\log$ that we extracted in \cite{Caucal:2022ulg} are given by
\begin{equation}
    s_0 = \ln\left(\frac{1}{z_1z_2R^2}\right)=\ln\left(\frac{2(1+\cosh(\Delta \eta_{12}))}{R^2}\right)\,,\quad    s_f=\ln\left(\frac{z_f\Pt^2}{z_1z_2c_0^2Q_f^2}\right)\,,
\label{eq:s0}
\end{equation}
up to powers of $R^2$ corrections.
Here $\Delta\eta_{12}\equiv\eta_1-\eta_2$, and $z_f,Q_f$ are factorization scales associated with the rapidity evolution of the WW gluon TMD discussed below.

\paragraph{Kinematically constrained rapidity evolution.}
An important result from the study in \cite{Caucal:2022ulg} is that consistent resummation of Sudakov logarithms to all orders requires 
that one employs a kinematically improved small $x$ renormalization group  (RG) evolution of the WW gluon TMD. This  RG equation reads, in integral form,
\begin{align}
&G^{ij}_{Y_f}(\bt,\bt')-G^{ij}_{Y_0}(\bt,\bt')\nonumber\\
&=\frac{\alpha_sN_c}{2\pi^2}\int_{Y_0}^{Y_f}\der Y\int\der^2\zt\Theta\left(Y_f-Y-\ln\left(\textrm{min}(\rzbt^2,\rzbpt^2)Q_f^2\right)\right)\mathbf{K}_{\rm DMMX}\otimes G^{ij}_{Y}(\bt,\bt')\,,
\label{wwe}
\end{align}
with 
\begin{equation}
Y_0=\ln(k_{\rm min}^-/q^-)=\ln(z_0)\,,\quad Y_f=\ln(k_f^-/q^-)=\ln(z_f)\,,
\end{equation}
respectively, the initial rapidity of the evolution and the rapidity factorization scale. The transverse momentum scale $Q_f^2$ should be taken of order $\textrm{max}(\Pt^2,Q^2)$ but was chosen to be arbitrary in  \cite{Caucal:2022ulg}. We will express $Q_f$ in this paper in terms of the $z_f$ and the kinematics of the dijet system. 

The kernel of the rapidity evolution of the WW gluon TMD (which we shall denote henceforth as $\mathbf{K}_{\rm DMMX}$) was computed explicitly from the Balitsky-JIMWLK hierarchy 
in \cite{Dominguez:2011gc}. The notation ``$\otimes$" is meant to remind the reader of the specific application of the universal JIMWLK kernel to the LO WW correlator, which generates additional correlators which are not WW-like. In other words, as shown in \cite{Dominguez:2011gc}, the rapidity evolution of the WW TMD does not have a closed form in the dense saturation regime. The main difference of our result with respect to the evolution found in  \cite{Dominguez:2011gc} lies in the $\Theta$-function term 
\begin{equation}
\Theta\left(Y_f-Y-\ln\left(\textrm{min}(\rzbt^2,\rzbpt^2)Q_f^2\right)\right)\,,
\label{eq:kin-constraint}
\end{equation}
which enforces lifetime (or $k_g^+$) ordering of successive gluon emissions from the projectile, on top of the usual projectile momentum $k_g^-$ ordering. Hence in the computation of Eqs.\,\eqref{eq:zeroth-moment-final} and \eqref{eq:second-moment-final}, the WW gluon TMDs should be evolved employing this kinematically improved RG evolution equation.

\paragraph{NLO coefficient function.} The virtual diagrams $\rm SE_2$, $\rm V_2$ and $\rm V_3$ (plus their $q\leftrightarrow \bar q$ counterparts) -- in which the gluon loop does not interact with the shockwave -- generate NLO hard factors that are distinct from $\Hcal^{\lambda,ij}_{\rm LO}$ and are denoted $\Hcal_{\rm NLO,1}$ and $\Hcal_{\rm NLO,2}$ in the formula Eq.\,\eqref{eq:zeroth-moment-final}. Exact expressions for these are given in the next section; their explicit computation is a key result of this paper. Other finite terms of $\mathcal{O}(\alpha_s)$ (not containing Sudakov or rapidity logs) are also present and given explicitly in Eqs.\,\eqref{eq:zeroth-moment-final} and \eqref{eq:second-moment-final}. They come from unresolved real gluon emission inside jet cones or soft gluon radiation outside the jet cones and are discussed further in Section~\ref{sub:real-soft}. In particular, the latter induces a sensitivity of the zeroth (second) moment to the linearly polarized (unpolarized) WW TMD which is a pure NLO effect, as shown in the fifth line of Eq.\,\eqref{eq:zeroth-moment-final} (Eq.~\eqref{eq:second-moment-final}). All of these terms, namely, NLO hard factors and finite terms from real gluon emissions, together comprise the NLO coefficient function.

\paragraph{The ``other" contribution.} The last term in Eq.\,\eqref{eq:zeroth-moment-final} labeled 
\begin{equation}
\der \sigma^{(0),\lambda=\rm L}_{\rm other}\,,
\end{equation}
is the contribution to the back-to-back dijet production of all the diagrams in which the emitted gluon interacts with the shockwave. These terms are particularly difficult to compute because they depend on a number of other color correlators besides the WW correlator. This is because a naive correlation expansion similar to Eq.\,\eqref{eq:correlation_limit_expansion} does not yield the WW gluon TMD. However, we will see that there exists an expansion for these contributions which correctly accounts for the leading power back-to-back limit and  where the leading term is the WW gluon TMD in coordinate space. The explicit calculation of this ``other" contribution is another key result of this paper. This is because it establishes the highly nontrivial result that the NLO back-to-back dijet cross-section, for 
$P_\perp \gg q_\perp, Q_s$ can be fully factorized into a NLO impact factor and the WW gluon TMD, providing a complete NLO proof of TMD factorization for this process. More precisely, the ``other" contribution also introduce new hard factors, labeled $\Hcal_{\rm NLO,3}$ and $\Hcal_{\rm NLO,4}$ in the next section, that contribute to the NLO coefficient function.

\section{NLO hard factors from virtual diagrams}
\label{sec:NLO-hard}
This section is dedicated to the calculation of the hard factors $\Hcal_{\rm NLO,1}$, $\Hcal_{\rm NLO,2}$ and those  that we will see emerge from the virtual diagrams $\rm SE_1$ and $\rm V_1$ included in the $\der \sigma^{(0),\lambda=\rm L}_{\rm other}$ term\footnote{A similar analysis can be carried out for the term $\der \sigma^{(2),\lambda=\rm L}_{\rm other}$ in Eq.\,\eqref{eq:second-moment-final}. We will not work this term out explicitly here but only quote the final result along with that for Eq.~\eqref{eq:zeroth-moment-final}.} of Eq.\,\eqref{eq:zeroth-moment-final}. 
We will obtain fully analytic expressions for these hard factors in terms of logarithms and dilogarithms of the ratio $Q/M_{q\bar q}$. The analytic results simplify considerably the otherwise challenging numerical computation of the small $x$ back-to back dijet cross-section at NLO.

On a more conceptual level, the calculation of the back-to-back limit of diagrams $\rm SE_1$ and $\rm V_1$ demonstrates that the full virtual cross-section can be factorized into a perturbative hard factor and the WW gluon TMD in back-to-back kinematics, which in combination with the results of the next section, shows factorization of the full NLO cross-section. As previously noted, this is a highly nontrivial result. The hard factors for the virtual diagrams in which the gluon scatters off the shockwave are meaningful if and only if the rapidity divergence that is subtracted is kinematically constrained. Thus this section provides supplementary evidence (to that provided by the ``right sign" of the Sudakov double logs \cite{Caucal:2022ulg,Taels:2022tza}) of the necessity to go beyond naive small $x$ evolution and to use the kinematic constrained evolution  discussed in the previous section. Further, our computation here enables us  to fix the arbitrary scale $Q_f$ in our kinematically improved rapidity evolution equation for the WW TMD, and therefore complete our calculation of the Sudakov single log coefficients in \cite{Caucal:2022ulg}.

\subsection{Final state vertex correction}

We begin with the hard factor $\Hcal_{\rm NLO,2}$ arising from the final state vertex diagram $\rm V_3$ in Fig.\,\ref{fig:NLO-dijet-all-diagrams}. The leading term in the correlation expansion of the CGC color correlator in this diagram gives the WW gluon TMD.
 In the back-to-back limit, the $\rm V_3\times LO$ (and its complex conjugate) contribution to the cross-section, with 
 its rapidity divergence subtracted,  factorizes as 
 \cite{Caucal:2022ulg}:
\begin{align}
\alpha_{\rm em}\alpha_se_f^2\deltatwo\left[\frac{(-\alpha_s)}{2\pi N_c}\Hcal_{\rm NLO,2}^{\lambda=\textrm {L},ij}(\Pt)\right]\int\frac{\der^2 \rbbpt}{(2\pi)^4}e^{-i\qt\cdot\rbbpt}\hat G^{ij}_{Y_f}(\rbbpt) \,.
\end{align}
One notes that its contribution is suppressed at large $N_c$. Thus this diagram does not contribute in the  large $N_c$ approximation.

The hard factor $\Hcal_{\rm NLO,2}^{\lambda=\textrm {L},ij}(\Pt)$ associated with this diagram reads \cite{Caucal:2022ulg}
\begin{align}
    &\Hcal_{\rm NLO,2}^{\lambda=\textrm{L},ij}(\Pt)\equiv   \frac{1}{2}\int\frac{\der^2\ut}{(2\pi)}\int\frac{\der^2\ut'}{(2\pi)}e^{-i\Pt\cdot\ruupt}\ut^i \ut'^j \Rcal^{\rm \lambda=\rm L}_{\rm LO}(\ut,\ut')\nonumber\\
    &\times \int_0^{z_1}\frac{\der z_g}{z_g}\left\{\frac{K_0(\bar Q_{\rm V3}u_\perp)}{K_0(\bar Qu_\perp)}\left[\left(1-\frac{z_g}{z_1}\right)^2\left(1+\frac{z_g}{z_2}\right)(1+z_g)e^{i\Pt\cdot\ut}K_0(-i\Delta_{\rm V3}u_\perp)\right.\right.\nonumber\\
    &\left.-\left(1-\frac{z_g}{z_1}\right)\left(1+\frac{z_g}{z_2}\right)\left(1-\frac{z_g}{2z_1}+\frac{z_g}{2z_2}-\frac{z_g^2}{2z_1z_2}\right)e^{i\frac{z_g}{z_1}\Pt\cdot\ut}\Jcal_{\odot}\left(\ut,\left(1-\frac{z_g}{z_1}\right)\Pt,\Delta_{\rm V3}\right)\right]\nonumber\\
    &\left.+\ln\left(\frac{z_g P_\perp u_\perp}{c_0z_1z_2}\right)\right\}+(1\leftrightarrow2) +c.c.\,,
    \label{eq:Hard_NLO2}
\end{align}
with 
\begin{align}
\bar Q_{\rm V3}^2&=z_1z_2(1-z_g/z_1)(1+z_g/z_2)Q^2\,,\label{eq:QV3}\\
\Delta_{\rm V3}^2&=(1-z_g/z_1)(1+z_g/z_2)\Pt^2\,.\label{eq:DV3}
\end{align}
In \cite{Caucal:2021ent}, we were unable to find a closed form analytic expression for the diagram $\rm V_3$ and our final result was expressed in terms of the function $\Jcal_{\odot}$ defined by
\begin{equation}
\Jcal_{\odot}(\rt,\Kt,\Delta)=\int\frac{\der^2 \lt}{(2\pi)}\frac{2\lt \cdot \Kt \ e^{i\lt \cdot \rt}}{\lt^2\left[(\lt-\Kt)^2-\Delta^2 - i \epsilon\right]}\,.
\label{eq:tough-integral}
\end{equation}
The main result of this section will be the explicit analytic evaluation of $\Hcal_{\rm NLO,2}^{\lambda=\textrm{L},ij}(\Pt)$.
As discussed in the Appendix J of \cite{Caucal:2021ent}, the $-i\epsilon$ prescription in the denominator, coming from the Feynman propagator of the quark or antiquark prior to the emission of the virtual gluon, is important to specify the integration contour since the denominator has a pole on the positive real axis of $|\lt-\Kt|$ when $|\lt-\Kt| = \Delta$.

In the formula for $\Hcal_{\rm NLO,2}^{\lambda=\textrm{L},ij}$, the last term proportional to 
\begin{equation}
    \ln\left(\frac{z_gP_\perp u_\perp}{c_0z_1z_2}\right)\,,\label{eq:slow-log}
\end{equation}
is the subtraction term of the leading $z_g\to0$ divergence of the integral, which ensures that the integrand inside the curly bracket is finite. We postpone the calculation of this term to the end of this subsection since its form can be inferred from its defining property of regularizing the slow ($z_g\to 0$) divergence.

Let us move now to the more complicated pieces. First, given the form of $\Rcal^{\lambda=\rm L}_{\rm LO}(\ut,\ut')$, one can obviously factorize the $\ut$ integral from the $\ut'$ one. The latter gives (cf Eq.\,\eqref{eq:app-LO-vector}):
\begin{align}
\int\frac{\der^2\ut'}{(2\pi)} e^{i\Pt\cdot\ut'}\ut'^jK_0(\bar Qu_\perp')&=\frac{2i\Pt^j}{(\Pt^2+\bar Q^2)^2}\,.
\end{align}
In the $\ut$ integral, the term proportional to $K_0(-i\Delta_{\rm V,3}u_\perp)$ does not contribute because the phase proportional to $\Pt\cdot\ut$ cancels and one is left with an integral which only depends on $\ut^i$ times a function of $\ut^2$. We are thus left with the computation of the integral
\begin{equation}
I_1^i\equiv\int\frac{\der^2 \ut}{(2\pi)}e^{-i\Pt\cdot\ut}\ut^i K _0(\bar Q_{\rm V3}\ut')e^{i\frac{z_g}{z_1}\Pt\cdot\ut}\Jcal_{\odot}\left(\ut,\left(1-\frac{z_g}{z_1}\right)\Pt,\Delta_{\rm V3}\right)\,,\label{eq:Iint-def}
\end{equation}
with 
\begin{align}
    \Hcal_{\rm NLO,2}^{\lambda=\textrm{L},ij}(\Pt)&=8z_1^3z_2^3Q^2\frac{i\Pt^j}{(\Pt^2+\bar Q^2)^2}\int_0^{z_1}\frac{\der z_g}{z_g}\left\{\left(1-\frac{z_g}{z_1}\right)\left(1+\frac{z_g}{z_2}\right)\right.\nonumber\\
    &\left.\times\left(1-\frac{z_g}{2z_1}+\frac{z_g}{2z_2}-\frac{z_g^2}{2z_1z_2}\right)(-I_1^i)-(z_g\to 0)\right\}+c.c.+(1\leftrightarrow2) \,,
\end{align}
where the $(z_g\to 0)$ term is the slow subtraction term given in   Eq.\,\eqref{eq:slow-log}.
As outlined in the beginning of this section, the integral $I_1^i$ can be computed analytically, even though $\Jcal_{\odot}$ does not have a closed analytic expression. This calculation is detailed in Appendix~\ref{app:NLO2}. In terms of $\Delta_{\rm V3}$, $\bar Q_{\rm V3}$ and $\Kt$ defined by Eqs.\,\eqref{eq:QV3}, \eqref{eq:DV3} and Eq.\,\eqref{eq:Kt-def}, our final result is
\begin{align}
I_1^i=-2i\Kt^i&\left\{\frac{1}{(\Delta_{\rm V3}^2+\bar Q_{\rm V3}^2)(\Kt^2+\bar Q_{\rm V3}^2)}+\frac{1}{2(\Delta_{\rm V3}^2+\bar Q_{\rm V3}^2)^2}\left[\ln\left(\frac{\Delta_{\rm V3}^2-\Kt^2}{\Kt^2+\bar Q_{\rm V3}^2}\right)\right.\right.\nonumber\\
&\left.\left.+\frac{\Delta_{\rm V3}^2}{\Kt^2}\ln\left(\frac{\bar Q_{\rm V3}^2(\Delta_{\rm V3}^2-\Kt^2)}{\Delta_{\rm V3}^2(\Kt^2+\bar Q_{\rm V3}^2)}\right)\right]-\frac{i\pi}{2(\Delta_{\rm V3}^2+\bar Q_{\rm V3}^2)^2}\right\}\,.\label{eq:IV3-final}
\end{align}
The imaginary part of the integral, which comes from the $-i\epsilon$ prescription in the contour integration for $\Jcal_{\odot}$, does not contribute to the total cross-section since the complex conjugate term is added to the hard factor in Eq.\,\eqref{eq:Hard_NLO2}. In principle, the ``$c.c.$" term should be added at the cross-section level and not in the hard factor. It is alright to do so since the Fourier transform of the WW gluon TMD is real. If the Fourier transform of the WW gluon TMD were instead complex, this imaginary part of $I_1^i$ would have contributed to the total cross-section.

\paragraph{Subtracting the slow gluon divergence.} The final step in the calculation of $\Hcal_{\rm NLO,2}^{\lambda=\textrm{L},ij}$ is to perform the remaining $z_g$ integral over the virtual gluon longitudinal momentum. As explained above, this integral is convergent provided the $z_g\to 0$ logarithmic divergence is subtracted off. In principle, one should then compute the subtraction term given by
\begin{align}
    \Hcal^{\lambda=\rmL,ij}_{\rm NLO,2,\rm slow}\equiv -8z_1^3z_2^3Q^2\frac{i\Pt^j}{(\Pt^2+\bar Q^2)^2}\int\frac{\der^2\ut}{(2\pi)}e^{-i\Pt\ut}\ut^i K_0(\bar Q u_\perp)\ln\left(\frac{z_g P_\perp u_\perp}{c_0z_1z_2}\right)\,.
\end{align}
Even though the $\ut$ integral can be performed, one can actually infer the final result from the $z_g\to 0$ limit of Eq.\,\eqref{eq:IV3-final}.
Inserting the values of $\Kt$, $\Delta_{\rm V3}$ and $\bar Q_{\rm V3}$, we find that
\begin{equation}
I_1^i=\frac{-2i\Pt^i}{(\Pt^2+\bar Q^2)^2}\left[1+\ln\left(\frac{z_g}{z_1z_2}\right)+\ln\left(\frac{\bar QP_\perp}{\Pt^2+\bar Q^2}\right)\right]+\mathcal{O}(z_g)\,.
\end{equation}
Hence, since by definition $\Hcal^{\lambda=\rmL,ij}_{\rm NLO,2,slow}$ is equal to the leading slow divergence of $\Hcal^{\lambda=\rmL,ij}_{\rm NLO,2}$, we have
\begin{equation}
    \Hcal^{\lambda=\rmL,ij}_{\rm NLO,2, slow}(\Pt)=-16 z_1^3z_2^3Q^2\frac{\Pt^i\Pt^j}{(\Pt^2+\bar Q^2)^4}\left[1+\ln\left(\frac{z_g}{z_1z_2}\right)+\ln\left(\frac{\bar QP_\perp}{\Pt^2+\bar Q^2}\right)\right]\,.\label{eq:HNLO2-slow}
\end{equation}
With this subtraction term, the hard factor can be expressed as a single integral over $z_g$:
\begin{align}
    \Hcal^{\lambda=\rmL,ij}_{\rm NLO,2}(\Pt)&=16z_1^3z_2^3\frac{Q^2\Pt^i\Pt^j}{(\Pt^2+\bar Q^2)^4}\times\int_0^{z_1}\frac{\der z_g}{z_g}\left\{2+2\ln\left(\frac{z_g}{z_1z_2}\right)-\frac{1}{2}\left(-\frac{z_g}{z_1}+\frac{2z_2+z_g}{z_2+z_g}\right)\right.\nonumber\\
&\times\left[\frac{2z_1(z_2+z_g)(1+\chi^2)}{z_2(z_1-z_g)+z_1(z_2+z_g)\chi^2}+\frac{2z_1(z_2+z_g)\ln(\chi)}{z_2(z_1-z_g)}+\frac{z_2(z_1-z_g)+z_1(z_2+z_g)}{z_2(z_1-z_g)}\right.\nonumber\\
&\left.\left.\times\ln\left(\frac{z_g}{z_2(z_1-z_g)+z_1(z_2+z_g)\chi^2}\right)\right]-2\ln\left(\chi+\frac{1}{\chi}\right)\right\}+(1\leftrightarrow 2)\,,
\end{align}
with the variable $\chi$ defined as
\begin{equation}
    \chi\equiv\frac{\bar Q}{P_\perp}=\frac{Q}{M_{q\bar q}}\,.
\end{equation}
The integral over $z_g$ is now perfectly regular. After a rather tedious calculation, the $z_g$ integral can be expressed in terms of logarithms and dilogarithms as 
\begin{align}
\Hcal_{\rm NLO,2}^{\lambda=\rmL,ij}(\Pt)&=\Hcal_{\rm LO}^{\lambda=\rmL,ij}(\Pt)\times\left\{\frac{\pi^2}{12}-\ln(\chi)-\ln(z_2)\ln(\chi)-\ln(z_2)\ln\left(\frac{z_1}{z_2}\right)\right.\nonumber\\
&+\frac{(z_1-3z_2)z_2+(1+8z_1z_2)\chi^2+z_1(z_2-3z_1)\chi^4}{2(z_2-\chi^2z_1)^2}\ln\left(\frac{z_2(1+\chi^2)}{\chi^2}\right)\nonumber\\
&+\frac{1}{2}\textrm{Li}_2\left(\frac{z_2-z_1\chi^2}{z_2}\right)+\frac{1}{2}\textrm{Li}_2\left(-\frac{z_1}{z_2}\right)-\frac{1}{2}\textrm{Li}_2\left(z_2-z_1\chi^2\right)-\frac{1}{2}\textrm{Li}_2\left(\frac{z_1\chi^2-z_2}{\chi^2}\right)\nonumber\\
&\left.-2\textrm{Li}_2\left(\frac{z_2-\chi^2z_1}{(1+\chi^2)z_2}\right)-\frac{(1+\chi^2)z_1}{z_2-\chi^2z_1}+(1\leftrightarrow 2)\right\}\,,
\label{eq:HNLO_final}
\end{align}
where the dilogarithm or Spence's function is defined by
\begin{align}
    \textrm{Li}_2(x)=-\int_0^x\frac{\ln(1-z)}{z}\der z\,,
\end{align}
for $x\le 1$.
Note that Eq.\,\eqref{eq:HNLO_final} is finite when $z_2-z_1\chi^2=0$ thanks to the cancellation of the single pole (which relies on the identity $z_1+z_2=1$), and that the arguments in the Spence's function are always smaller than 1.

\subsection{Initial state self-energy and vertex corrections}
\label{sub:V2-SE2}

The virtual diagrams $\rm SE_2$ and $\rm V_2$ also contribute to the back-to-back limit through another hard factor labeled $\Hcal^{\lambda=\rmL,ij}_{\rm NLO,1}(\Pt)$ in Eq.\,\eqref{eq:zeroth-moment-final}, whose definition is~\cite{Caucal:2022ulg}:
\begin{equation}
    \Hcal_{\rm NLO,1}^{\lambda=\textrm{L},ij}(\Pt)\equiv\frac{1}{2}\int\frac{\der^2\ut}{(2\pi)}\int\frac{\der^2\ut'}{(2\pi)}e^{-i\Pt\cdot\ruupt}\ut^i\ut'^j\Rcal_{\rm LO}^\lambda(\ut,\ut')\ln(\Pt^4\ut^2\ut'^2)\,.\label{eq:hard-NLO1}
\end{equation}
To compute this hard factor, one notices that it is related to $\Hcal^{\lambda=\rmL,ij}_{\rm NLO,2, slow}(\Pt)$ since one has
\begin{equation}
    \Hcal_{\rm NLO,1}^{\lambda=\textrm{L},ij}(\Pt)=-4 \Hcal^{\lambda=\rmL,ij}_{\rm NLO,2, slow}(\Pt)+4\ln\left(\frac{z_g}{c_0z_1z_2}\right)\Hcal_{\rm LO}^{\lambda=\rmL,ij}(\Pt)\,.
    \label{eq:NLO1-0}
\end{equation}
Using the result Eq.\,\eqref{eq:HNLO2-slow}, one gets the simple expression
\begin{equation}
    \Hcal_{\rm NLO,1}^{\lambda=\textrm{L},ij}(\Pt)=\Hcal_{\rm LO}^{\lambda=\rmL,ij}(\Pt)\times\left\{4-4\ln\left(\frac{\chi}{c_0}\right)-4\ln\left(1+\frac{1}{\chi^2}\right)\right\}\,,\label{eq:NLO1-final}
\end{equation}
where we recall that $\chi=Q/M_{q\bar q}$.
We have obtained this expression based on our knowledge of the slow gluon divergence of $\Hcal_{\rm NLO,2}^{\lambda=\textrm{L},ij}(\Pt)$ but we note that one can also obtain it by a direct computation of the integral, as shown in Appendix~\ref{app:NLO1}.

\subsection{Self-energy crossing the shockwave}

We turn now to the discussion of the back-to-back limit of virtual diagrams in which the gluon scatters off the shockwave, namely the self-energy $\rm SE_1$ and the vertex correction $\rm V_1$ diagrams. Their respective contributions are included in the term $\der \sigma^{(0),\lambda=\rm L}_{\rm other}$ because they depend on color correlators which do not ``naturally" reduce to the WW gluon TMD. However, we will see upon closer inspection that their back-to-back limit does depend only on the WW gluon TMD, again up to $q_\perp/P_\perp$ power corrections.

We first consider diagram $\rm SE_1$, the expression for which was obtained in \cite{Caucal:2021ent}. For a longitudinally polarized photon, the diagram $\rm SE_1$ times the complex conjugate of the leading order diagram reads
\begin{align}
    \der\sigma^{\lambda=\rm L}_{\rm SE_1}&= \frac{\alpha_{\rm em}e_f^2N_c\deltatwo}{(2\pi)^6} \int\der^8\Xt e^{-i\ktone\cdot\rxxtp-i\kttwo\ryytp} 8z_1^3z_{2}^3Q^2K_0(\bar Qr_{x'y'})\nonumber\\
    &\times \frac{\alpha_s}{\pi}\int_0^{z_1}\frac{\der z_g}{z_g}\left(1-\frac{z_g}{z_1}+\frac{z_g^2}{2z_1^2}\right)\nonumber\\
    & \times \int\frac{\der^2\zt}{\pi} \frac{1}{\rzxt^2}\left[e^{-i\frac{z_g}{z_1}\ktone \cdot \rzxt} K_0(QX_{\rm V}) \Xi_{\rm NLO,1}-e^{-\frac{\rzxt^2}{\rxyt^2e^{\gamma_E}}}K_0(\bar Qr_{xy}) C_F\Xi_{\rm LO}\right]\nonumber\\
    &-\der\sigma^{\lambda=\rm L}_{\rm SE_1,slow} \,,\label{eq:SE1-full} 
\end{align}
where the color correlator $\Xi_{\rm NLO,1}$ is  given by
\begin{align}
   \Xi_{\rm NLO,1}&=\frac{1}{N_c}\left \langle \textrm{Tr}[t^aV(\xt)V^\dagger(\zt)t_aV(\zt)V^\dagger(\yt)-C_F][V(\yt')V^\dagger(\xt')-1] \right\rangle_Y\,.\label{eq:Xi-NLO1}
\end{align}
Note that this contribution to the cross-section depends on the integral over the transverse coordinate $\zt$ representing the position where the gluon crosses the shockwave. The formula for $\rm SE_1$ also involves the $q\bar qg$ effective dipole size $X_{\rm V}$ defined by
\begin{equation}
    X_{\rm V}^2=z_{2}(z_1-z_g)\rxyt^2+z_g(z_1-z_g)\rzxt^2+z_{2}z_g\rzyt^2\,.
\end{equation}
The square bracket in Eq.~\eqref{eq:SE1-full} contains two terms whose difference regulates the UV divergence $\rzxt\to0$ of the self-energy. This counterterm is discussed at length in \cite{Caucal:2021ent}.
Finally, the slow divergence subtraction term regularizes the $z_g\to 0$ divergence of the $z_g$ integral and will be discussed towards the end of this subsection.

\begin{figure}[H]
    \centering
    \includegraphics[width=0.8\textwidth]{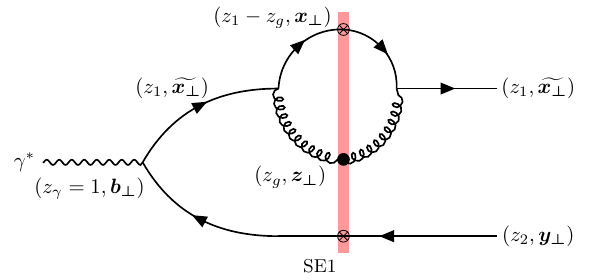}
    \caption{A diagrammatic representation motivating the change of variables in Eq.\,\eqref{eq:SE1-b2b-variable}. $\widetilde{\xt}$ represents the transverse location of the quark after reabsorbing the gluon (and also before emitting it), in contrast to $\xt$ which is the transverse position of the quark after emitting the gluon. $\bt$ represents the transverse location of the photon. $\ut= \widetilde{\xt} - \yt$ (not shown) is the relative vector between quark and anti-quark.}
    \label{fig:virtualSE_crossingSW}
\end{figure}

In order to obtain the back-to-back limit of the expression Eq.\,\eqref{eq:SE1-full}, one must absorb the phase 
\begin{equation}
    e^{-i\frac{z_g}{z_1}\ktone\cdot\rzxt}\,,\label{eq:SE1-phase}
\end{equation}
into an overall phase factor that resembles the leading order one
\begin{equation}
    \der^8\Xttilde e^{-i\Pt\cdot\ruupt-i\qt\rbbpt}\,,
\end{equation}
with a suitable change of variables. One possible such change of variables is
\begin{align}
    \ut &=\left(1-\frac{z_g}{z_1}\right)\xt-\yt+\frac{z_g}{z_1}\zt \,,\nonumber \\
    \bt&=(z_1-z_g)\xt+z_2\yt+z_g\zt \,, \nonumber \\
    \rt&=\left(1-\frac{z_g}{z_1}\right)(\zt-\xt) \,.\label{eq:SE1-b2b-variable}
\end{align}
There is a natural physical interpretation for this change of variables in terms of the transverse location of the parton in the diagram. The transverse coordinate $\ut$ is the difference between the transverse locations of the quark after reabsorbing the gluon (and also before emitting it) $\widetilde{\xt}=(1-z_g/z_1)\xt+z_g/z_1\zt$ and the transverse location of the antiquark at the position $\yt$ where it interacts with the shockwave. In a similar fashion, the transverse location of the photon $\bt$ can be expressed as $\bt=z_1\widetilde{\xt}+z_2\yt$ as in the leading order case Eqs.\,\eqref{eq:LO-b2b-variable}. Finally, the variable $\rt$ corresponds, up to the factor $1-z_g/z_1$, to the transverse separation between the quark and the gluon at the shockwave (see Fig.\,\ref{fig:virtualSE_crossingSW}). 

Performing the change of variables given by Eqs.\,\eqref{eq:SE1-b2b-variable} in  Eq.\,\eqref{eq:SE1-full} (with $\ut'$ and $\bt'$ defined as in the LO case Eq.\,\eqref{eq:LO-b2b-variable} since the transverse coordinates in the complex conjugate are those of the LO diagram) we get
\begin{align}
   &\der\sigma^{\lambda=\rm L}_{\rm SE_1}= \frac{\alpha_{\rm em}e_f^2N_c\deltatwo}{(2\pi)^6} \int\der^8\Xttilde e^{-i\Pt\cdot\ruupt-i\qt\rbbpt} 8z_1^3z_{2}^3Q^2K_0(\bar Qu_\perp') \nonumber\\
   &\times \int_0^{z_1}\frac{\der z_g}{z_g}\left(1-\frac{z_g}{z_1}+\frac{z_g^2}{2z_1^2}\right)  \nonumber\\
    & \times \frac{\alpha_s}{\pi}\int\frac{\der^2\rt}{\pi}\frac{1}{\rt^2}\left[ K_0\left(\bar Q\sqrt{\ut^2+\omega\rt^2}\right)\widetilde\Xi_{\rm NLO,1}- e^{-\frac{\rt^2}{\ut^2e^{\gamma_E}}}K_0(\bar Qu_\perp) C_F\Xi_{\rm LO} \right]\nonumber\\
    &-\der\sigma^{\lambda=\rm L}_{\rm SE_1,slow}\,,\label{eq:SE1-new-variables}
\end{align}
where $\widetilde{\Xi}_{\rm NLO,1}$ represents the color correlator $\Xi_{\rm NLO,1}$ expressed as a function of $\ut,\bt,\rt,\ut'$ and $\bt'$.
The variable $\omega$ appearing in one of the Bessel functions is 
defined to be 
\begin{equation}
    \omega = \frac{z_g}{z_2(z_1-z_g)}\,.
\end{equation}
Notice that in the UV subtraction term, we have used the LO change of variables given by Eqs.\,\eqref{eq:LO-b2b-variable} as this term is not multiplied by the phase Eq.\,\eqref{eq:SE1-phase}. 

\paragraph{Correlation expansion of $\widetilde\Xi_{\rm NLO,1}$.} In Eq.\,\eqref{eq:SE1-new-variables}, it is not yet manifest that the back-to-back limit $P_\perp\gg q_\perp$ of diagram $\rm SE_1$ only depends on the WW gluon TMD since Eq.\,\eqref{eq:SE1-new-variables} depends on the color correlator $\Xi_{\rm NLO,1}$. To see this, one must evaluate $\widetilde{\Xi}_{\rm NLO,1}$ as a function of $\ut$, $\bt$ and $\rt$ using
\begin{align}
    \xt&=\bt+z_2\ut-\frac{z_g}{z_1-z_g}\rt\,, \nonumber \\
    \yt&=\bt-z_1\ut\,, \nonumber \\
    \zt&=\bt+z_2\ut+\rt \,.
\end{align}
Expanding $\widetilde{\Xi}_{\rm NLO,1}$, to get the leading power in $q_\perp/P_\perp$ or $Q_s/P_\perp$, it is sufficient to expand the correlator to leading order in $\ut$, $\ut'$ and $\rt$. In Appendix\,\ref{app:power-suppressed} we show that any higher order terms in the Taylor series are power suppressed in the ratio $q_\perp/P_\perp$ or $Q_s/P_\perp$. Hence even though one integrates over $\rt$, the back-to-back limit of the diagram $\rm SE_1$ is controlled by values of $\rt$ such that $r_\perp\sim u_\perp\sim 1/P_\perp$ - namely, gluons with large transverse momentum. The Taylor series expansion of the color correlator $\widetilde{\Xi}_{\rm NLO,1}$ around $\bt$ (dubbed the ``correlation expansion" throughout this paper) gives to first finite order,
\begin{align}
    \widetilde{\Xi}_{\rm NLO,1}
    &=\left[C_F\ut^i\ut'^j+\left(\frac{N_c}{2}+\frac{1}{2N_c}\frac{z_g}{(z_1-z_g)}\right)\rt^i\ut'^j\right]\nonumber\\
    &\times\frac{1}{N_c}\left\langle\textrm{Tr} V(\bt)\partial^iV^\dagger(\bt)\partial^j V(\bt')V^\dagger(\bt')\right\rangle_Y\,.\label{eq:NLO1-b2b}
\end{align}
One recognizes the WW color correlator in the second line. Hence as conjectured, the product of the diagram $\rm SE_1$ with the LO diagram can be factorized into a hard factor and the WW gluon TMD:
\begin{align}
    \der\sigma^{\lambda=\rm L}_{\rm SE_1}=\alpha_{\rm em}\alpha_se_f^2\deltatwo\left[\frac{\alpha_sC_F}{\pi}\Hcal_{\rm NLO,3}^{\lambda=\rmL,ij}(\Pt)\right]\int\frac{\der^2\rbbpt}{(2\pi)^4}e^{-i\qt\cdot\rbbpt}\hat{G}^{ij}_{Y_f}(\rbbpt)+\mathcal{O}\left(\frac{q_\perp}{P_\perp},\frac{Q_s}{P_\perp}\right) \,,
\end{align}
with the hard factor $\rm NLO,3$ defined as
\begin{align}
    \Hcal_{\rm NLO,3}^{\lambda=\rmL,ij}(\Pt)&\equiv8z_1^2z_2^3Q^2\int\frac{\der^2\ut}{(2\pi)}\frac{\der^2\ut'}{(2\pi)} \ut^i\ut'^ je^{-i\Pt\cdot\ruupt}K_0\left(\bar Q u_\perp'\right)\int_0^{z_1}\frac{\der z_g}{z_g}\left(1-\frac{z_g}{z_1}+\frac{z_g^2}{2z_1^2}\right)\nonumber\\
    &\times\int\frac{\der^2\rt}{(2\pi)}\frac{1}{\rt^2}\left[ K_0\left(\bar Q\sqrt{\ut^2+\omega\rt^2}\right)-e^{-\frac{\rt^2}{\ut^2e^{\gamma_E}}}K_0(\bar Qu_\perp)\right]\nonumber\\
    &-\Hcal_{\rm NLO,3,slow}^{\lambda=\rmL,ij}(\Pt)\,.\label{eq:HNLO3-def}
\end{align}
The term proportional to $\rt^i\ut'^j$ that appears in  Eq.\,\eqref{eq:NLO1-b2b} does not contribute since it gives an integral of the form 
\begin{equation}
    \int\frac{\der^2\rt}{(2\pi)}\rt^i f(\rt^2)=0\,,
\end{equation}
that vanishes for any function $f$. The rapidity subtraction term labeled  $\Hcal_{\rm NLO,3,slow}^{\lambda=\rmL,ij}(\Pt)$ ensures that the $z_g$ integral is integrable near $z_g=0$ and is obtained by the same expansion of the color correlator in the rapidity subtraction term $\der\sigma^{\lambda=\rm L}_{\rm SE_1,slow}$. As we shall see, this subtraction term is given by the kinematically constrained slow gluon limit of diagram $\rm SE_1$.

\paragraph{Calculation of $\Hcal_{\rm NLO,3}^{\lambda=\rmL,ij}(\Pt)$.} 
Similarly to the cases of the $\rm NLO,1$ and $\rm NLO,2$ hard factors, the hard factor associated with $\rm SE_1$ can be computed analytically. Since the correlation expansion assumes $u_\perp\sim r_\perp$, one cannot make any approximation in the transverse coordinate integral over $\rt$ so we have to evaluate it as is. The $\ut'$ integral is straightforward from Eq.\,\eqref{eq:app-LO-vector}:
\begin{align}
    &\Hcal_{\rm NLO,3}^{\lambda=\rmL,ij}(\Pt)=16z_1^2z_2^3Q^2\frac{(-\Pt^j)}{\Pt^2+\bar Q^2}\int_0^{z_1}\frac{\der z_g}{z_g}\left(1-\frac{z_g}{z_1}+\frac{z_g^2}{2z_1^2}\right)\partial^i_{\Pt}I_2-\Hcal_{\rm NLO,3,slow}^{\lambda=\rmL,ij}(\Pt)\,,
\end{align}
where the scalar integral $I_2$ is defined by
\begin{align}
    I_2&\equiv\int\frac{\der^2\ut}{(2\pi)}e^{-i\Pt\cdot\ut}\int\frac{\der^2\rt}{(2\pi)}\frac{1}{\rt^2}\left[ K_0\left(\bar Q\sqrt{\ut^2+\omega\rt^2}\right)-e^{-\frac{\rt^2}{\ut^2e^{\gamma_E}}}K_0(\bar Qu_\perp)\right] \label{eq:Lint-def}\\
    &=\frac{1}{2(\Pt^2+\bar Q^2)}\ln\left(\frac{\Pt^2+\bar Q^2}{c_0\omega\bar Q^2}\right)\,.
\end{align}
Details of this result are given in Appendix~\ref{app:NLO3}. We quote here the final result for $\Hcal_{\rm NLO,3}^{\lambda=\rmL,ij}$ before integrating over $z_g$, which reads
\begin{align}
\Hcal_{\rm NLO,3}^{\lambda=\rmL,ij}(\Pt)&=16z_1^2z_2^3Q^2\frac{\Pt^i\Pt^j}{(\Pt^2+\bar Q^2)^4}\int_0^{z_1}\frac{\der z_g}{z_g}\left(1-\frac{z_g}{z_1}+\frac{z_g^2}{2z_1^2}\right)\nonumber\\
    &\times\left[-1+\ln\left(\frac{1+\chi^2}{\chi^2}\right)-\ln\left(\frac{z_g}{z_2(z_1-z_g)}\right)\right]-\Hcal_{\rm NLO,3,slow}^{\lambda=\rmL,ij}(\Pt) \,,
    \label{eq:HNLO3-int}
\end{align}
with $\chi=Q/M_{q\bar q}$, as previously.

\paragraph{Subtracting the slow gluon divergence.} The $z_g$ integral is double logarithmically divergent as $z_g\to 0$, but 
this divergence should be cured by the slow subtraction term $\der\sigma^{\lambda=\rm L}_{\rm SE_1,slow}$ in Eq.\,\eqref{eq:SE1-new-variables} that we have not specified so far. 

A first attempt would be to use the subtraction term used in \cite{Caucal:2021ent} that enables one to obtain the first step in the JIMWLK evolution of the LO cross-section between $z_0$ and $z_f$. This naive subtraction term is given by setting $z_g=0$ everywhere but in the $1/z_g$ prefactor. One then gets:
\begin{align}
   &\der\sigma^{\lambda=\rm L}_{\rm SE_1, slow}= \frac{\alpha_{\rm em}e_f^2N_c\deltatwo}{(2\pi)^6} \int\der^8\Xttilde e^{-i\Pt\cdot\ruupt-i\qt\rbbpt} 8z_1^3z_{2}^3Q^2K_0(\bar Qu_\perp') \nonumber\\
   &\times\frac{\alpha_s}{\pi} \int_{z_0}^{z_f}\frac{\der z_g}{z_g}  \int\frac{\der^2\rt}{\pi}\frac{1}{\rt^2}\left[ K_0\left(\bar Qu_\perp\right)\widetilde\Xi_{\rm NLO,1}- e^{-\frac{\rt^2}{\ut^2e^{\gamma_E}}}K_0(\bar Qu_\perp) C_F\Xi_{\rm LO} \right]\,,
\end{align}
with 
\begin{equation}
    \widetilde\Xi_{\rm NLO,1}=\Xi_{\rm NLO,1}(\bt+z_2\ut,\bt-z_1\ut,\bt+z_2\ut+\rt)\,.
\end{equation}
Expanding $\widetilde\Xi_{\rm NLO,1}$ as above, we get for the hard factor 
\begin{align}
   \Hcal_{\rm NLO,3,slow}^{\lambda=\rmL,ij}(\Pt)&= \int\frac{\der^2\ut}{(2\pi)}\frac{\der^2\ut'}{(2\pi)} e^{-i\Pt\cdot\ruupt} 8z_1^3z_{2}^3Q^2\ut^i\ut'^jK_0(\bar Qu_\perp')K_0\left(\bar Qu_\perp\right) \nonumber\\
   &\times \int_0^{z_f}\frac{\der z_g}{z_g}  \int\frac{\der^2\rt}{\pi}\frac{1}{\rt^2}\left[ 1- e^{-\frac{\rt^2}{\ut^2e^{\gamma_E}}} \right]\,.\label{eq:SE1-naive-slow}
\end{align}
This naive subtraction term is problematic for two reasons. Firstly, it results in the integral
\begin{equation}
    \int\frac{\der^2\rt}{(2\pi)}\frac{1}{\rt^2}\left[1-e^{-\frac{\rt^2}{\ut^2e^{\gamma_E}}}\right]\,,
\end{equation}
which is IR divergent when $r_\perp$ goes to infinity. It means that one is not allowed to expand the color correlator $\widetilde\Xi_{\rm NLO,1}$ inside the $\rt$ integral. Secondly, even if we assume that the $\rt$ integral is naturally cut by a scale of order say $1/Q_s$, the subtraction term would be still proportional to a \textit{single} logarithm of the $z_0$ cut-off, which prevents the complete cancellation of the double logarithmic divergence in Eq.\,\eqref{eq:HNLO3-int}.

The solution to this problem was already outlined in Section~\ref{sub:b2b-NLO} and points independently to the kinematically constrained rapidity evolution of the WW gluon TMD discussed there. Thus not only is the kinematic constraint crucial to obtain the correct sign for the double logarithmic Sudakov suppression of the back-to-back dijet cross-section as we showed previously in \cite{Caucal:2022ulg},  it is also mandatory to obtain a well defined hard factor for the virtual graphs where the gluon scatters off the shockwave. This is a nontrivial consistency check of our results for back-to-back dijets in the CGC, but hardly a surprise. This is because the original motivation of the existence of a  kinematic constraint followed from examining the spacetime structure of the  diagrams $\rm SE_1$ and $\rm V_1$ as $z_g\to 0$ \cite{Caucal:2022ulg}.
(See  \cite{Beuf:2014uia} for similar arguments in the context of fully inclusive DIS.) 

Indeed, for the NLO contributions in which the gluon crosses the shockwave, $z_g\propto \omega$ and $\rt$ are coupled in the NLO light-cone wavefunction:
\begin{equation}
    K_0\left(\bar Q\sqrt{\ut^2+\omega \rt^2}\right)\,.
\end{equation}
Thus, in order to recover the expected LO wavefunction $K_0(\bar Q u_\perp)$ in the slow gluon limit $z_g \to 0$, we must require the auxiliary condition:
\begin{equation}
    z_g \rt^2\ll z_1 z_2 \ut^2 \sim \frac{z_1z_2}{\Pt^2}\,,\label{eq:heuristic-kin-const}
\end{equation}
which we refer to as the kinematic constraint. 

In more concrete and mathematical terms, the constraint Eq.\,\eqref{eq:kin-constraint} applied to the rapidity evolution of the WW gluon TMD translates into the improved subtraction term for the diagram $\rm SE_1$:
\begin{align}
   &\der\sigma^{\lambda=\rm L}_{\rm SE_1, slow}= \frac{\alpha_{\rm em}e_f^2N_c\deltatwo}{(2\pi)^6} \int\der^8\Xttilde e^{-i\Pt\cdot\ruupt-i\qt\rbbpt} 8z_1^3z_{2}^3Q^2K_0(\bar Qu_\perp')\frac{\alpha_s}{\pi} \nonumber\\
   &\times \int_{z_0}^{z_f}\frac{\der z_g}{z_g}  \int\frac{\der^2\rt}{\pi}\frac{1}{\rt^2}\left[ K_0\left(\bar Qu_\perp\right)\widetilde\Xi_{\rm NLO,1}\Theta\left(\frac{z_f}{z_gQ_f^2}-\rt^2\right)- e^{-\frac{\rt^2}{\ut^2e^{\gamma_E}}}K_0(\bar Qu_\perp) C_F\Xi_{\rm LO} \right]\,,\label{eq:SE1-slow}
\end{align}
where $Q_f$ is \textit{a priori} an arbitrary transverse momentum scale of order $\Pt$, based on the argument leading to Eq.\,\eqref{eq:heuristic-kin-const}.
Expanding $\widetilde\Xi_{\rm NLO,1}$ to first non trivial order to extract its component proportional to the WW gluon TMD, we get the subtraction term of the hard factor $\Hcal^{\lambda=\rmL,ij}_{\rm NLO,3}$:
\begin{align}
   \Hcal_{\rm NLO,3,slow}^{\lambda=\rmL,ij}(\Pt)&=\int\frac{\der^2\ut}{(2\pi)}\frac{\der^2\ut'}{(2\pi)} 8z_1^3z_{2}^3Q^2\ut^i\ut'^jK_0(\bar Qu_\perp')K_0\left(\bar Qu_\perp\right) \nonumber\\
   &\times \int_0^{z_f}\frac{\der z_g}{z_g}  \int\frac{\der^2\rt}{\pi}\frac{1}{\rt^2}\left[\Theta\left(\frac{z_f}{z_gQ_f^2}-\rt^2\right)- e^{-\frac{\rt^2}{\ut^2e^{\gamma_E}}} \right]\,.\label{eq:HNLO3-slow}
\end{align}
The integral over $\rt$ is now convergent both in the UV and in the infrared:
\begin{equation}
    \int\frac{\der^2\rt}{(2\pi)}\frac{1}{\rt^2}\left[\Theta\left(\frac{z_f}{z_gQ_f^2}-\rt^2\right)-e^{-\frac{\rt^2}{\ut^2e^{\gamma_E}}}\right]=\frac{1}{2}\ln\left(\frac{z_f\Pt^2}{Q_f^2 z_g}\right)-\frac{1}{2}\ln(\Pt^2\ut^2) \,.
\end{equation}
The resulting integral over $\ut$ of the second term in this expression is given by Eq.\,\eqref{eq:app-LO-vector-ln} in the appendix. At the end of the day,  we get
\begin{align}
\Hcal_{\rm NLO,3,slow}^{\lambda=\rmL,ij}(\Pt)&= \alpha_{\rm em}\alpha_se_f^2\deltatwo \Hcal_{\rm LO}^{ij,\lambda=\textrm{L}}(\Pt)\nonumber\\
&\times \int_0^{z_f}\frac{\der z_g}{z_g}\left[\ln\left(\frac{z_f\Pt^2}{Q_f^2 z_g}\right)-2+2\ln\left(\frac{\Pt^2+\bar Q^2}{c_0\bar Q P_\perp}\right)\right]\,.
\label{eq:HNLO3-slow-final}
\end{align}
We observe that the $k_g^-$ or $z_g$ dependence of the kinematic constraint turns the single logarithmic slow divergence into a double logarithmic one. Therefore, the use of a kinematically constrained rapidity evolution solves both of the problematic issues mentioned after Eq.\,\eqref{eq:SE1-naive-slow} resulting from   ``naive" leading logarithmic BFKL evolution.  

This is by no means the end of the story and indeed results in a further satisfying outcome. The counterterm $\Hcal_{\rm NLO,3,slow}^{\lambda=\rmL,ij}(\Pt)$ must cancel both double and single logarithmic divergences of $\Hcal_{\rm NLO,3}^{\lambda=\rmL,ij}(\Pt)$ as $z_g$ goes to 0. This requirement for a convergent $z_g$ integral entirely specifies the scale $Q_f$ which was treated as an arbitrary scale in our previous work. Comparing Eq.~\eqref{eq:HNLO3-slow-final} with Eq.\,\eqref{eq:HNLO3-int} for $\Hcal_{\rm NLO,3}^{\lambda=\rmL,ij}(\Pt)$, one finds that cancellation of the $z_g$ divergences requires that 
\begin{equation}
   z_f=ec_0^2z_1z_2\frac{Q_f^2}{\Pt^2+\bar Q^2}\,.\label{eq:Qf-identity}
\end{equation}
Since typically $z_f\sim z_1z_2$, one sees that indeed, $Q_f$ must be of order $\textrm{max}(Q,P_\perp)$.
Of course, this identity is specific to our implementation of the kinematically constrained rapidity evolution of the WW gluon distribution in coordinate space. In this sense, it is a scheme-dependent statement. However the ratio $z_f/Q_f^2$ is scheme-independent since it only depends on the kinematics of the dijet process. Therefore the constraint Eq.\,\eqref{eq:Qf-identity} may lead to scheme-independent results, one of them being the value of the Sudakov single logarithmic coefficient $s_f$ defined in Section~\ref{sub:b2b-NLO}. We shall come back to this point in  Section~\ref{sec:summary}.

\paragraph{Final result.} With the counterterm in Eq.\,\eqref{eq:HNLO3-slow-final}, and the constraint Eq.\,\eqref{eq:Qf-identity}, the third NLO hard factor Eq.\,\eqref{eq:HNLO3-int} is now finite since the $z_g$ integral becomes regular for $z_g=0$. The calculation of this integral is straightforward and gives
\begin{align}
    \Hcal_{\rm NLO,3}^{\lambda=\rmL,ij}(\Pt)&=\Hcal^{\lambda=\rmL,ij}_{\rm LO}(\Pt)\left[\frac{1}{2}-\frac{\pi^2}{6}-\frac{3}{4}\ln\left(\frac{z_2(1+\chi^2)}{\chi^2}\right)\right]\nonumber\\
    &+\Hcal^{\lambda=\rmL,ij}_{\rm LO}(\Pt)\int_{z_f}^{z_1}\frac{\der z_g}{z_g}\left[-1+\ln\left(\frac{1+\chi^2}{\chi^2}\right)-\ln\left(\frac{z_g}{z_2z_1}\right)\right]\,.\label{eq:HNLO3-final}
\end{align}
The integral in the second line is the leftover contribution from the  subtraction of the slow gluon phase space between $z_f$ and $z_1$. It  does not need to be computed as it will cancel with the $C_F$ dependent slow gluon counterterm of diagram $\rm V_1$, as we shall see in the next section. Without this important cancellation, the integral in the second line would generate double logarithms of the rapidity factorization scale $z_f$, spoiling rapidity factorization. 

Finally, the diagram $\rm SE_1'$ obtained by quark-antiquark exchange ($z_1\leftrightarrow z_2$), and the complex conjugate terms must be 
added together to obtain the final self-energy contributions of gluons crossing the shockwave to the back-to-back dijet cross-section.

\paragraph{Order of magnitude of neglected terms.} To summarize our discussion thus far, we showed that the leading term in the correlation expansion of $\rm SE_1\times LO$ can be factorized into a perturbative hard factor and a nonperturbative WW TMD distribution. Our method relied  on

(i) finding suitable transverse coordinate integration variables 
such that $\ut-\ut'$ is conjugate to $\Pt$ and likewise $\bt-\bt'$ is conjugate to $\qt$, as is the case for the LO cross-section, resulting in the coordinate phase space factor
\begin{equation}
\der^8\Xttilde e^{-i\Pt\cdot\ruupt-i\qt\cdot\rbbpt}\,.
\label{eq:LO-phases}
\end{equation}
(ii) Expanding the CGC operators inside the transverse coordinate integration, to first order around $\xt=\yt=\zt=\bt$ in the limit $|\rt|, |\ut|\ll |\bt|$ where $\rt$ is typically conjugate to the transverse momentum of the virtual gluon. While it is natural to justify the small $|\ut|$ expansion from the phases Eq.\,\eqref{eq:LO-phases}, one may wonder why one is allowed to truncate the expansion to first order in $\rt$ as well, to arrive at the back-to-back limit. 
Hence to prove TMD factorization of all virtual diagrams in the leading power approximation, we must also demonstrate that the higher order terms in the correlation expansion of the CGC operator $\Xi_{\rm NLO,1}$, are power suppressed either by $q_\perp/P_\perp$ or $Q_s/P_\perp$. 

As a matter of fact, this can be understood from a simple dimensional analysis argument. The kinematic constraint from the NLO light cone wave-function (the Bessel function) imposes $r_\perp$ and $u_\perp$ to be of comparable sizes, $r_\perp\sim u_\perp$. Hence, after Fourier transform, each additional $r_\perp$ power leads to a $1/P_\perp\sim 1/Q$ suppression. We provide in Appendix~\ref{app:power-suppressed-virtual} a full mathematically rigorous derivation showing that this is indeed the case.

\subsection{Vertex correction crossing the shockwave}

The back-to-back limit of the vertex correction with gluon interacting with the shockwave, labeled $\rm V_1$ in Fig.\,\ref{fig:NLO-dijet-all-diagrams} can be addressed in a similar fashion. In particular, this diagram can also be factorized into a hard part and the WW gluon TMD. The most important part of this calculation will be to cross-check that the same value of $Q_f$ as the one extracted from diagram $\rm SE_1$ in Eq.\,\eqref{eq:Qf-identity} must be used in the kinematically constrained rapidity subtraction term to ensure a finite hard factor for the vertex diagram $\rm V_1$.

This diagram was computed in full generality within the CGC EFT in \cite{Caucal:2022ulg}. For a longitudinally polarized virtual photon, we found 
\begin{align}
   &\der\sigma^{\lambda=\rm L}_{\rm V1}= \frac{\alpha_{\rm em}e_f^2N_c\deltatwo}{(2\pi)^6} \int\der^8\Xt e^{-i\ktone\cdot\rxxtp-i\kttwo\ryytp} 8z_1^3z_{2}^3Q^2K_0(\bar Qr_{x'y'})\nonumber\\
   &\times \frac{(-\alpha_s)}{\pi}\int_0^{z_1}\frac{\der z_g}{z_g} \left(1-\frac{z_g}{z_1}\right)\left(1+\frac{z_g}{z_2}\right)\left(1-\frac{z_g}{2z_1}-\frac{z_g}{2(z_2+z_g)}\right)\nonumber\\
    & \times \int\frac{\der^2\zt}{\pi}\frac{\rzxt\cdot\rzyt}{\rzxt^2\rzyt^2}e^{-i\frac{z_g}{z_1}\ktone \cdot \rzxt} K_0(QX_{\rm V})\Xi_{\rm NLO,1}- \der\sigma^{\lambda=\rm L}_{\rm V1,slow}\,,\label{eq:V1}
\end{align}
for the product between $\rm V_1$ and the complex conjugate of the leading order amplitude. As the phases that appear in this expression are identical to those present in Eq.\,\eqref{eq:SE1-full} for $\rm SE_1$, we shall use the same change of variables given in Eq.\,\eqref{eq:SE1-b2b-variable}. Another motivation for using the same values of $\ut$, $\bt$ and $\rt$ comes from the fact that these transverse coordinates have the same physical meaning as those for the self-energy crossing the shockwave.

After this change of variables, one can write the cross-section as 
\begin{align}
       &\der\sigma^{\lambda=\rm L}_{\rm V1}= \frac{\alpha_{\rm em}e_f^2N_c\deltatwo}{(2\pi)^6} \int\der^8\Xttilde e^{-i\Pt\cdot\ruupt-i\qt\cdot\rbbpt} 8z_1^3z_{2}^3Q^2K_0(\bar Qu_\perp')\nonumber\\
       &\times\int_0^{z_1}\frac{\der z_g}{z_g}\left(1-\frac{z_g}{z_1}\right)\left(1+\frac{z_g}{z_2}\right)\left(1-\frac{z_g}{2z_1}-\frac{z_g}{2(z_2+z_g)}\right)\frac{z_1}{z_1-z_g} \nonumber\\
    & \times \frac{(-\alpha_s)}{\pi}\int\frac{\der^2\rt}{\pi}\frac{\rt\cdot(\rt+\ut)}{\rt^2(\rt+\ut)^2} K_0(\bar Q\sqrt{\ut^2+\omega \rt^2})\widetilde\Xi_{\rm NLO,1}-\der\sigma^{\lambda=\rm L}_{\rm V1,slow}\,,
\end{align}
with $\omega=z_g/(z_2(z_1-z_g))$. The extra factor $z_1/(z_1-z_g)$ in the second line comes from the Jacobian of the change of coordinates. The back-to-back limit of this expression can now be obtained as previously from the ``correlation expansion" of the color correlator $\widetilde\Xi_{\rm NLO,1}$, keeping the leading terms in $\ut$, $\ut'$ and $\rt$ in the regime $u_\perp\sim u_\perp'\sim r_\perp$. We find for the leading term, 
\begin{align}
    \widetilde\Xi_{\rm NLO,1}
    &=\left[C_F\ut^i\ut'^j+\left(\frac{N_c}{2}+\frac{1}{2N_c}\frac{z_g}{(z_1-z_g)}\right)\rt^i\ut'^j\right]\nonumber\\
    &\times\frac{1}{N_c}\left\langle\textrm{Tr} V(\bt)\partial^iV^\dagger(\bt)\partial^j V(\bt')V^\dagger(\bt')\right\rangle\,.
    \label{eq:XiNLO1-expansion}
\end{align}
This time however the term proportional to $\rt^i$ does not vanish since the $\rt$ integral depends on the scalar product $\rt\cdot\ut$ and not only on $\rt^2$.
Thanks to this expansion, the leading term can be written in a factorized form up to corrections of order $q_\perp/P_\perp$ or $Q_s/P_\perp$:
\begin{align}
    \der\sigma^{\lambda=\rm L}_{\rm V_1}=\alpha_{\rm em}\alpha_se_f^2\deltatwo\left[\frac{(-\alpha_s)}{\pi}\Hcal_{\rm NLO,4}^{\lambda=\rmL,ij}(\Pt)\right]\int\frac{\der^2\rbbpt}{(2\pi)^4}e^{-i\qt\cdot\rbbpt}\hat{G}^{ij}_{Y_f}(\rbbpt)+\mathcal{O}\left(\frac{q_\perp}{P_\perp},\frac{Q_s}{P_\perp}\right)\,,
\end{align}
with the hard factor defined to be
\begin{align}
\Hcal^{\lambda=\rmL,ij}_{\rm NLO,4}(\Pt)&\equiv16z_1^3z_{2}^3Q^2\frac{(-\Pt^j)}{(\Pt^2+\bar Q^2)^2} \int_0^{z_1}\frac{\der z_g}{z_g}\left(1+\frac{z_g}{z_2}\right)\left(1-\frac{z_g}{2z_1}-\frac{z_g}{2(z_2+z_g)}\right) \nonumber\\
       &\times\left[C_F\partial^i_{\Pt}I_3+\left(\frac{N_c}{2}+\frac{1}{2N_c}\frac{z_g}{(z_1-z_g)}\right)(-iI_4^i)\right]-\Hcal^{\lambda=\rmL,ij}_{\rm NLO,4,slow}(\Pt)\label{eq:HNLO-4-def} \,,
\end{align}
and the integrals $I_{3}$ and $I_{4}^i$ given by
\begin{align}
    I_{3}&=\int\frac{\der^2\ut}{(2\pi)}\int\frac{\der^2\rt}{(2\pi)}e^{-i\Pt\cdot\ut}\frac{\rt\cdot(\rt+\ut)}{\rt^2(\rt+\ut)^2} K_0\left(\bar Q\sqrt{\ut^2+\omega \rt^2}\right)\,,\label{eq:I3-def}\\
      I_{4}^i&=\int\frac{\der^2\ut}{(2\pi)}\int\frac{\der^2\rt}{(2\pi)}e^{-i\Pt\cdot\ut}\rt^i\frac{\rt\cdot(\rt+\ut)}{\rt^2(\rt+\ut)^2} K_0\left(\bar Q\sqrt{\ut^2+\omega \rt^2}\right)\,.\label{eq:I4-def}
\end{align}
As for the case of $\rm SE_1$, we cannot make any approximations in these integrals since the correlation expansion of $\widetilde\Xi_{\rm NLO,1}$ assumes $r_\perp\sim u_\perp$. In particular, $\partial^i_{\Pt}I_3$ and $I_{4}^i$ are both ``leading power" since they both fall like $1/P_\perp^3$ at large $P_\perp$. This confirms the importance of the second term in the correlation expansion Eq.\,\eqref{eq:XiNLO1-expansion} of $\widetilde\Xi_{\rm NLO,1}$. On the other hand, any higher order corrections to Eq.\,\eqref{eq:XiNLO1-expansion} give a result vanishing at least like $1/P_\perp^4$. Thus such contributions are suppressed in the back-to-back limit using the same rationale as in the previous subsection. 

Remarkably, the two integrals $I_{3}$ and $I_{4}^i$ can be computed analytically using tricks similar to those employed in the calculation of the hard factor in $\rm SE_1$.  A detailed computation of these integrals is provided in Appendix \ref{app:NLO4}, and we only quote here the final results:
\begin{align}
I_{3}&=\frac{1}{2(\Pt^2+\bar Q^2)}\ln\left(\frac{\Pt^2+(1+\omega)\bar Q^2}{\omega\bar Q^2}\right)\,,\label{eq:Jcal-final}\\
I_{4}^i&=-\frac{i\Pt^i}{(\Pt^2+\bar Q^2)^2}\left[\frac{\Pt^2+\bar Q^2}{\Pt^2+(1+\omega)\bar Q^2}+\frac{1}{2}\ln\left(\frac{\omega \bar Q^2}{\Pt^2+(1+\omega)\bar Q^2}\right)\right]\,.\label{eq:Kcal-final}
\end{align}
Plugging these expressions into Eq.\,\eqref{eq:HNLO-4-def}, we obtain a representation of $\Hcal_{\rm NLO,4}$ as a single integral over $z_g$.

\paragraph{Subtracting the kinematically constrained slow divergence.} Similarly to $\rm SE_1$, this $z_g$ integral has both double and single logarithmic divergences at $z_g=0$, but is regularized by the counterterm $\Hcal^{\lambda=\rmL,ij}_{\rm NLO,4,slow}(\Pt)$ obtained by extracting the piece proportional to the WW gluon TMD in the subtraction term $ \der\sigma^{\lambda=\rm L}_{\rm V1,slow}$. The latter should include the kinematic constraint as discussed in the previous subsection:
\begin{align}
   \der\sigma^{\lambda=\rm L}_{\rm V1,slow}&= \frac{\alpha_{\rm em}e_f^2N_c\deltatwo}{(2\pi)^6} \int\der^8\Xttilde e^{-i\Pt\cdot\ruupt-i\qt\rbbpt} 8z_1^3z_{2}^3Q^2K_0(\bar Q u_\perp)K_0(\bar Qu_\perp')\,\nonumber\\
   &\times \frac{(-\alpha_s)}{\pi}\int_0^{z_f}\frac{\der z_g}{z_g} \int\frac{\der^2\rt}{\pi}\frac{\rt\cdot(\rt+\ut)}{\rt^2(\rt+\ut)^2}\Theta\left(\frac{z_f}{z_gQ_f^2}-\rt^2\right)\widetilde\Xi_{\rm NLO,1}(\bt,\ut,\rt)\,,\label{eq:V1-slow}
\end{align}
with the color correlator evaluated at the transverse coordinates
\begin{equation}
\widetilde\Xi_{\rm NLO,1}(\bt,\ut,\rt)=\Xi_{\rm NLO,1}(\bt+z_2\ut,\bt-z_1\ut,\bt+z_2\ut+\rt)\,.
\end{equation}
As we have discussed, the $\Theta$-function in Eq.\,\eqref{eq:V1-slow} is the only modification as compared to the ``naive" rapidity subtraction term giving leading-log BFKL/BK/JIMWLK evolution.

Expanding around $\bt$ to extract the WW gluon TMD, we get the result for the subtraction term for slow gluons
\begin{align}
\Hcal^{\lambda=\rmL,ij}_{\rm NLO,4,slow}(\Pt)&\equiv16z_1^3z_{2}^3Q^2\frac{(-\Pt^j)}{(\Pt^2+\bar Q^2)^2} \int_0^{z_f}\frac{\der z_g}{z_g}\left[C_F\partial^i_{\Pt}I_{3,s}+\frac{N_c}{2}(-iI_{4,s}^i)\right]\,,\label{eq:HNLO-4-slow}
\end{align}
with the integrals $I_{3,s}$ and $I_{4,s}^i$ given by
\begin{align}
    I_{3,s}&=\int\frac{\der^2\ut}{(2\pi)}\int\frac{\der^2\rt}{(2\pi)}e^{-i\Pt\cdot\ut}K_0(\bar Q u_\perp)\frac{\rt\cdot(\rt+\ut)}{\rt^2(\rt+\ut)^2}\Theta\left(\frac{z_f}{z_g Q_f^2}-\rt^2\right) \,,\label{eq:I3s-def} \\
    I_{4,s}^i&=\int\frac{\der^2\ut}{(2\pi)}\int\frac{\der^2\rt}{(2\pi)}e^{-i\Pt\cdot\ut}K_0(\bar Q u_\perp)\rt^i\frac{\rt\cdot(\rt+\ut)}{\rt^2(\rt+\ut)^2} \Theta\left(\frac{z_f}{z_g Q_f^2}-\rt^2\right) \,.\label{eq:I4s-def}
\end{align}
As outlined at the beginning of this subsection, it is crucial for the consistency of our kinematically improved evolution that the same choice of $Q_f$ as for $\rm SE_1$ gives a well-defined $z_g$ integral for $\rm V_1$. The calculation of the two integrals $I_{3,s}$ and $I_{4,s}^i$ is also detailed in Appendix \ref{app:NLO4}. They read, up to powers of $z_g$ corrections:
\begin{align}
    I_{3,s}&=\frac{1}{2(\Pt^2+\bar Q^2)}\left[\ln\left(\frac{z_f\Pt^2}{z_g Q_f^2}\right)+2\ln\left(\frac{\Pt^2+\bar Q^2}{c_0P_\perp\bar Q}\right)\right]+\mathcal{O}(z_g) \,,\label{eq:I3s}\\
    I_{4,s}^i&=\frac{-i\Pt^i}{(\Pt^2+\bar Q^2)^2}\left[\frac{3}{2}-\frac{1}{2}\ln\left(\frac{z_f\Pt^2}{z_g Q_f^2}\right)-\ln\left(\frac{\Pt^2+\bar Q^2}{c_0P_\perp\bar Q}\right)\right]+\mathcal{O}(z_g) \,.\label{eq:I4s}
\end{align}
Comparing these expressions with Eqs\,\eqref{eq:Jcal-final} and \eqref{eq:Kcal-final}, one verifies that the kinematically constrained slow gluon subtraction  term $\der\sigma_{\rm V1,slow}^{\lambda=\rmL}$ gives precisely the counterterm $\Hcal^{\lambda=\rmL,ij}_{\rm NLO,4,slow}(\Pt)$ that makes the $z_g$ integral in $\Hcal^{\lambda=\rmL,ij}_{\rm NLO,4}(\Pt)$ convergent, provided we choose 
\begin{equation}
   z_f=ec_0^2z_1z_2\frac{Q_f^2}{\Pt^2+\bar Q^2}\,.\label{eq:Qf-identity-2}
\end{equation}
The relation between $z_f$ and $Q_f$ can be seen to be identical to that derived from $\rm SE_1$ in Eq.\,\eqref{eq:Qf-identity}.

\paragraph{Final result.} With the inclusion of the subtraction term  for slow gluons $\Hcal^{\lambda=\rmL,ij}_{\rm NLO,4,slow}(\Pt)$ that we just computed, the remaining $z_g$ integral in Eq.\,\eqref{eq:HNLO-4-def} is convergent.  The calculation of this integral is rather tedious. In the end, our final result for the hard factor associated with $\rm V_1$ is, (with $\chi=\bar Q/P_\perp=Q/M_{q\bar q}$):
\begin{align}
      &\Hcal^{\lambda=\rmL,ij}_{\rm NLO,4}(\Pt)=\Hcal_{\rm LO}^{ij,\lambda=\textrm{L}}(\Pt)\times\left\{\frac{N_c}{2}\left[-\frac{1}{8}+\frac{1}{8(z_2-z_1\chi^2)}-\frac{1}{2}\textrm{Li}_2\left(\frac{z_2-z_1\chi^2}{z_2(1+\chi^2)}\right)\right.\right.\nonumber\\
       &\left.+\left(-\frac{3}{8}+\frac{z_2-2\chi^2+z_1\chi^4}{8(z_2-z_1\chi^2)^2}\right)\ln\left(\frac{z_2(1+\chi^2)}{\chi^2}\right)\right]+\frac{1}{2N_c}\left[-\frac{\pi^2}{24}-\frac{3}{8}+\frac{3}{8(z_2-z_1\chi^2)}\right.\nonumber\\
       &\left.+\textrm{Li}_2\left(\frac{z_2-z_1\chi^2}{z_2(1+\chi^2)}\right)\left.+\left(\frac{9}{8}-\frac{3z_2+2z_2\chi^2+z_1\chi^4}{8(z_2-z_1\chi^2)^2}\right)\ln\left(\frac{z_2(1+\chi^2)}{\chi^2}\right)+\frac{1}{4}\textrm{Li}_2\left(\frac{z_1\chi^2-z_2}{\chi^2}\right)\right]\right\}\nonumber\\
       &+\Hcal^{\lambda=\rmL,ij}_{\rm LO}(\Pt)C_F\int_{z_f}^{z_1}\frac{\der z_g}{z_g}\left[-1+\ln\left(\frac{1+\chi^2}{\chi^2}\right)-\ln\left(\frac{z_g}{z_2z_1}\right)\right]\nonumber\\
       &+\Hcal^{\lambda=\rmL,ij}_{\rm LO}(\Pt)\frac{N_c}{2}\left[\ln\left(\frac{z_1}{z_f}\right)-\frac{1}{4}\ln^2\left(\frac{z_1}{z_f}\right)-\frac{1}{2}\ln\left(\frac{z_1}{z_f}\right)\ln\left(\frac{z_2(1+\chi^2)}{\chi^2}\right)\right]\,.\label{eq:HNLO4-final}
\end{align}
The last two lines come from the slow gluon phase space between $z_f$ and $z_1$ (respectively $z_2$ for the quark-antiquark exchanged diagram $\rm V_1'$) which is not subtracted off since Eq.\,\eqref{eq:HNLO-4-slow} only subtracts the phase space below the factorization scale $z_f$. In particular, one notices that the $C_F$ term (left unintegrated in $z_g$) cancels against the corresponding term for diagram $\rm SE_1$ (the second line in Eq.\eqref{eq:HNLO3-final}) in such a way that the $z_f$ dependent piece of the sum of $C_F\Hcal_{\rm NLO,3}-\Hcal_{\rm NLO,4}$ is proportional to the color factor $N_c$, in agreement with the overall color factor of the evolution equation of the WW gluon TMD.

In the last line, one notices a double logarithmic dependence upon the factorization scale $z_f$ which will cancel in our final result with the $\ln^2(Q_f^2/\Pt^2)$ term in Eq.\,\eqref{eq:zeroth-moment-final} once $Q_f$ is expressed in terms of $z_f$ and the kinematics of the dijet process thanks to Eq.\,\eqref{eq:Qf-identity-2}. The absence of $\ln^2(z_f)$ dependence in the impact factor is another important feature of TMD factorization at NLO at small $x_{\rm Bj}$.

\section{Soft contributions from real diagrams}
\label{sec:NLO_reals}

In the previous section, we discussed the virtual contributions to the back-to-back NLO cross-section and demonstrated that they factorize into  hard factors and the WW gluon TMD in the back-to-back limit $P_\perp\gg q_\perp, Q_s$. We explicitly computed each of the hard factors obtaining analytical results for them. 
Further, we showed that the leading log rapidity evolution of the WW TMD distribution must satisfy a kinematic constraint corresponding to lifetime ordering of slow gluons. Not least, we were able to obtain a unique transverse momentum factorization scale $Q_f$ that only depends on the external kinematic variables of the back-to-back cross-section and on the rapidity factorization scale $z_f$. 

This section is dedicated to real contributions to the back-to-back NLO cross-section. We review the results from \cite{Caucal:2022ulg} for the contribution to the back-to-back cross-section from soft gluon radiation emitted from quark/antiquark after the interaction with the shockwave in the amplitude and the complex conjugate amplitude. Furthermore, we revisit the real emissions where the gluon interacts with the shockwave either in the amplitude or in the complex conjugate amplitude and we argue that they give power suppressed contributions in back-to-back kinematics.

\subsection{Final state soft gluon radiation: Leading power Sudakov logarithms and $\mathcal{O}(\alpha_s)$ corrections }
\label{sub:real-soft}

Amongst all the real diagrams gathered in Fig.\,\ref{fig:NLO-dijet-all-diagrams}, only the subset with gluon emissions after the shockwave contribute to the back-to-back dijet cross-section at NLO and leading power in $q_\perp/P_\perp$. These contributions are labeled $\rm R_2\times R_2$ and $\rm R_2\times R_2'$, plus those obtained after quark-antiquark exchange.

The diagram $\rm R_2\times R_2$ contributes in two ways to the back-to-back dijet cross-section. Firstly, when the real gluon is emitted inside the quark jet cone, the phase space integration is divergent because of the QCD collinear singularity but this divergence cancels with the virtual cross-section. The finite leftover contribution is included in Eq.\,\eqref{eq:zeroth-moment-final}. It also contributes via soft gluons emitted outside the jet cone as \cite{Caucal:2022ulg}
\begin{align}
 \der\sigma_{\rm R_2\times R_2,soft}&=\alpha_{\rm em}\alpha_se_f^2\deltatwo\Hcal^{\lambda,ij}_{\rm LO}(\Pt)\int\frac{\der^2\rbbpt}{(2\pi)^4} e^{-i\qt\cdot\rbbpt}\hat G^{ij}_{Y_f}(\bt,\bt')\nonumber\\
    &\times \frac{\alpha_sC_F}{\pi}\int_0^1\frac{\der\xi}{\xi}\left[1-e^{-i\xi\Pt\cdot\rbbpt}\right]\ln\left(\frac{\Pt^2\rbbpt^2R^2\xi^2}{c_0^2}\right)\,.\label{eq:R2R2-soft-b2b} 
\end{align}
The $R$ dependence of this result comes from the out-of-cone condition which obviously depends on the jet radius.

Similarly, the diagram $\rm R_2\times R_2'$ contributes in the soft regime via the term
\begin{align}
        \der\sigma_{\rm R_2\times R_2',soft}&=\alpha_{\rm em}\alpha_se_f^2\deltatwo\Hcal^{\lambda,ij}_{\rm LO}(\Pt)\int\frac{\der^2\rbbpt}{(2\pi)^4} e^{-i\qt\cdot\rbbpt}\hat G^{ij}_{Y_f}(\bt,\bt')\nonumber\\
    &\times \frac{\alpha_s}{2\pi N_c}\int_0^1\frac{\der\xi}{\xi}\left[1-e^{-i\xi\Pt\cdot\rbbpt}\right]\ln\left(\frac{\Pt^2\rbbpt^2\xi^2}{z_2^2c_0^2}\right)\,. \label{eq:R2R2'-soft-b2b}
\end{align}
Since this diagram has no collinear divergence, the in-cone phase space is power of $R^2$ suppressed and to leading order in $R$, the soft out-of-cone component does not depend on the jet radius.

Both the expressions above display a dependence on the WW gluon TMD but the phase $e^{-i\xi \Pt\cdot\rbbpt}$ and the logarithm couple the $\Pt$ and $\rbbpt$ dependence. The $\xi$ integral was calculated in \cite{Caucal:2022ulg}. It gives Sudakov double and single logarithms which are resummed in the Sudakov form factor $\mathcal{S}(\Pt^2,\rbbpt^2)$ in Eq.\,\eqref{eq:zeroth-moment-final}. 

It also contributes finite pieces to the coefficient function and these are included in Eq.\,\eqref{eq:zeroth-moment-final}. Soft gluon radiation has an important effect on azimuthal anisotropies because of the scalar product $\Pt\cdot\rbbpt$ in the phase. Therefore the finite terms coming from soft gluons emitted after interacting with the shockwave, both in the amplitude and in the complex conjugate amplitude, depend on the harmonics of the Fourier decomposition of the dijet cross-section with respect to the azimuthal angle between $\Pt$ and $\qt$. This Fourier decomposition is given by 
\begin{equation}
    \frac{\der \sigma^{\gamma_{\lambda}^{*}+A\to \textrm{dijet}+X}}{ \der^2 \Pt \der^2 \qt \der \eta_1 \der \eta_{2}}= \der \sigma^{(0),\lambda}(P_\perp,q_\perp,\eta_1,\eta_2)+2\sum_{n=1}^{\infty} \der \sigma^{(n),\lambda}(P_\perp,q_\perp,\eta_1,\eta_2)\cos(n\phi)\,,
\end{equation}
with $\phi$ the azimuthal angle between $\Pt$ and $\qt$. After absorbing the Sudakov logarithms into the Sudakov form factor, the finite pieces coming from Eq.\,\eqref{eq:R2R2-soft-b2b} and Eq.\,\eqref{eq:R2R2'-soft-b2b} contributing to the coefficient function (and multiplying the linearly polarized WW distribution) take the form 
\begin{align}
\der \sigma^{(0),\lambda=\rm L}_{\rm R_2\times R_2, fin}&=\alpha_{\rm em}\alpha_s e_f^2\deltatwo\Hcal_{\rm LO}^{0,\lambda=\rm L}(\Pt)\frac{\alpha_sC_F}{\pi}\left[\frac{1}{2}+\ln(R)\right]\nonumber\\
    &\times \int\frac{\der^2\rbbpt}{(2\pi)^4}e^{-i\qt\cdot\rbbpt}\hat h^0_{Y_f}(\rbbpt)\,,\label{eq:c0b2b-final}\\
    \der \sigma^{(0),\lambda=\rm L}_{\rm R_2\times R_2',fin.}&=\alpha_{\rm em}\alpha_s e_f^2\deltatwo\Hcal_{\rm LO}^{0,\lambda=\rm L}(\Pt)\frac{\alpha_s}{2\pi N_c}\left[\frac{1}{2}-\ln(z_2)\right] \nonumber\\
    &\times\int\frac{\der^2\rbbpt}{(2\pi)^4}e^{-i\qt\cdot\rbbpt}\hat h^0_{Y_f}(\rbbpt)\,,\label{eq:c0_R2R2'_b2b-final}
\end{align}
for the azimuthally averaged cross-section. These contributions, and their quark-antiquark exchange counterparts, are included in Eq.\,\eqref{eq:zeroth-moment-final}. For the $\langle \cos(2\phi)\rangle$ anisotropy, the corresponding results are
\begin{align}
    \der \sigma^{(2),\lambda=\rm L}_{\rm R_2\times R_2, fin.}&=\alpha_{\rm em}\alpha_s e_f^2\deltatwo\Hcal_{\rm LO}^{0,\lambda=\rm L}(\Pt)\frac{\alpha_sC_F}{\pi}\int\frac{\der^2\rbbpt}{(2\pi)^4}e^{-i\qt\cdot\rbbpt}\frac{\cos(2\theta)}{2}\nonumber\\
    &\times \left\{\hat h^0_{Y_f}(\rbbpt)\left[-\frac{5}{8}-\frac{1}{2}\ln(R)\right]+\hat G^0_{Y_f}(\rbbpt)\left[1+2\ln(R)\right]\right\}\,,
    \label{eq:c2b2b-final}\\
    \der \sigma^{(2),\lambda=\rm L}_{\rm R_2\times R_2',fin.}&=\alpha_{\rm em}\alpha_s e_f^2\deltatwo\Hcal_{\rm LO}^{0,\lambda=\rm L}(\Pt)\frac{\alpha_s}{2\pi N_c} \int\frac{\der^2\rbbpt}{(2\pi)^4}e^{-i\qt\cdot\rbbpt}\frac{\cos(2\theta)}{2}\nonumber\\
    &\times\left\{\hat h^0_{Y_f}(\rbbpt)\left[-\frac{5}{8}+\frac{1}{2}\ln(z_2)\right]+\hat G^0_{Y_f}(\rbbpt)\left[1-2\ln(z_2)\right]\right\}\,,\label{eq:c2_R2R2'_b2b-final}
\end{align}
with $\theta$ the angle between $\qt$ and $\rbbpt$ in polar coordinates.
The finite terms for higher harmonics coming from soft gluon radiation have been computed in Appendix E of \cite{Caucal:2022ulg}.

To sum up, real soft gluon radiation emitted after the shockwave contributes to the back-to-back dijet cross-section to leading power and their contribution is proportional to the WW gluon TMD. They also induce azimuthal anisotropies in such a way that the azimuthally averaged cross-section depends on the linearly polarized WW TMD and the $\langle \cos(2\phi)\rangle$ anistropy is sensitive to the unpolarized WW TMD. As shown in Section 5.1 of \cite{Caucal:2022ulg}, the contribution from hard gluons with transverse momenta $k_{g\perp} \gtrsim z_g P_\perp$ to diagrams $\rm R_2\times R_2$ and $\rm R_2\times R_2'$ is power suppressed in the back-to-back limit.

\subsection{Real emissions crossing the shockwave: Beyond leading power $\mathcal{O}(\alpha_s)$ corrections}

In order to prove factorization of the back-to-back dijet cross-section at NLO in terms of the WW gluon TMD, we must also demonstrate that the other real diagrams, in which the gluon crosses the shockwave either in the amplitude or in the complex conjugate amplitude are power suppressed in the back-to-back limit. 

The main difficulty with these diagrams comes from their dependence on CGC correlators which do not naturally collapse to the WW gluon TMD as the leading order one does in Eq.\,\eqref{eq:xiLO-def}. This was also apparently the case for the virtual diagrams in which the gluon scatters off the shockwave; however as we discussed in Section~\ref{sec:NLO-hard}, there exists a correlation expansion that enables one to isolate the leading power term which depends only on the WW gluon TMD. 

This correlation expansion relies on the hierarchy of transverse scales $u_\perp\sim r_\perp\ll b_\perp$, where typically $u_\perp\sim 1/P_\perp$, $r_\perp\sim 1/|\kgt|$ and $b_\perp\sim 1/q_\perp$. In particular, for virtual graphs with gluons crossing the shockwave, this expansion gives a mathematically well-defined impact factor thanks to the kinematic constraint. Effectively, the kinematic constraint enables one to clearly distinguish gluons with hard momenta, of order of $P_\perp$, that contribute to the impact factor, from gluons with momenta of order $Q_s$, $q_\perp$, that control the WW gluon TMD and its evolution at small-$x$. Hence, the contribution to the NLO back-to-back impact factor of virtual diagrams with gluons crossing the shockwave is dominated by \textit{hard} gluons with transverse momentum of order of $P_\perp$ (such gluons are not kinematically forbidden in back-to-back kinematics since the gluon is virtual). For real emissions, one can similarly expand the color correlator. Imposing back-to-back kinematics requires $|\kgt|$ to be smaller than $q_\perp$, or in terms of conjugate transverse coordinates, $r_\perp$ larger than $b_\perp$. One may then question the validity of the correlation expansion. However, the kinematic constraint, which is naturally enforced in real graphs with a gluon crossing the shockwave through the argument of the Bessel function, enforces that $|\kgt|\gtrsim \sqrt{z_g}P_\perp\gg q_\perp$, i.e. $r_\perp\ll b_\perp$. Therefore, the correlation expansion remains well defined but the lack of overlapping phase space between real gluons satisfying both the kinematic and back-to-back constraints leads to power suppression of these diagrams\footnote{If one is interested in  configurations of three hard jets instead of back-to-back dijets, this argument does not apply. As a matter of fact, the study in \cite{Altinoluk:2020qet} shows that the photoproduction of three \textit{hard} jets in a ``mercedes-like" shape can be factorized in terms of the WW gluon TMD at leading power, including contributions from real diagrams with a gluon scattering off the shockwave.}. In Appendix~\ref{app:power-suppressed-real}, we provide a more mathematically grounded demonstration showing that diagrams $\rm R_1\times \rm R_2$, $\rm R_1'\times R_2$, $\rm R_1\times R_1$ and $\rm R_1'\times R_1$ are indeed power suppressed in the back-to-back limit.

\section{Weizs\"{a}cker-Williams gluon TMD factorization at NLO from the CGC}
\label{sec:WW-TMD}

In this section, we will put together all of the results of the previous sections for the NLO back-to-back dijet cross-section that we have shown can be expressed as remarkably simple factorized expressions for the convolution of the NLO impact factor and the WW TMD gluon distribution, with the former expressed as a product of a 
universal soft factor and a coefficient function.  The WW distribution obeys a kinematically constrained JIMWLK rapidity evolution equation. 

In Section~\ref{sec:summary}, we will discuss the structure of these results, and outline subtle features that emerge when one considers the rapidity evolution of the WW TMD distribution. In addition, we show that one needs to further reorganize our results using an appropriate choice of a hard factorization scale of the Sudakov form factor to obtain physically sensible cross-sections. 
Our final NLO results for all moments of the back-to-back dijet cross-section are compactly summarized in Section~\ref{sec:summary1}. 

\subsection{Structure and reorganization of TMD factorized results}
\label{sec:summary}
We first combine the analytic results of the previous sections for the functions $\Hcal_{\rm NLO,1}$ and  $\Hcal_{\rm NLO,2}$ with those that as we have seen greatly simplify the ``other" contribution $\der \sigma^{(0),\lambda=\rm L}_{\rm other}$ in 
Eq.\,\eqref{eq:zeroth-moment-final} into two new hard factors $\Hcal_{\rm NLO,3}$ and $\Hcal_{\rm NLO,4}$. For the purpose for a discussion of the choice of the hard scale, we shall first write the fixed order $\mathcal{O}(\alpha_s)$ result without assuming Sudakov exponentiation; we will expand the Sudakov form factor in Eq.\,\eqref{eq:zeroth-moment-final} to first order in $\alpha_s$.
For jets defined with a generalized $k_t$ algorithm, we find that our previous result in Eq.\,\eqref{eq:zeroth-moment-final} becomes 
\begin{align}
 \der \sigma^{(0),\lambda=\rm L}&=\alpha_{\rm em}\alpha_s e_f^2\deltatwo\Hcal_{\rm LO}^{0,\lambda=\rm L}\int\frac{\der^2\rbbpt}{(2\pi)^4}e^{-i\qt\cdot\rbbpt}\hat G^0_{Y_f}(\rbbpt)\left\{1-\frac{\alpha_sN_c}{4\pi}\ln^2\left(\frac{\Pt^2\rbbpt^2}{c_0^2}\right)\right.\nonumber\\
 &\left.-\frac{\alpha_s}{\pi}\left[C_Fs_0-N_c s_f\right]\ln\left(\frac{\Pt^2\rbbpt^2}{c_0^2}\right)+\frac{\alpha_sN_c}{2\pi}f^{\lambda=\rmL}_1(\chi,z_f)+\frac{\alpha_s}{2\pi N_c}f^{\lambda=\rmL}_2(\chi)\right\}\nonumber\\
    &+\alpha_{\rm em}\alpha_s e_f^2\deltatwo\Hcal_{\rm LO}^{0,\lambda=\rm L}\int\frac{\der^2\rbbpt}{(2\pi)^4}e^{-i\qt\cdot\rbbpt}\hat h^0_{Y_f}(\rbbpt)\left\{\frac{\alpha_sN_c}{2\pi}\left[1+\ln(R^2)\right]\right.\nonumber\\
    &\left.+\frac{\alpha_s}{2\pi N_c}\left[-\ln(z_1z_2R^2)\right]\right\}\,,
\label{result_xsect_L}
\end{align}
with $\chi=\bar Q/P_\perp=Q/M_{q\bar q}$
We have classified the NLO coefficient function by color structure, thanks to the two functions $f_1^{\lambda=\rmL}$ and $f_2^{\lambda=\rmL}$, given by
\begin{align}
    f^{\lambda=\rmL}_1(\chi,z_f)&=7-\frac{3\pi^2}{2}-\frac{3}{2}\ln\left(\frac{z_1z_2R^2}{\chi^2}\right)-\ln(z_1)\ln(z_2)+2\ln\left(\frac{(1+\chi^2)z_f}{z_1z_2}\right)\nonumber\\
    &-\ln(1+\chi^2)\ln\left(\frac{1+\chi^2}{z_1z_2}\right)+\left\{\textrm{Li}_2\left(\frac{z_2-z_1\chi^2}{z_2(1+\chi^2)}\right)-\frac{1}{4(z_2-z_1\chi^2)}\right.\nonumber\\
    &\left.+\frac{(1+\chi^2)(z_2(2z_2-z_1)+z_1(2z_1-z_2)\chi^2)}{4(z_2-z_1\chi^2)^2}\ln\left(\frac{z_2(1+\chi^2)}{\chi^2}\right)+(1\leftrightarrow2)\right\}     \,,   \label{f1_def} \\
    f^{\lambda=\rmL}_2(\chi)&=-8+\frac{19\pi^2}{12}+\frac{3}{2}\ln(z_1z_2R^2)-\frac{3}{4}\ln^2\left(\frac{z_1}{z_2}\right)-\ln(\chi)\nonumber\\
    +&\left\{\frac{1}{4(z_2-z_1\chi^2)}+\frac{(1+\chi^2)z_1(z_2-(1+z_1)\chi^2)}{4(z_2-z_1\chi^2)^2}\ln\left(\frac{z_2(1+\chi^2)}{\chi^2}\right)\right.\nonumber\\
    &\left.+\frac{1}{2}\textrm{Li}_2(z_2-z_1\chi^2)-\frac{1}{2}\textrm{Li}_2\left(\frac{z_2-z_1\chi^2}{z_2}\right) +(1\leftrightarrow2)\right\}  \,,  \label{f2_def}
\end{align}
for a longitudinally polarized virtual photon. These two functions also depend on $z_1$, $z_2$ and $R$ but we choose here to leave manifest only the dependence upon the dimensionless ratio $Q/M_{q\bar q}$ and the rapidity factorization scale $z_f$.

Because of the $\bar Q^2$ factor included in $\Hcal_{\rm LO}^{0,\lambda=\rmL}(\Pt)$, the photo-production limit $Q^2\to 0$ of the cross-section is zero although the functions $f_1^{\lambda=\rmL}$ and $f_2^{\lambda=\rmL}$ diverge logarithmically as $\chi\to 0$:
\begin{align}
  f_1^{\lambda=\rmL}(\chi,z_f)&\underset{\chi\ll1}{=}\frac{1}{4z_1z_2}\ln(\chi^2)+\mathcal{O}(1)\\
  f_2^{\lambda=\rmL}(\chi)&\underset{\chi\ll1}{=}-\frac{1}{4z_1z_2}\ln(\chi^2)+\mathcal{O}(1)
\end{align}
Albeit finite, the $\chi\to 0$ limit or equivalently the $M_{q\bar q}^2\gg Q^2$ limit, leads to an unphysical negative NLO cross-section when $\alpha_s|\ln(\chi)|\sim 1$. This unphysical behaviour, resulting from imposing an additional hierarchy of scales $M_{q\bar q}^2\gg Q^2$ on top of the $W^2\gg Q^2$ and $P_\perp\gg q_\perp$ conditions, points towards the need for a more sytematic study of higher order corrections or some additional resummation of $\ln(\chi)$ logarithms. This will be also discussed below in the opposite regime $Q^2\gg M_{q\bar q}^2$.

For the record, and for later convenience, we can also define the NLO hard factor associated with virtual gluon emission and unresolved real emissions (those inside the jet cone):
\begin{align}
    \Hcal_{\rm NLO}^{ij,\lambda=\rmL}(\Pt)\equiv\frac{\alpha_s}{\pi}\times\Hcal_{\rm LO}^{ij,\lambda=\rmL}(\Pt)\times \left[\frac{N_c}{2}f_1^{\lambda=\rmL}(\chi,z_f)+\frac{1}{2N_c}f_2^{\lambda=\rmL}(\chi)\right]\label{eq:H_NLO_final}\,.
\end{align}
This hard factor gives the NLO finite pieces $f_1$ and $f_2$ in Eq.\,\eqref{result_xsect_L} after contraction with the WW gluon TMD $\hat{G}^{ij}_{Y_f}(\rbbpt)$. The second term in Eq.\,\eqref{result_xsect_L}, which is proportional to the linearly polarized WW gluon TMD, comes from real soft gluon emissions out of the jet cone, as explained in Sec.\,\ref{sub:real-soft}.

\paragraph{Rapidity factorization scale $z_f$ dependence.} The NLO coefficient function has a remaining rapidity factorization scale $z_f$ dependence. Remarkably, this dependence enters through the $z_f$ term in the finite piece only, without double logarithms of $z_f$ nor $Q_f$. Indeed, the transverse momentum scale $Q_f$ introduced in the kinematically constrained rapidity evolution of the WW gluon TMD has been simplified thanks to the relation 
\begin{equation}
    \frac{z_f}{Q_f^2}=\frac{ec_0}{M_{q\bar q}^2+Q^2}\,.
    \label{eq:Qf-relation-summary}
\end{equation}
As noted previously, this constraint follows from our significantly more detailed analysis here of the virtual diagrams in the back-to-back limit, and should be employed when evolving the TMDs $\hat G_{Y_f}^{ij}$ from some initial scale $Y_0=\ln(z_0)$ up to the factorization scale $Y_f=\ln(z_f)$ with the evolution equation given by Eq.\,\eqref{wwe}. Eq.\,\eqref{eq:Qf-relation-summary} also enables one to simplify the single logarithmic coefficient $s_f$ in the Sudakov form factor,  defined in Eq.\,\eqref{eq:s0}. Plugging the identity Eq.\,\eqref{eq:Qf-relation-summary} into $s_f$, we find 
\begin{align}
s_f&=1-\ln\left(1+\frac{Q^2}{M_{q\bar q}^2}\right)\,,
\end{align}
and note thus that $s_f$ does not depend anymore on the rapidity factorization scale $z_f$. Our study enables us to unambiguously fix the value of the Sudakov logarithms to single logarithmic accuracy. Eq.\,\eqref{eq:Qf-relation-summary} is also consistent in the photo-production limit since $z_f/Q_f^2$ then goes to $ec_0/M_{q\bar q}^2$.

With regard to the $z_f$ dependence of the result, one might be tempted to derive the $Y_f=\ln(z_f)$ evolution equation of the WW gluon TMD by requiring that the cross-section should be independent of $Y_f$ up to $\mathcal{O}(\alpha_s^2)$ sub-leading corrections. Then from Eq.\,\eqref{result_xsect_L}, one would obtain the linear RG equation for the WW gluon TMD:
\begin{equation}
    0=\frac{\partial \hat G^{ij}_{Y_f}}{\partial Y_f}+\frac{\alpha_sN_c}{\pi}\hat G^{ij}_{Y_f}+\mathcal{O}(\alpha_s^2)\,.
    \label{eq:wwe-linear}
\end{equation}
This is however not the particular evolution equation Eq.\,\eqref{wwe} mentioned in the first section  that was derived employing the 
JIMWLK equation in \cite{Dominguez:2011gc} (see also \cite{Balitsky:2015qba}). Since we noted there that this equation does not generate a closed form expression, clearly something is missing here.

The resolution lies on the observation that the kinematic constraint interpolates between full non-linear evolution in Eq.\,\eqref{wwe} when $z_g \ll 1$ to the linear evolution in Eq.\,\eqref{eq:wwe-linear} when $z_g \sim \mathcal{O}(1)$. As we will see below, this is because the correlation limit expansion is only valid in the NLO impact factor where $z_g \sim \mathcal{O}(1)$, whereas it breaks down in the small-$x$ enhanced terms where $z_g \ll 1$. In deriving our impact factor, we implicitly assume $z_f \sim \mathcal{O}(1)$; thus all the non-linearities are absorbed into the small-$x$ evolution ensuring the impact factor factorizes from the WW distribution.

To illustrate this, we can formally write the fixed order NLO cross-section as a sum of the LO cross-section,  the NLO impact factor and the $\mathcal{O}(\alpha_s\ln(x_0/x_f))$ slow contributions:
\begin{align}
    \der\sigma_{\rm NLO}=\der\sigma_{\rm LO}+\alpha_s\int_0^\infty\frac{\der z_g}{z_g}\left[I(z_g)-I_{\rm k.c.,slow}(z_g)\Theta(z_f-z_g)\right]+\alpha_s\int_{z_0}^{z_f}\frac{\der z_g}{z_g}I_{\rm k.c.,slow}(z_g)\,,\label{eq:subtraction-scheme}
\end{align}
where $I_{\rm k.c.,slow}$ incorporates the slow gluon divergence including the kinematic constraint.
The second term in the expression on the r.h.s corresponds to the impact factor. Note that $I(z_g)$ implicitly contains step functions that bound the $z_g$ integral from above.
%(like $\Theta(z_1-z_g)$, etc).  
The third term on the r.h.s is the large $\ln(z_f/z_0)\sim \ln(x_0/x_f)$ rapidity logarithm. At this step, the cross-section is indeed obviously independent of $z_f$. Now we can schematically write the correlation expansion as 
\begin{align}
    I(z_g)&=I^{(0)}(z_g)+I^{(1)}(z_g)+...+I^{(n)}(z_g)+...\,,\\
    I_{\rm k.c, slow}(z_g)&=I_{\rm k.c, slow}^{(0)}(z_g)+I_{\rm k.c, slow}^{(1)}(z_g)+...+I_{\rm k.c, slow}^{(n)}(z_g)+...\,,
    \label{eq:corr-exp-Islow}
\end{align}
where $I^{(0)}$ represents the leading power contribution proportional to the WW gluon TMD and the terms $I^{(n)}$ are suppressed by $(q_\perp/P_\perp)^n$. By construction, the kinematically constrained 
slow gluon counterterms $I_{\rm k.c, slow}^{(n)}(z_g)$ cancel the $z_g\to0$ divergence order by order in the correlation expansion. Hence in the impact factor term, the leading power term in the back-to-back limit is simply
\begin{align}
    \int_0^\infty\frac{\der z_g}{z_g}\left[I^{(0)}(z_g)-I_{\rm k.c, slow}^{(0)}(z_g)\Theta(z_f-z_g)\right]\,.
\end{align}
This is the leading contribution in the correlation expansion computed analytically in this paper from all virtual diagrams and from real soft emissions emitted after the shockwave interaction with the quark-antiquark dipole.

However the correlation expansion breaks down for the third term in the r.h.s of Eq.\,\eqref{eq:subtraction-scheme} when $z_0$ is small and all $I_{\rm k.c, slow}^{(i)}$ terms contribute equally. This is because the correlation expansion relies on $r_\perp\sim u_\perp\ll b_\perp,b_\perp'$, with each additional term in Eq.\,\eqref{eq:corr-exp-Islow} having an extra power of $r_\perp$ or $u_\perp$. Therefore $I_{\rm k.c, slow}^{(n)}(z_g)$ behaves parametrically like (for $n\ge 1$)
\begin{align}
    I_{\rm k.c, slow}^{(n)}&\sim\alpha_{\rm em}\alpha_s\Hcal_{\rm LO}^{\lambda,ij}(\Pt)\int\frac{\der^2\rt}{(2\pi)}\frac{r_\perp^n}{\rt^2}\Theta\left(\frac{z_f}{z_gQ_f^2}-\rt^2\right)\times \langle O^{(n)}(\qt)\rangle_{Y_f}\,,\\
    &\sim \alpha_{\rm em} \alpha_s\Hcal_{\rm LO}^{\lambda,ij}(\Pt)\frac{1}{n}\left(\frac{z_f}{z_gQ_f^2}\right)^{n/2}\times\langle O^{(n)}(\qt)\rangle_{Y_f}\,,\\
    &\sim\alpha_{\rm em} \alpha_s\Hcal_{\rm LO}^{\lambda,ij}(\Pt)\frac{1}{n}\left(\frac{z_f}{z_g}\right)^{n/2}\left(\frac{q_\perp^2}{Q_f^2}\right)^{n/2}\times G^{ij}_{Y_f}(\qt)\,,
\end{align}
with $O^{(n)}$ an operator such that $\langle O^{(n)}(\qt)\rangle_{Y_f}\sim \alpha_sq_\perp^n G^{ij}_{Y_f}(\qt)$.
Since $Q_f\sim P_\perp$, one may be tempted to say that the larger $n$ is, the more it is suppressed by powers of $P_\perp$. Yet the kinematic constraint, represented by the step function which couples the $\rt$ and $z_g$ dependence, makes each term more divergent in the $z_g\to0$ limit.
One is thus forced to keep the full $ I_{\rm k.c, slow}$ integrand in the phase space between $z_0$ and $z_f$, leading to
\begin{align}
    \der\sigma_{\rm NLO}&=\der\sigma_{\rm LO}+\alpha_s\int_0^\infty\frac{\der z_g}{z_g}\left[I^{(0)}(z_g)- I_{\rm k.c, slow}^{(0)}(z_g)\Theta(z_f-z_g)\right]+\mathcal{O}\left(\frac{q_\perp}{P_\perp}\right)\nonumber\\
    &+\alpha_s\int_{z_0}^{z_f}\frac{\der z_g}{z_g} I_{\rm k.c, slow}(z_g) \,.
    \label{eq:schematic_NLO_zf_indep}
\end{align}
The first two terms on the r.h.s correspond to the result Eq.\,\eqref{result_xsect_L}, while the last one gives the first step of the rapidity evolution of the WW gluon TMD. It is now clear that the full cross-section, as written, is not strictly $z_f$ independent. Hence the evolution equation for the TMD in the LO cross-section should be inferred from the form of the last term in Eq.\,\eqref{eq:schematic_NLO_zf_indep}. Since this term is precisely the first step of the kinematically constrained JIMWLK evolution equation for the WW distribution (Eq.\,\eqref{wwe}), as demonstrated in \cite{Caucal:2021ent,Caucal:2022ulg}, the all order resummation of small-$x$ logarithms should be performed using this evolution equation. 

If one insists on writing the impact factor such that one gets Eq.\,\eqref{wwe} from the $z_f$ independence of the NLO cross-section up to $\mathcal{O}(\alpha_s^2)$ corrections (as one must), we need to add the  to the result in Eq.\,\eqref{result_xsect_L}, the term 
\begin{align}
    +\int_{z_f}^\infty\frac{\der z_g}{z_g}\left[I_{\rm k.c, slow}(z_g)-I_{\rm k.c, slow}^{(0)}(z_g)\right]\,.
\end{align}
Although this term is superficially power suppressed in the back-to-back limit for $z_f\sim 1$ (as it should be), it allows us to recover the non-linear evolution of the WW gluon TMD Eq.\,\eqref{wwe} by taking the $Y_f=\ln(z_f)$ derivative of the NLO cross-section.

When $z_0$ is not small ($x_{\rm Bj}\sim 1$), it makes sense to truncate the correlation expansion of $I_{\rm k.c.,slow}$ to first order. In this case, the $Y_f$ dependence of the WW gluon TMD should be given by the linear equation Eq.\,\eqref{eq:wwe-linear}.
However this evolution equation (given in Eq.\,\eqref{eq:wwe-linear}) is not meaningful since when $x_{\rm Bj}\sim 1$ there is no need for small $x$ resummation in the first place.

\paragraph{Jet definition.} The NLO coefficient function displays a $\ln(R)$ dependence which is the remnant of the infrared and collinear safe jet definition of the final state. The formula in Eq.\,\eqref{result_xsect_L} is valid for jets defined using the generalized $k_t$ algorithm. (These include the anti-$k_t$ algorithm \cite{Cacciari:2008gp} and the Cambridge/Aachen algorithm \cite{Dokshitzer:1997in,Wobisch:1998wt}.) For such jet-cone algorithms, one should subtract from the NLO hard factor the quantity \cite{Kang:2016mcy}
\begin{equation}
    \Delta_{k_t}=\frac{\alpha_sC_F}{\pi}\left[3-\frac{\pi^2}{3}-3\ln(2)\right]\,.
\end{equation}
We leave for future phenomenological study the computation of the NLO coefficient function and soft factor for dihadrons instead of dijets. In that case, the NLO coefficient function would not depend on $R$ but rather on the factorization scale $\mu_F$ corresponding to the evolution of the dihadron fragmentation function \cite{Bergabo:2022tcu}.

\paragraph{The regime $Q^2\gg M_{q\bar q}^2$.} Besides Sudakov logarithms which are resummed via the Sudakov form factor, the impact factor may potentially contain large logarithms that spoil the perturbative convergence of the cross-section or lead to unphysical results such as  negative cross-sections. One example is the $\ln(R)$ term which becomes large for jets defined with small jet radii. The resummation of such logs can be performed using ``micro-jet" evolution \cite{Dasgupta:2014yra} or jet functions in the Soft Collinear Effective Field Theory (SCET) approach \cite{Kang:2016mcy}.

Another potentially problematic logarithm involves the terms containing $\ln(\chi)$  (with  $\chi=Q/M_{q\bar q}$), when $\chi$ becomes large. Indeed, $\chi$ can in principle be arbitrarily large for a given $P_\perp^2$, while maintaining the back-to-back constraint\footnote{This would correspond to the kinematic domain $P_\perp^2 \ll {\bar Q}^2 \ll W^2$, where $W^2 = (q+P_n)^2$.}. 
In this $\chi\gg 1$ limit, the NLO hard factor behaves as 
\begin{equation}
  \frac{\Hcal_{\rm NLO}^{ij,\lambda=\rmL}}{\Hcal_{\rm LO}^{ij,\lambda=\rmL}}\underset{\chi\to \infty}{\sim} \frac{\alpha_sN_c}{2\pi}\left[-4\ln^2(\chi)+2\left(\frac{7}{2}+\ln(z_1z_2)\right)\ln(\chi)\right]-\frac{\alpha_s}{2\pi N_c}\left[1+\ln(z_1z_2)\right]\ln(\chi)\,.
\end{equation}
When $\chi\gg 1$, the NLO cross-section becomes negative due to the dominant $-4\ln^2(\chi)$ term in the impact factor. Such negative double logarithms of the ratio between the hard scale in the Sudakov logarithms (here $P_\perp$) and the photon virtuality $Q$ also appear in lepton-jet correlation in DIS \cite{Liu:2018trl} or in back-to-back heavy quark pair production in DIS \cite{Zhang:2017uiz} within the TMD framework.

In our calculation, the origin of this large $-\ln^2(\chi)$ contribution to the NLO cross-section can be traced to our choice of the hard scale $\mu_h^2$ in the Sudakov logarithms in Eq.\,\eqref{result_xsect_L} to be $\Pt^2$. Indeed, the double logarithm is a leftover of the kinematically improved small-$x$ evolution given by the last term of the third line in Eq.\,\eqref{eq:zeroth-moment-final}, namely
\begin{equation}
    -\frac{\alpha_s N_c}{2\pi}\ln^2\left(\frac{Q_f^2c_0^2}{\Pt^2}\right)\,.
\end{equation}
Since $Q_f^2\sim Q^2$ when $\chi\gg1$, this is the negative double logarithm we were looking for. The presence of $\Pt^2$ in the denominator of the argument of this double logarithm comes from the choice of $\Pt^2$ as the hard scale in the Sudakov double logarithm.

Having identified the origin of this term, we note that we have the freedom to make a different choice for the hard scale in the Sudakov evolution to ensure that the cross-section is positive definite and the coefficient function remains of order $\mathcal{O}(\alpha_s)$ for this scale choice. We will therefore change  the hard scale in the Sudakov resummation to 
\begin{equation}
    \mu_h^2=\Pt^2+\bar Q^2\,,
\end{equation}
instead of $\Pt^2$. In doing so, we ensure that the contribution to the large logarithm from the phase space $\bar Q^2\sim \mu_h^2\gg \Pt^2$ is  resummed in the Sudakov factor. 
 
Introducing this choice of hard scale $\mu_h$ inside the Sudakov double and single logarithms of the first line in Eq.\,\eqref{result_xsect_L}, 
we can rewrite the NLO $\mathcal{O}(\alpha_s)$ Sudakov logarithms
\begin{equation}
-\frac{\alpha_sN_c}{4\pi}\ln^2\left(\frac{\Pt^2\rbbpt^2}{c_0^2}\right)-\frac{\alpha_s}{\pi}\left[C_F s_0-N_cs_f\right]\ln\left(\frac{\Pt^2\rbbpt^2}{c_0^2}\right)\,,
\end{equation}
as 
\begin{align}
  &-\frac{\alpha_sN_c}{4\pi}\ln^2\left(\frac{\mu_h^2\rbbpt^2}{c_0^2}\right)+\frac{\alpha_s}{\pi}\left[C_F\ln(z_1z_2R^2)+N_c-\frac{N_c}{2}\ln\left(1+\chi^2\right)\right]\ln\left(\frac{\mu_h^2\rbbpt^2}{c_0^2}\right)\nonumber\\
  &-\frac{\alpha_sN_c}{4\pi}\ln^2\left(\frac{\Pt^2}{\mu_h^2}\right)+\frac{\alpha_sC_F}{\pi}\ln(z_1z_2R^2)\ln\left(\frac{\Pt^2}{\mu_h^2}\right)+\frac{\alpha_sN_c}{\pi}\ln\left(\frac{e\Pt^2}{\Pt^2+\bar Q^2}\right)\ln\left(\frac{\Pt^2}{\mu_h^2}\right)\,.
\end{align}
We can now put the terms in the second line above in the impact factor while the terms in the first line can be exponentiated into the Sudakov form factor as
\begin{equation}
\widetilde{\mathcal{S}}(\mu_h^2,\rbbpt^2)=\exp\left(-\int_{c_0^2/\rbbpt^2}^{\mu_h^2}\frac{\der\mu^2}{\mu^2}\frac{\alpha_s(\mu^2)N_c}{\pi}\left[\frac{1}{2}\ln\left(\frac{\mu_h^2}{\mu^2}\right)+\frac{C_F}{N_c}s_0-\tilde{s}_f\right]\right) \,,\label{eq:Sudakov-final0}
\end{equation}
with 
\begin{equation}
    s_0 = \ln\left(\frac{2(1+\cosh(\Delta \eta_{12}))}{R^2}\right)+\mathcal{O}(R^2)\,,\quad \tilde{s}_f=1-\frac{1}{2}\ln\left(\frac{\Pt^2+\bar Q^2}{\Pt^2}\right)\,.
    \label{eq:s02}
\end{equation}

\begin{figure}[H]
    \centering
    \includegraphics[width=0.49\textwidth]{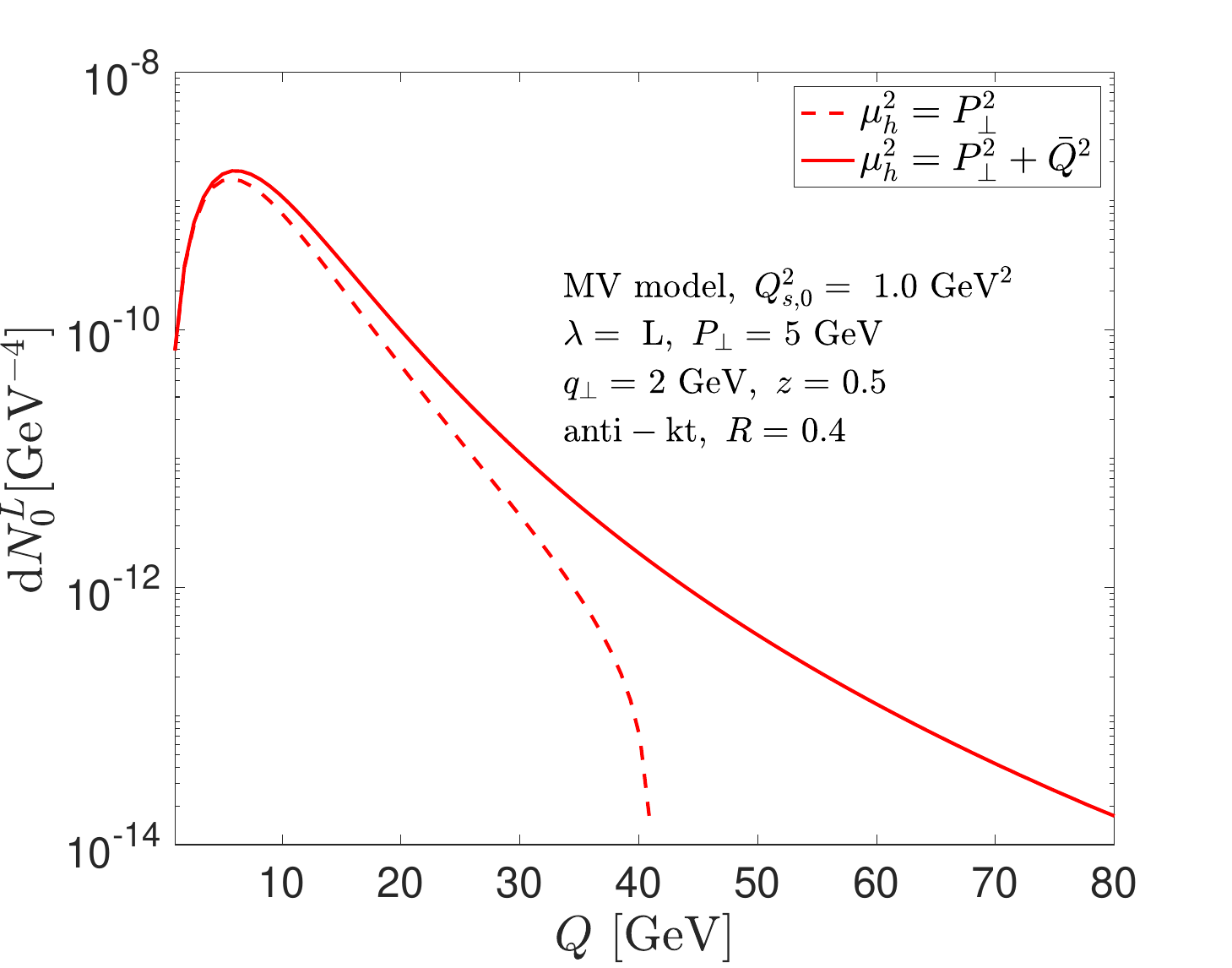}
    \caption{Differential yield as a function of the photon virtuality $Q$ for two different choices of the hard scale in the Sudakov form factor. Both curves are full NLO results.\label{fig:Qxsection-hardscale}}
\end{figure}

The modified hard factor can similarly be expressed in terms of the functions $\tilde f_1$ and $\tilde f_2$:
\begin{align}
    \tilde f_1^{\lambda=\rmL}(\chi,z_f)&=f_1^{\lambda=\rmL}(\chi,z_f)+\frac{3}{2}\ln^2(1+\chi^2)-\ln(z_1z_2R^2)\ln(1+\chi^2)-2\ln(1+\chi^2)\,,
    \label{tilde_f1_def}\\
       \tilde f_2^{\lambda=\rmL}(\chi)&=f_2(\chi)+\ln(z_1z_2R^2)\ln(1+\chi^2)\,.
       \label{tilde_f2_def}
\end{align}
With this modified Sudakov form factor, the behaviour of the NLO hard factor at large $\chi\gg 1$ becomes
\begin{align}
       \frac{\widetilde{\mathcal{H}}_{\rm NLO}^{ij,\lambda=\rmL}}{\mathcal{H}_{\rm LO}^{ij,\lambda=\rmL}}\underset{\chi\to \infty}{\sim}\frac{\alpha_sN_c}{2\pi}\left[\frac{1}{2}\ln^2(\chi^2)+\left(\frac{3}{2}-\ln(R^2)\right)\ln(\chi^2)\right]+\frac{\alpha_s}{4\pi N_c}\left[-1+\ln(z_1z_2R^4)\right]\ln(\chi^2) \,,
\end{align}
where the coefficient of the double logarithm $\ln^2(\chi)$ is now positive. Hence by changing the hard scale of the Sudakov form factor, we  were able to obtain a cross-section that is positive definite even in the regime $\chi\gg 1$ ($Q^2\gg M_{q\bar q}^2$). For an estimate of the effect of this change of scale, we refer the reader to Fig.\,\ref{fig:Qxsection-hardscale} in Section~\ref{sec:numerics}. 

Although positive definite in the regime $\chi\gg 1$, the NLO coefficient function still has double and single logarithms of $\chi$ which may spoil the perturbative convergence of the cross-section when enforcing the additional constraint $Q^2\gg M_{q\bar q}^2$. A two loop calculation of the NNLO coefficient function would help to clarify this issue, or even better, stabilize the cross-section as observed in other processes \cite{Becher:2009th,Becher:2013vva}.

\begin{figure}[H]
    \centering
    \includegraphics[width=0.49\textwidth]{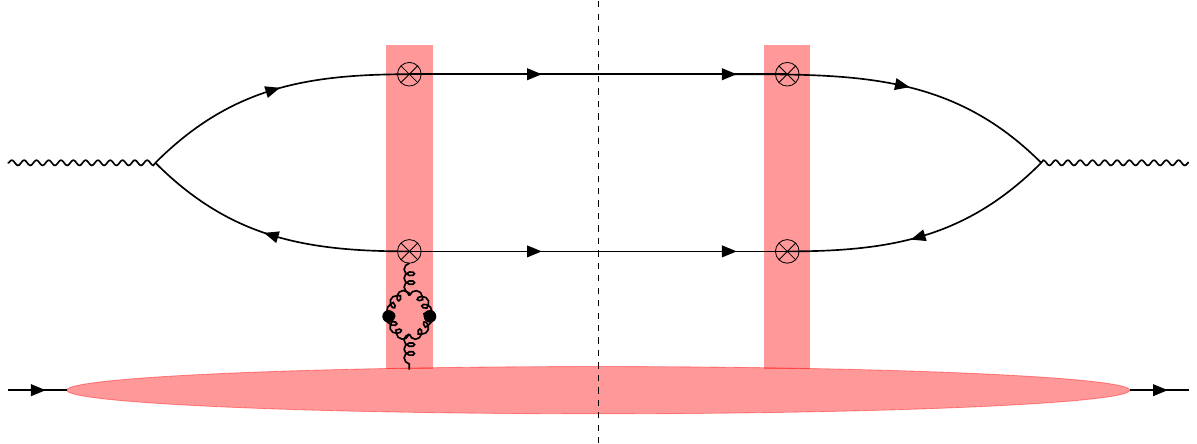}
    \caption{Diagram from the one-loop correction
to the classical background field times the leading order
amplitude. The gluon polarization tensor in the background field can also be inserted in the CGC effective vertex of the quark propagator. These two diagrams plus their complex conjugate counterpart build the one-loop correction to the classical WW gluon TMD.}
    \label{fig:1loop_classical}
\end{figure}

\paragraph{Anomalous dimension of the WW gluon TMD.} Besides the NLO diagrams computed in sections \ref{sec:NLO-hard} and \ref{sec:NLO_reals} which include all $\mathcal{O}(\alpha_s)$ quantum corrections obtained using the small fluctuation propagator (CGC effective vertices) in the classical background field, there is an additional contribution in the CGC EFT coming from the one-loop correction to the classical background field $A^{+,a}_{\rm cl}$ itself \cite{Ayala:1995hx,Gelis:2006yv,Gelis:2006cr,Gelis:2019yfm}. The diagram associated with this contribution is displayed in Fig.\,\ref{fig:1loop_classical}. As shown in \cite{Ayala:1995hx}, this diagram has a UV divergence, which is the same as the standard UV divergence of the gluon vacuum polarization diagram at one-loop, namely
\begin{equation}
     \hat G^{(1)}_{Y}(\rbbpt)=-\alpha_s\beta_0\left[\frac{1}{\varepsilon}+\textrm{finite}\right]\hat G^{(0)}_{Y}(\rbbpt)\,,\label{eq:1loop-WW}
\end{equation}
in dimensional regularization, with $\beta_0=(11N_c-2n_f)/(12\pi)$ and $\hat G^{(1)}$ the one-loop correction to the (classical) WW gluon TMD $\hat G^{(0)}$. 

This UV divergence is simply removed by replacing the bare coupling constant in the definition Eq.\,\eqref{eq:WWTMD} of the WW gluon TMD by the renormalized one \cite{Ayala:1995hx}. The TMD then acquires an additional scale dependence $\mu$ given by 
\begin{align}
    \hat G^{ij}_{Y}(\rbbpt,\mu)&\equiv\frac{-2}{\alpha_s(\mu)}\left\langle\Tr\left[V(\bt) \left(\partial^iV^\dagger(\bt) \right) V(\bt') \left(\partial^jV^\dagger(\bt') \right)\right]\right\rangle_{Y}\,.
    \label{eq:WWTMD-R}
\end{align}
One may then ask what the natural scale $\mu$ should be. The answer to this question depends on the hard scale which is probed by the process to be considered. At the level of the WW gluon TMD alone, $\mu$ should be set by the saturation scale or $q_\perp$. Since the WW gluon TMD only depends on $\rbbpt$ (assuming a homogeneous target in the transverse plane), a reasonable choice would be $\mu^2=c_0^2/\rbbpt^2$, where the $c_0$ constant is rather arbitrary here. Such ambiguities can be absorbed into the initial condition for the WW gluon TMD.

However, for the back-to-back hard dijet process, the natural hard scale is rather $\mu_h\sim P_\perp$, and therefore, $\alpha_s$ in the NLO corrections should run at this specific value. The phase space between the ``initial" scale $c_0^2/\rbbpt^2$ of the TMD and the hard scale of the process leads to an additional Sudakov single logarithm, as we now demonstrate. The one-loop running coupling satisfies
\begin{equation}
    \frac{1}{\alpha_s(\mu_h^2)}=\frac{1}{\alpha_s(c_0^2/\rbbpt^2)}\left[1+\alpha_s(\mu_h^2)\beta_0\ln\left(\frac{\mu_h^2\rbbpt^2}{c_0^2}\right)+\mathcal{O}(\alpha_s^2)\right]\,,
\end{equation}
so that 
\begin{equation}
    \hat G^{ij}_{Y}(\rbbpt,\mu_h)=\left[1+\alpha_s(\mu_h^2)\beta_0\ln\left(\frac{\mu_h^2\rbbpt^2}{c_0^2}\right)+\mathcal{O}(\alpha_s^2)\right]\hat G^{ij}_{Y}(\rbbpt,c_0^2/\rbbpt^2)
\end{equation}
Plugging this expression for the WW gluon TMD renormalized at the hard scale $\mu_h$ into the LO cross-section Eq.\,\eqref{eq:diff_xsec_TMD_LO} gives a NLO correction to the impact factor, which displays a single Sudakov logarithm with coefficient $\beta_0$. This Sudakov single log comes physically from the large phase space for the logarithmic running of the coupling between $P_\perp$ and $q_\perp\sim 1/r_{bb'}$.

From the one-loop QCD $\beta$-function and Eq.\,\eqref{eq:WWTMD-R}, one can readily deduce the $\mu$ evolution equation that enables one to resum this single logarithmic correction:
\begin{equation}
    \frac{\partial \hat G_{Y_f}(\rbbpt,\mu)}{\partial \ln(\mu)}=\alpha_s\beta_0 \times \hat G_{Y_f}(\rbbpt,\mu)\,.
\end{equation}
Integrating this equation between the natural initial evolution scale $c_0^2/\rbbpt^2$ of the TMD and the hard scale $\mu_h^2$ results in the following exponential
\begin{equation}
    \hat G_{Y_f}(\rbbpt,\mu_h^2)=\exp\left(\int_{c_0^2/\rbbpt^2}^{\mu_h^2}\frac{\der \mu^2}{\mu^2}\alpha_s\beta_0\right)\hat G_{Y_f}(\rbbpt,c_0^2/\rbbpt^2)\,,\label{eq:WW-anomalous-dim}
\end{equation}
which can be included into the Sudakov form factor as it provides a single logarithmic correction of the same order as the $s_0$ or $s_f$ term in Eq.\,\eqref{eq:Sudakov-final0}. The somewhat arbitrary choices of factorization scale $\mu_h$ and initial scale $c_0^2/\rbbpt^2$ is equivalent to a choice of scheme to remove the UV divergence of the diagram in Fig.\,\ref{fig:1loop_classical}. In practice, one can gauge the scheme dependence by varying the renormalization scale $\mu_h$ by a factor of two around our central choice $\mu_h^2=P_\perp^2+\bar Q^2$ in our final expressions given in section~\ref{sec:summary1}.
%rapidity factorization scale $z_f$ which separates the contribution from the one-loop correction to the classical field that goes into the NLO impact factor with the contribution that goes into the rapidity evolution of the TMD. This is analogous to the CSS formalism where the finite part of the virtual corrections to the gluon TMD depends on the ratio $\mu^2/\zeta_c^2$ \cite{Xiao:2017yya} with $\zeta_c$ the rapidity regulator in the so-called Collins-2011 scheme \cite{Collins:2011zzd}.

Including in Eq.\,\eqref{eq:Sudakov-final0} the anomalous dimension of the WW gluon TMD Eq.\,\eqref{eq:WW-anomalous-dim} leads to the following Sudakov form factor:
\begin{equation}
\widetilde{\mathcal{S}}(\mu_h^2,\rbbpt^2)=\exp\left(-\int_{c_0^2/\rbbpt^2}^{\mu_h^2}\frac{\der\mu^2}{\mu^2}\frac{\alpha_s(\mu^2)N_c}{\pi}\left[\frac{1}{2}\ln\left(\frac{\mu_h^2}{\mu^2}\right)+\frac{C_F}{N_c}s_0-\tilde{s}_f-\frac{\pi\beta_0}{N_c}\right]\right) \,.\label{eq:Sudakov-final}
\end{equation}
The presence of the $\beta_0$ term is in agreement with the calculation done in \cite{Zhou:2018lfq} within the CSS formalism. However our result differs from the Sudakov factor obtained in conventional collinear factorization \cite{Xiao:2017yya,Hatta:2021jcd} by the $1$ term in the single log coefficient $\tilde s_f$. This difference (which arises in our case from the kinematically constrained small-$x$ evolution for the WW gluon TMD) deserves further investigation.

\subsection{Final results for moments of the NLO back-to-back dijet cross-section}
\label{sec:summary1}
Following the extended discussion in Section~\ref{sec:summary}, and the subsequent reorganization of our results, we can now finally write down the principal results of this paper for the moments of the NLO back-to-back longitudinally polarized dijet cross-section. For the zeroth moment of the azimuthal anisotropy, we obtain 
\begin{align}
 \der \sigma^{(0),\lambda=\rm L}&=\alpha_{\rm em}\alpha_s e_f^2\deltatwo\Hcal_{\rm LO}^{0,\lambda=\rm L}\left\{1+\frac{\alpha_s(\mu_h)}{\pi}\left[\frac{N_c}{2}\tilde{f}^{\lambda=\rmL}_1(\chi,z_f)+\frac{1}{2\pi N_c}\tilde{f}^{\lambda=\rmL}_2(\chi)\right]\right\}\nonumber\\
 &\times\int\frac{\der^2\rbbpt}{(2\pi)^4}e^{-i\qt\cdot\rbbpt}\hat G^0_{Y_f}(\rbbpt)\,\widetilde{\mathcal{S}}(\mu_h^2,\rbbpt^2)\nonumber\\
    &+\alpha_{\rm em}\alpha_s e_f^2\deltatwo\Hcal_{\rm LO}^{0,\lambda=\rm L}\frac{\alpha_s(\mu_h)}{\pi}\left\{\frac{N_c}{2}\left[1+\ln(R^2)\right]+\frac{1}{2 N_c}\left[-\ln(z_1z_2R^2)\right]\right\}\nonumber\\
    &\times\int\frac{\der^2\rbbpt}{(2\pi)^4}e^{-i\qt\cdot\rbbpt}\hat h^0_{Y_f}(\rbbpt)\,\widetilde{\mathcal{S}}(\mu_h^2,\rbbpt^2)+\mathcal{O}\left(\frac{q_\perp^2}{P_\perp^2},\frac{Q_s^2}{P_\perp^2},R^2,\alpha_s^2\right)\,.
\label{eq:result_xsect_L-summary}
\end{align}
The function $\tilde{f}_1^{\lambda=\rmL}$ is  given by Eqs.\,\eqref{f1_def} and \eqref{tilde_f1_def} while and $\tilde{f}_2^{\lambda=\rmL}$ is given by  \eqref{f2_def} and \eqref{tilde_f2_def}. The Sudakov form factor is given by Eq.\,\eqref{eq:Sudakov-final}. In writing this expression, we exponentiated our fixed order results for the Sudakov double and single logarithms to all orders presuming that CSS evolution can be extended to small $x$. A formal proof of this exponentiation to all orders does not exist at small $x$ and is an important topic for future research. 

In a similar fashion, by collecting the various terms in Eq.\,\eqref{eq:second-moment-final} computed in Section~\ref{sec:NLO-hard}, we obtain the analytic expression (modulo the Fourier transform of the WW gluon TMDs) for the $\langle\cos(2\phi)\rangle$ anisotropy:
\begin{align}
 \der \sigma^{(2),\lambda=\rm L}&=\alpha_{\rm em}\alpha_s e_f^2\deltatwo\Hcal_{\rm LO}^{0,\lambda=\rm L}\left\{1+\frac{\alpha_s(\mu_h)}{\pi}\left[\frac{N_c}{2}\tilde{g}_1(\chi,z_f)+\frac{1}{2N_c}\tilde{g}_2(\chi)\right]\right\}\nonumber\\
 &\times\int\frac{\der^2\rbbpt}{(2\pi)^4}e^{-i\qt\cdot\rbbpt} \frac{\cos (2\theta)}{2} \hat h^0_{Y_f}(\rbbpt)\,\widetilde{\mathcal{S}}(\mu_h^2,\rbbpt^2) \nonumber\\
    &+\alpha_{\rm em}\alpha_s e_f^2\deltatwo\Hcal_{\rm LO}^{0,\lambda=\rm L}\frac{\alpha_s(\mu_h)}{\pi}\left\{N_c\left[1+\ln(R^2)\right]-\frac{1}{ N_c}\ln(z_1z_2R^2)\right\}\nonumber\\
    &\times\int\frac{\der^2\rbbpt}{(2\pi)^4}e^{-i\qt\cdot\rbbpt} \frac{\cos (2\theta)}{2} \hat G^0_{Y_f}(\rbbpt)\,\widetilde{\mathcal{S}}(\mu_h^2,\rbbpt^2) +\mathcal{O}\left(\frac{q_\perp^2}{P_\perp^2},\frac{Q_s^2}{P_\perp^2},R^2,\alpha_s^2\right)\,,
\label{result_xsect_L2}
\end{align}
where
\begin{align}
    \tilde{g}_1(\chi,z_f) &= \tilde{f}_1(\chi,z_f) -\frac{5}{4} -\ln(R)\,, \nonumber \\
    \tilde{g}_2(\chi) &= \tilde{f}_2(\chi) + \frac{1}{2} \ln(z_1 z_2 R^2)\,.
\end{align}
For completeness, we recall here the results obtained in \cite{Caucal:2022ulg} for higher even harmonic coefficients. They arise solely from real soft gluon radiation as virtual corrections do not generate\footnote{This is because the tensor structure of the NLO hard factor in Eq.\,\eqref{eq:H_NLO_final} is identical to that of the LO expression.} $\langle \cos(n\phi)\rangle $ anisotropies for $n\ge 4$.  They read
\begin{align}
     \der \sigma^{(n=2p),\lambda=\rm L}&= \alpha_{\rm em}\alpha_s e_f^2\deltatwo\Hcal_{\rm LO}^{0,\lambda=\rm L}\int\frac{\der^2\rbbpt}{(2\pi)^4}e^{-i\qt\cdot\rbbpt}\cos(n\theta) \hat G^0_{Y_f}(\rbbpt)\widetilde{\mathcal{S}}(\mu_h^2,\rbbpt^2)\nonumber\\
     &\times\frac{\alpha_s(\mu_h)(-1)^{p+1}}{n\pi}\left\{2N_c\left(\mathfrak{H}(p)-\frac{1}{n}\right)+2C_F\ln(R^2)-\frac{1}{N_c}\ln(z_1z_2)\right\}\nonumber\\
     &+\alpha_{\rm em}\alpha_s e_f^2\deltatwo\Hcal_{\rm LO}^{0,\lambda=\rm L}\int\frac{\der^2\rbbpt}{(2\pi)^4}e^{-i\qt\cdot\rbbpt}\cos(n\theta) \hat h^0_{Y_f}(\rbbpt)\widetilde{\mathcal{S}}(\mu_h^2,\rbbpt^2)\nonumber\\
     &\times\frac{\alpha_s(\mu_h)(-1)^p}{n^2-4}\left\{N_c\left((n+2)\mathfrak{H}(p-1)+(n-2)\mathfrak{H}(p+1)-\frac{2(n^2+4)}{n^2-4}\right)\right.\nonumber\\
     &\left.+n\left(2C_F\ln(R^2)-\frac{1}{N_c}\ln(z_1z_2)\right)\right\}+\mathcal{O}\left(\frac{q_\perp^2}{P_\perp^2},\frac{Q_s^2}{P_\perp^2},R^2,\alpha_s^2\right)\,,
     \label{eq:R2R2b2b-cn-final}
\end{align}
where $\mathfrak{h}(p)= \sum_{k=1}^p \frac{1}{k}$ is the p$^{\rm th}$ harmonic number.

Our results significantly extend those obtained previously in \cite{Mueller:2012uf,Mueller:2013wwa,Xiao:2017yya,Hatta:2020bgy,Hatta:2021jcd}. Results for the  back-to-back dijet cross-section for transversely polarized virtual photons will be presented in a separate work. 

\section{Numerical analysis of the results in Section~\ref{sec:summary1}}
\label{sec:numerics}

In this section, we will provide a brief numerical study of the results summarized in Section~\ref{sec:summary1}. Our aim here is to illustrate qualitatively the relative magnitude of Sudakov suppression and the NLO coefficient function to the impact factor. In our numerical study we will employ the McLerran-Venugopalan (MV) model to compute the Weizsäcker-Williams (WW) gluon distribution. We will present a more detailed examination of the interplay of Sudakov suppression and the energy evolution of the WW gluon distribution in a subsequent publication. 

For simplicity, we shall neglect the impact parameter dependence of the nuclear target; the differential cross-section is therefore proportional to the effective area $S_\perp$ of the target. We will consider here the differential yield
\begin{align}
    \frac{\der N^{\lambda=\mathrm{L}}}{ \der^2 \ktone \der^2 \kttwo \der \eta_1 \der \eta_{2}} = \frac{1}{S_\perp} \frac{\der \sigma^{\lambda=\mathrm{L}}}{ \der^2 \ktone \der^2 \kttwo \der \eta_1 \der \eta_{2}} \,,
\end{align}
which we can decompose into Fourier modes with respect to $\phi$ (the angle between $\Pt$ and $\qt$) as
\begin{align}
    \frac{\der N^{\lambda=\mathrm{L}}}{ \der^2 \ktone \der^2 \kttwo \der \eta_1 \der \eta_{2}} = \der N^{(0),\lambda=\mathrm{L}} \left[1 + 2 \sum_{n=1}^{\infty} v^{\lambda=\mathrm{L}}_{2n} \cos(2n\phi) \right] \,.
\end{align}
We will present results for the azimuthally averaged differential yield $\der N^{(0),\lambda=\mathrm{L}}$, the elliptic anisotropy $v^{\lambda=\mathrm{L}}_{2}$, and the quadrangular anisotropy $v^{\lambda=\mathrm{L}}_{4}$. Before presenting our numerical results we will first discuss the nonperturbative modeling of the WW gluon TMD at small-$x$ and nonperturbative contributions to the Sudakov form factor. 

\subsection{The WW gluon TMD in the MV model} 
While the rapidity evolution of the WW TMD satisfies the kinematically constrained JIMWLK renormalization group equation, the initial conditions for the evolution are intrinsically nonperturbative.  
To compute the WW gluon TMD, we will rely on a Gaussian approximation which allows one to express any correlator of Wilson lines (or their derivatives) in terms of a nontrivial funciton of the two point dipole correlator of Wilson lines $S_{Y}(\rbbpt)$ \cite{Blaizot:2004wv,Dumitru:2011vk,Iancu:2011nj}.  Within this approximation, the WW gluon TMD $\hat G_{Y}^{ij}(\rbbpt)$  can be expressed as \cite{Metz:2011wb,Dominguez:2011br,Boussarie:2021ybe}
\begin{align}
    \hat G^{ij}(\rbbpt)=\frac{2C_FS_\perp}{\alpha_s}\frac{\partial^i\partial ^j\Gamma(\rbbpt)} {\Gamma(\rbbpt)}\left[1-\exp\left(-\frac{C_A}{C_F}\Gamma(\rbbpt)\right)\right] \,,
\end{align}
with 
\begin{equation}
    \Gamma(\rbbpt)=-\ln\left(S(\rbbpt)\right) \,.
\end{equation}
For brevity, we dropped the rapidity dependence $Y$ in the dipole and the WW correlators as we only consider the MV model in our numerical analysis.

In our numerical study,  we will use the MV model parametrization 
\begin{equation}
    S(\rbbpt)=\exp\left[-\frac{\rbbpt^2Q_{s0}^2}{4}\ln\left(\frac{1}{m|\rbbpt| }+e\right)\right] \,,
\end{equation}
where we choose $Q_{s0}^2=1.0$ GeV$^2$ and a nonperturbative scale $m =0.241$ GeV of order $\Lambda_{\rm QCD}$ (see also \cite{Dumitru:2020gla,Dumitru:2021tvw}) as a proxy for a large nucleus.

\subsection{Modeling the Sudakov form factor} The Sudakov form factor is computed using Eq.\,\eqref{eq:Sudakov-final} with the two loop running coupling. Our default choice for the hard scale is $\mu_h^2=P_\perp^2+\bar Q^2$. The analytic expressions are provided in appendix~\ref{app:sudakov}. When computing the Sudakov factor with running coupling, one has to make sure that the QCD Landau pole does not belong to the $\mu$ integration range. In other words, we will  ensure that 
\begin{align}
   \rbbpt^2 < \frac{c_0^2}{\mu_h^2}\exp\left(\frac{1}{\beta_0 \alpha_s(\mu_h^2)} \right) \,.
\end{align}
In the TMD literature, there is a widely used prescription to tame the nonperturbative contributions from large distances \cite{Collins:1984kg}. The lower bound of the $\mu^2$ integration in Eq.\,\eqref{eq:Sudakov-final} is changed from $c_0^2/\rbbpt^2$ to $c_0^2/\rbbpt^{*2}$ where
\begin{align}
    \rbbpt^{*2}=\frac{\rbbpt^{2}}{1+\rbbpt^{2}/r_{\mathrm{max}}^2} \,.
    \label{eq:star_prescription}
\end{align}
In order to compensate for the missing contribution at large $\rbbpt^{2}$, one includes a nonperturbative Sudakov contribution following \cite{Sun:2014dqm,Prokudin:2015ysa}
\begin{align}
    \mathcal{S}(\mu_h^2,\rbbpt^2) \to \mathcal{S}(\mu_h^2,\rbbpt^{*2}) \,\mathcal{S}_{\mathrm{NP}}(\mu_h^2,\rbbpt^2) \,,
    \label{eq:sud-NP}
\end{align}
where
\begin{align}
    \mathcal{S}_{\mathrm{NP}}(\mu_h^2,\rbbpt^2) = \exp \left\{-2 \left[ \frac{g_1}{2} \rbbpt^2 + \frac{1}{4} \frac{g_2}{2} \ln\left( \frac{\mu_h^2}{Q_0^2} \right) \ln\left( \frac{\rbbpt^2}{\rbbpt^{*2}} \right) \right] \right\} \,,
\end{align}
with the overall factor of $2$ in the exponent due to the two quark jets in the final state. We use the same values as in \cite{Sun:2014dqm,Prokudin:2015ysa} (see also \cite{Stasto:2018rci,Benic:2022ixp} in the context of small-$x$ physics)  for $g_1= 0.212$ GeV$^2$, $g_2=0.84$, $Q_0^2= 2.4$ GeV$^2$, and $r_{\mathrm{max}} = 1.5$ GeV$^{-1}$, which are values constrained from fits to experimental data.

\subsection{Numerical results}

The NLO impact factor has two contributions:
(i) the soft Sudakov factor containing double and single Sudakov logs, and
(ii) the $\mathcal{O}(\alpha_s)$ coefficient function which includes the NLO hard factor as well as finite pieces from soft gluon emissions outside the jet cone.

\begin{figure}[H]
    \centering
    \includegraphics[width=0.49\textwidth]{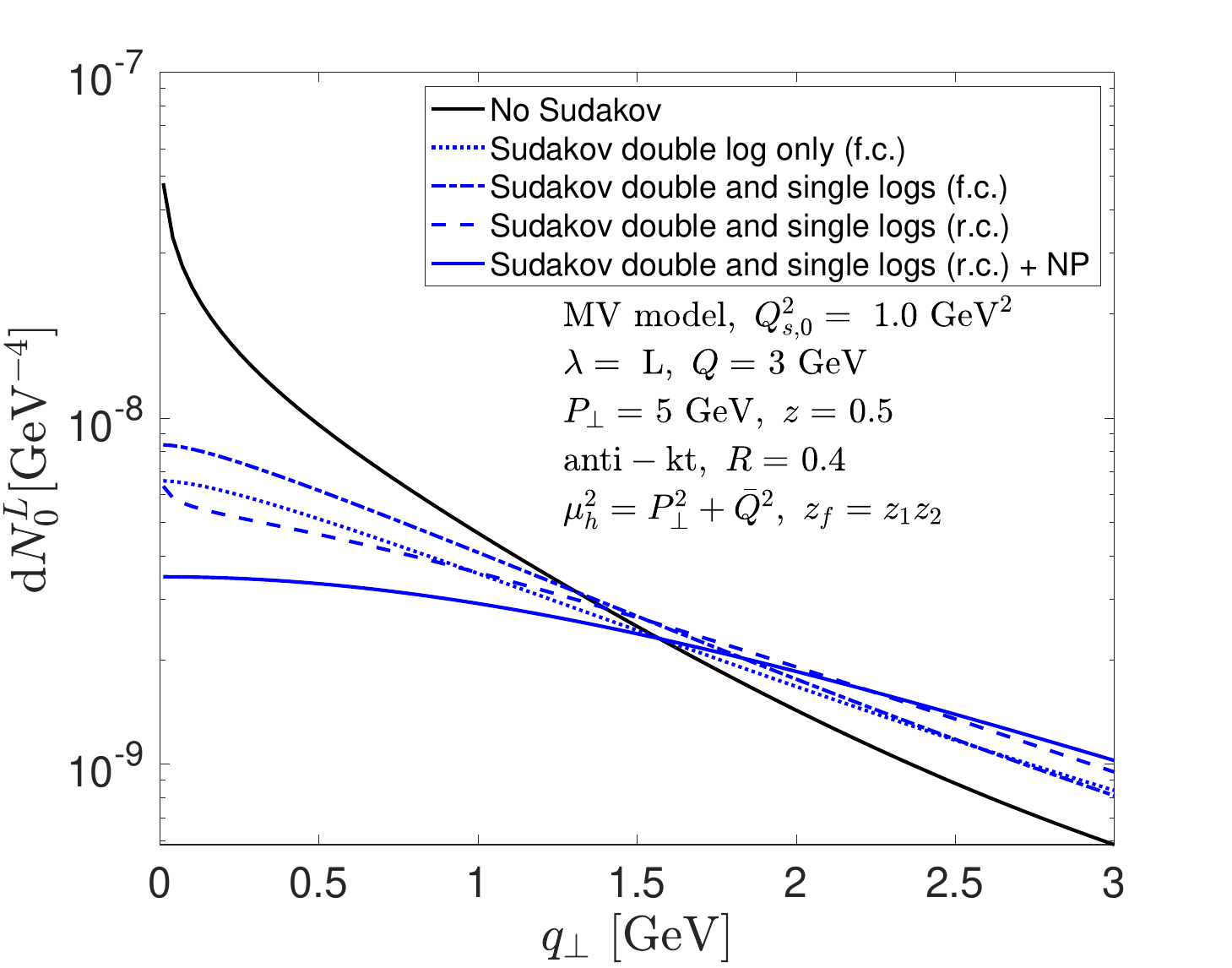}
    \includegraphics[width=0.49\textwidth]{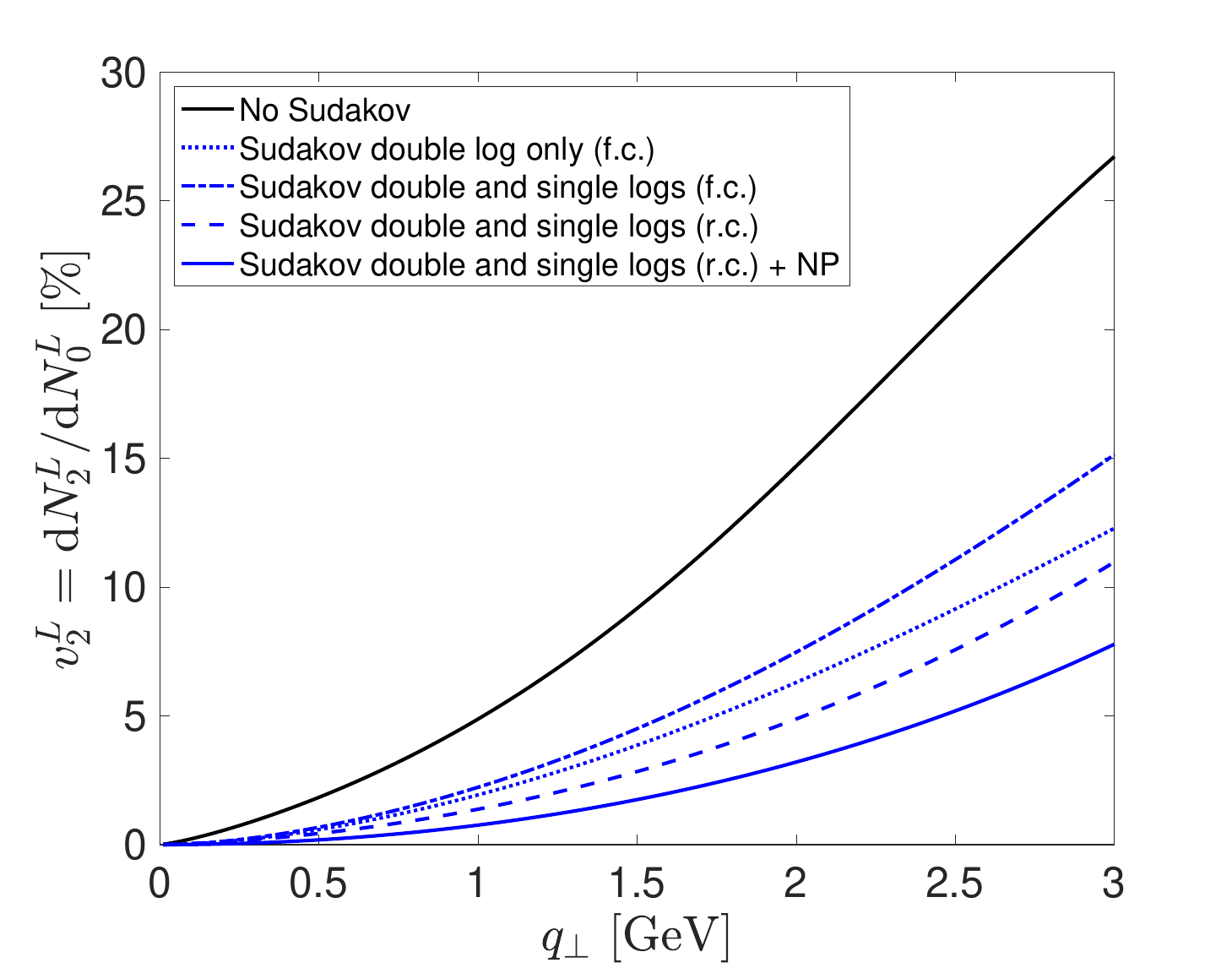}    

    \caption{Azimuthally averaged differential yield (left panel) and elliptic anisotropy (right panel) as a function of $q_\perp$. The black solid line represents the leading order (LO) baseline.  Blue curves are computed using LO impact factor + contributions to the NLO impact factor from the Sudakov factor alone and exclude 
    %the NLO finite pieces
    contributions from the NLO coefficient function. In the captions, f.c denotes fixed coupling, r.c running coupling, and NP the nonperturbative contributions to the Sudakov factor. See text for further explanation.}
    \label{fig:Sudakov}
\end{figure}

We begin this section by first studying the impact of the Sudakov form factor on the azimuthally averaged cross-section and the elliptic anisotropy as a function of $q_\perp$. Previous phenomenological studies for dijet/dihadron production in DIS at small $x$ \cite{Zheng:2014vka,vanHameren:2021sqc,Zhao:2021kae} only considered the Sudakov double logs for fixed strong coupling and ignored the nonperturbative contribution to the Sudakov form factor. In  Fig.\,\ref{fig:Sudakov}, 
we will illustrate the effect of Sudakov suppression beyond the double log approximation. For this purpose, we shall examine the LO impact factor + contributions to the NLO impact factor containing Sudakov logarithms (and excluding the NLO coefficient function):
\begin{align}
    \der \sigma^{(0),\lambda=\rm L}&=\alpha_{\rm em}\alpha_s e_f^2\deltatwo\Hcal_{\rm LO}^{0,\lambda=\rm L}  \int\frac{\der^2\rbbpt}{(2\pi)^4}e^{-i\qt\cdot\rbbpt}\hat G^0_{Y_f}(\rbbpt)\,\widetilde{\mathcal{S}}(\mu_h^2,\rbbpt^2) \,,
\label{eq:result_xsect_L-SudakovOnly}
\end{align}
and
\begin{align}
    \der \sigma^{(2),\lambda=\rm L}=\alpha_{\rm em}\alpha_s e_f^2\deltatwo\Hcal_{\rm LO}^{0,\lambda=\rm L}  \int\frac{\der^2\rbbpt}{(2\pi)^4}e^{-i\qt\cdot\rbbpt} \frac{\cos (2\theta)}{2} \hat h^0_{Y_f}(\rbbpt)\,\widetilde{\mathcal{S}}(\mu_h^2,\rbbpt^2) \,.
\label{result_xsect_L2_SudakovOnly}
\end{align}
We plot Eqs.\,\eqref{eq:result_xsect_L-SudakovOnly} and \eqref{result_xsect_L2_SudakovOnly} with four different contributions to $\widetilde{\mathcal{S}}(\mu_h^2,\rbbpt^2)$: (i) double log with fixed coupling, (ii) double and single logs with fixed coupling, (iii) double and single logs with running coupling, and (iv) double and single logs with running coupling and with non-perturbative Sudakov factor. For the fixed coupling case, we evaluate the  coupling at the hard scale $\mu_h$. For reference, we also include the LO impact factor which is obtained by setting the Sudakov form factor to 1 in Eqs.\,\eqref{eq:result_xsect_L-SudakovOnly} and \eqref{result_xsect_L2_SudakovOnly}.

Firstly, as also noticed in \cite{Zheng:2014vka,vanHameren:2021sqc,Zhao:2021kae}, the Sudakov form factor including double logarithms alone leads to a significant suppression of the yield at small $q_\perp$.
For out choice of kinematics and jet radii, adding the contribution of the Sudakov single log results in enhancement of the cross-section and its elliptic anisotropy as its coefficient has the opposite sign to that of the double log\footnote{The coefficient of the Sudakov single log depends on the kinematics and the jet radii. For smaller jet radii, we observe that the single log results in further suppression of the cross-section.}. In addition, we observe a significant suppression of the cross-section when we switch from fixed coupling $\alpha_s(\mu_h^2)$ to the running coupling (at two loops) $\alpha_s(\mu^2)$. Since $ \alpha_s(\mu_h^2) < \alpha_s(\mu^2)$ when $\mu_h^2 > \mu^2$, soft gluon emission is enhanced in the running coupling case and results in a stronger Sudakov suppression. The effect of the running of the coupling can be understood in more detail by expanding  Eqs.\,\eqref{eq:2loop-resummed}:
\begin{align}
    J_1 = \underbrace{\frac{1}{2} \alpha_s L^2 + \frac{1}{3}\beta_0 \alpha_s^2 L^3 + \frac{1}{4}\beta_0^2 \alpha_s^3 L^4}_{\mathrm{LL}} + \underbrace{\frac{1}{3}\beta_1 \alpha_s^3 L^3}_{\mathrm{NLL}} + \mathcal{O}(\alpha_s^4)\,,
    \label{eq:Sudakov_I1_expansion}
\end{align}
\begin{align}
    J_2 = \underbrace{\alpha_s L + \frac{1}{2} \beta_0 \alpha_s^2 L^2 + \frac{1}{3} \alpha_s^3 L^3}_{\mathrm{NLL}} + \underbrace{\frac{1}{2} \beta_1 \alpha_s^3 L^2 }_{\mathrm{N^2LL}}+ \mathcal{O}(\alpha_s^4)\,,
    \label{eq:Sudakov_I2_expansion}
\end{align}
where $L = \ln\left( \frac{\mu_h^2 \rbbpt^2}{c_0^2} \right) $. The fixed coupling case accounts only for the first terms in Eqs.\,\eqref{eq:Sudakov_I1_expansion} and \eqref{eq:Sudakov_I2_expansion}. On the other hand, the running of the coupling at two loops contains all leading logs (LL) $\alpha_s^n L^{n+1}$, next-to-leading-logs (NLL) $\alpha_s^n L^{n}$, as well as a subset of the N$^2$LL terms. These contributions are responsible for the stronger Sudakov suppression observed in Fig.\,\ref{fig:Sudakov}. Lastly, adding the non-perturbative contribution to the Sudakov factor further suppresses the cross-section.
Since the effects plotted are large, to understand the interplay between saturation and Sudakov suppression it is therefore crucial that both double and single logs as well as the running of the coupling and the non-perturbative contribution to the Sudakov form factor are taken into account in phenomenological studies.

\begin{figure}[H]
    \centering
    \includegraphics[width=0.49\textwidth]{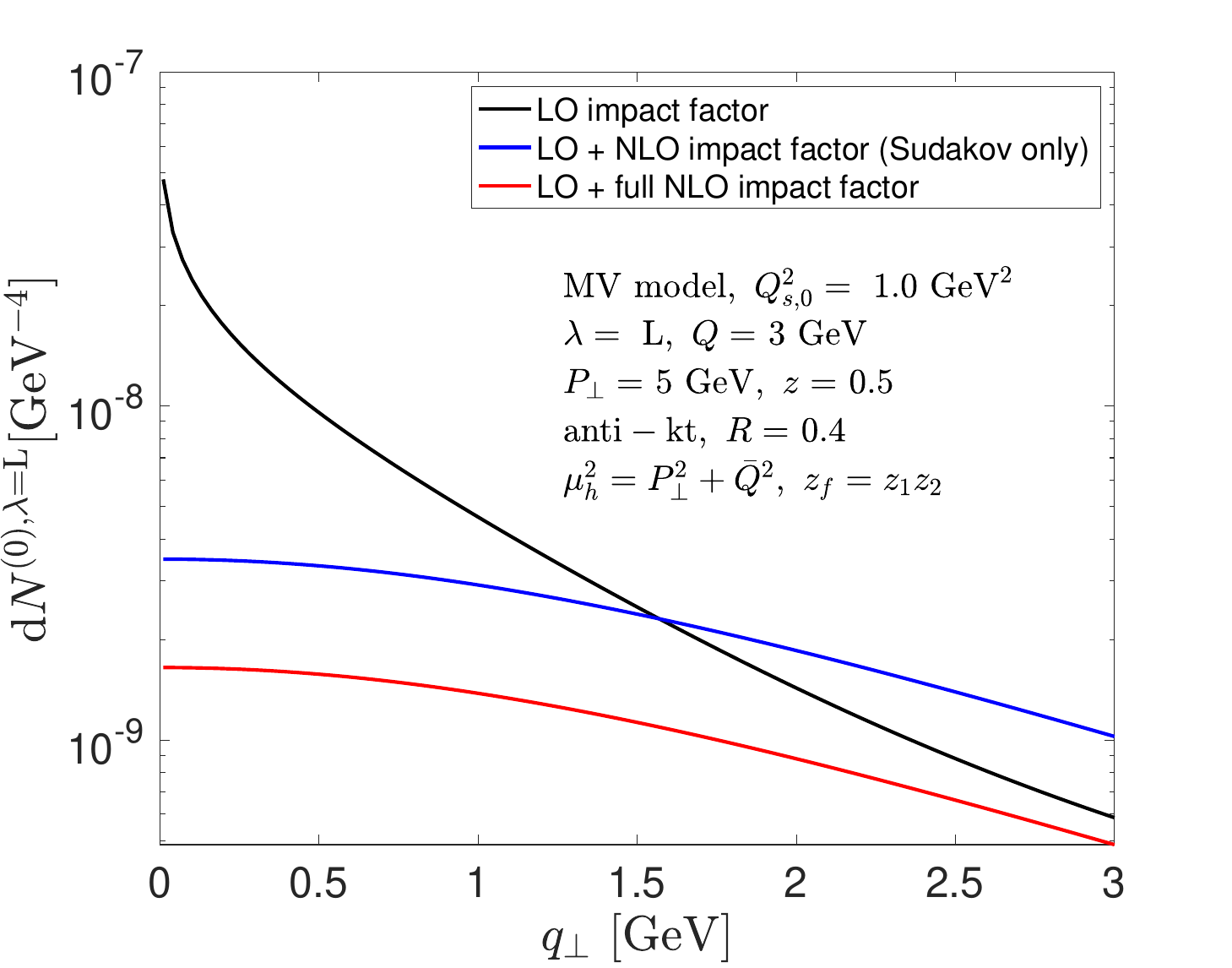}
    \includegraphics[width=0.49\textwidth]{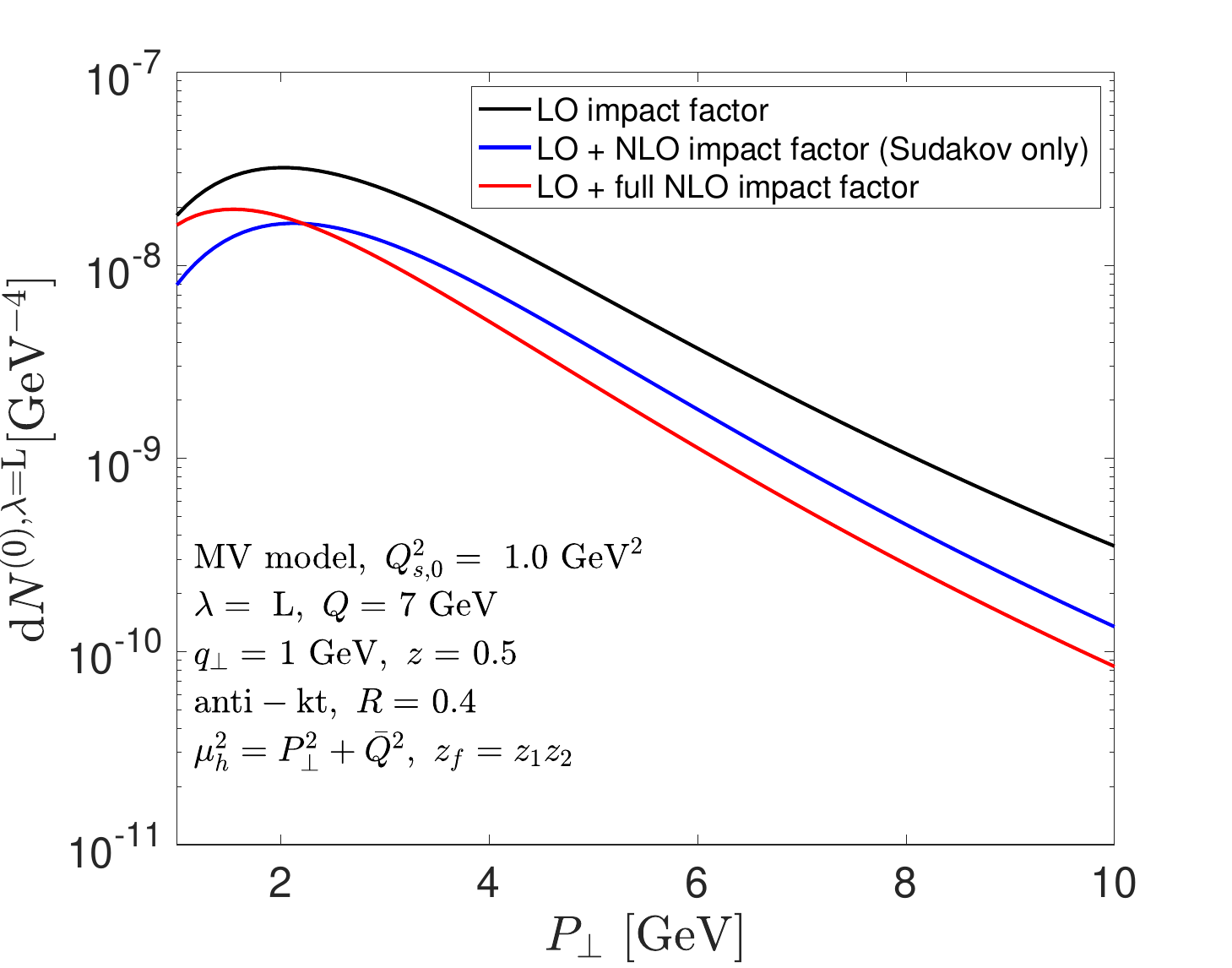}
    
    \caption{Azimuthally averaged differential yield as a function of $q_\perp$ (left panel) and as a function of $P_\perp$ (right panel). Black line represents the leading order result. Blue line includes the contributions to the NLO impact factor from Sudakov logarithms. Red curve is the LO + NLO result where the latter includes the soft factor times the coefficient function.
    %and hard factors (Sudakov and finite $\mathcal{O}(\alpha_s)$ pieces respectively).
    }
    \label{fig:dijet_NLO}
\end{figure}

In Fig.\,\ref{fig:dijet_NLO}, we turn our attention to the  $\mathcal{O}(\alpha_s)$ NLO coefficient function that multiplies the soft factor. We compute the azimuthally averaged yield as a function of $q_\perp$ (in the left panel) and as a function of $P_\perp$ (in the right panel). We display the results employing the LO impact factor, the LO + NLO contributions to the impact factor that contain Sudakov logarithms (see Eq.\,\eqref{eq:result_xsect_L-SudakovOnly}), and the LO + full NLO impact factor which contains both Sudakov and the finite $\mathcal{O}(\alpha_s)$ coefficient function in Eq.\,\eqref{eq:result_xsect_L-summary}. The size of the NLO coefficient function are strongly dependent on $P_\perp$ or more precisely, on the ratio $\chi = \bar{Q}/P_\perp$ as expected from the factorization formula in Eq.\,\eqref{eq:result_xsect_L-summary}. At small values of $\chi$ the NLO finite pieces enhance the differential yield while they decrease it at large $\chi$. As noted in \cite{Caucal:2022ulg}, the finite NLO terms for the azimuthally averaged cross-section have a contribution from linearly polarized gluons due to their interplay with soft gluon radiation. This contribution is rather small in the saturation region $q_\perp \lesssim Q_s$ since the relative contribution from linearly polarized gluons in the WW distribution is suppressed. At the end of day, the NLO coefficient function is nearly  independent of $q_\perp$ and leads to a constant shift of the cross-section as a function of $q_\perp$ as shown on Fig.\,\ref{fig:dijet_NLO}-left when going from the blue to the red curve.

\begin{figure}[H]
    \centering
    \includegraphics[width=0.49\textwidth]{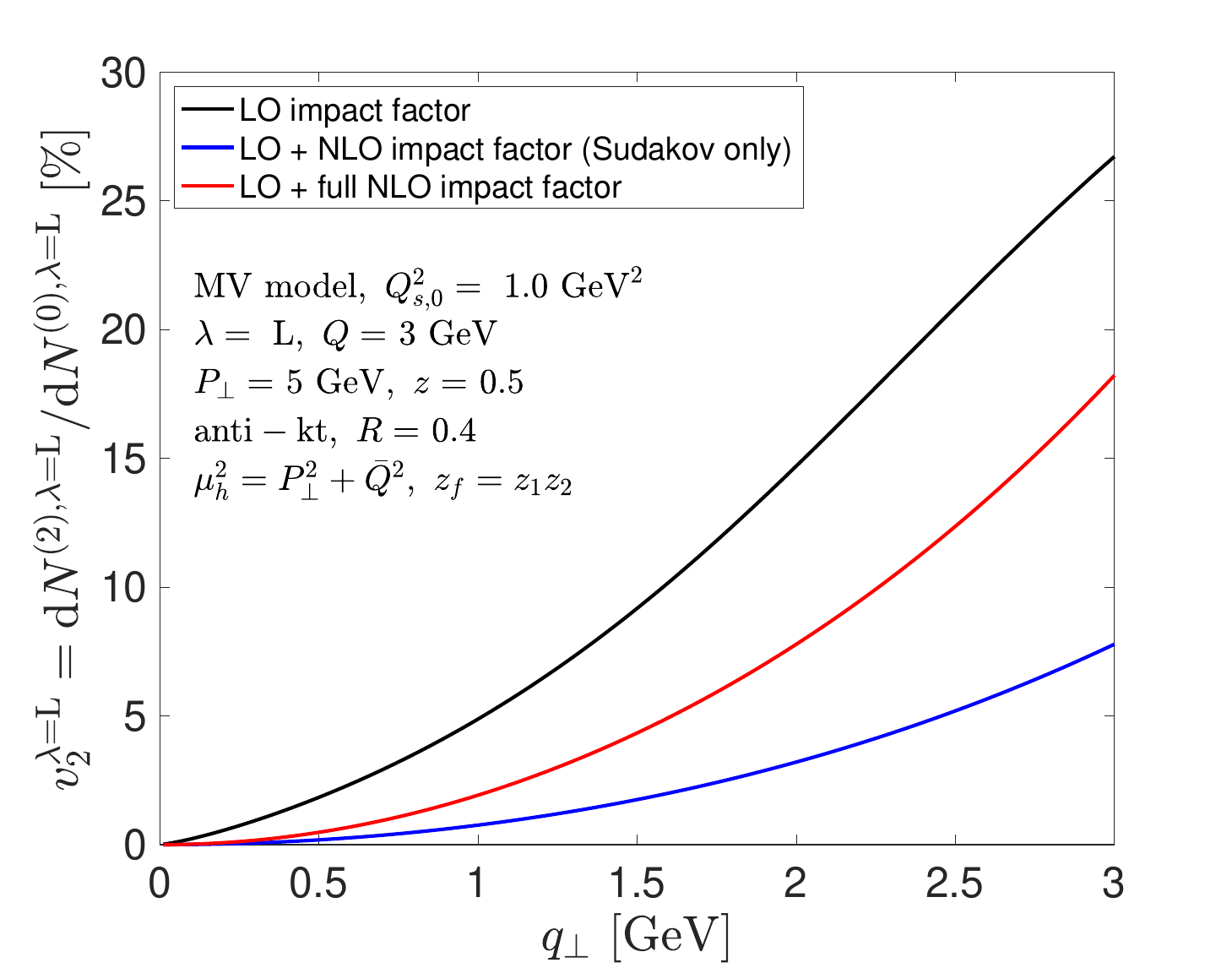}
    \includegraphics[width=0.49\textwidth]{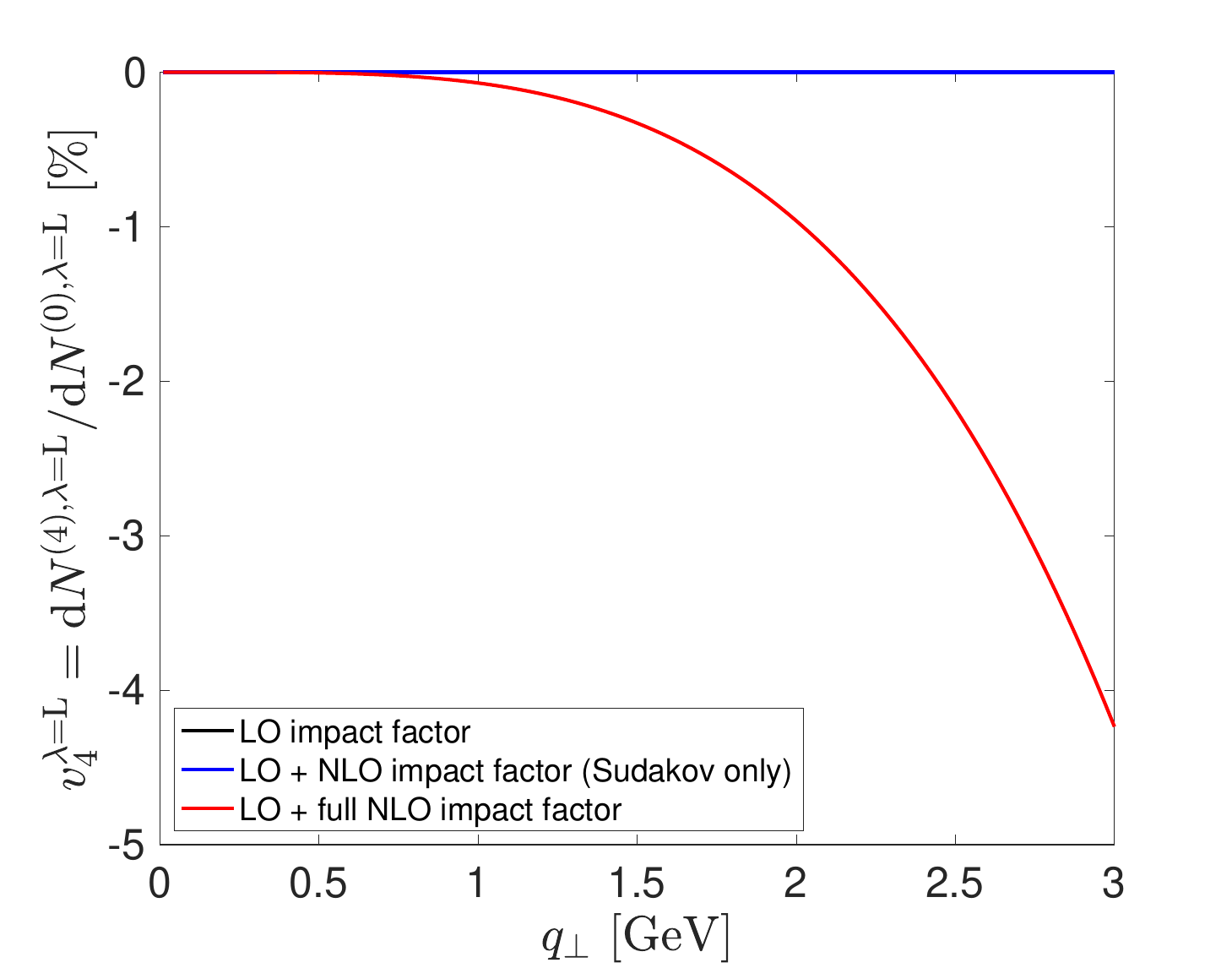}
    \caption{Elliptic (left panel) and quadrangular (right panel) anisotropies as a function of $q_\perp$. Black line represents the leading order result. Blue line includes the contributions to the NLO impact factor from Sudakov logarithms. Red curve is the full LO+NLO (Sudakov times coefficient function) result.
    %finite $\mathcal{O}(\alpha_s)$ terms
    }
    \label{fig:dijet_NLO_anisotropies}
\end{figure}

Next we study the elliptic and quadrangular anisotropies as a function of the momentum imbalance $q_\perp$ in Fig.\,\ref{fig:dijet_NLO_anisotropies}. At LO, as seen from Eq.\,\eqref{eq:LOb2b-L}, the elliptic anisotropy is proportional to the ratio between linearly polarized to unpolarized gluons which grows with $q_\perp$. The Sudakov factor significantly decreases the elliptic anisotropy. This effect is nontrivial since $v_2$ is defined as a ratio and one expects therefore a partial cancellation of the Sudakov suppression. However  the unpolarized and linearly polarized WW gluon TMDs are not identical and thus the Sudakov factor affects them differently. On the other hand, the effect of the NLO coefficient function is to enhance the elliptic anisotropy, as it now acquires a contribution from unpolarized gluons. This can be seen from the last two lines of Eq.\,\eqref{result_xsect_L2} and physically corresponds to the effect of soft gluon emissions that are near to but yet outside the jet cone.  These emissions tend to preferentially align the imbalance $\qt$ parallel (or anti-parallel) to the transverse jet direction $\Pt$ \cite{Hatta:2021jcd} (see also \cite{Tong:2022zwp} for the case of lepton-jet correlations). The NLO finite pieces also produce a small quadrangular anisotropy \eqref{eq:R2R2b2b-cn-final} which is completely absent in the leading order and in the LO + ``Sudakov only" result.

We close this section by studying the $Q$ dependence of the azimuthally averaged differential yield for two different choices of the hard factorization scale $\mu_h$ in the Sudakov factor. As discussed in Section \ref{sec:summary}, different choices of $\mu_h$ also lead to different NLO coefficient function. A common choice of the factorization scale is $\mu_h^2 = \Pt^2$ \cite{Mueller:2013wwa,Zheng:2014vka,Zhang:2021tcc}. However, as we discussed in Section \ref{sec:summary}, we find that the associated NLO finite piece develops large negative logs at large $Q$, resulting in 
an unphysical negative yield. A more natural choice is the combination $\mu_h^2 = \Pt^2 + \bar{Q}^2$ which renders the yield positive. As expected, these scheme choices do not affect the $Q\lesssim P_\perp$ region.

\section{Summary and outlook}
\label{sec:Conclusions}

In this paper, we  performed the first complete analysis of the back-to-back limit of the NLO dijet impact factor at small Bjorken $x_{\rm Bj}$ in the CGC effective field theory. Our study significantly extends the results obtained in our previous paper \cite{Caucal:2022ulg} where the emphasis was placed on Sudakov double and single logarithms appearing in the NLO impact factor and the interplay of these contributions with rapidity evolution. It was believed in \cite{Caucal:2022ulg} that NLO hard factor contributions not containing Sudakov logarithms were only factorizable in the strict dilute TMD limit of $P_\perp\gg q_\perp\gg Q_s$. Our principal result here is that the back-to-back limit of the NLO impact factor can indeed be written in a fully factorized form to all orders in $Q_s/q_\perp$ and up to power corrections in $q_\perp/P_\perp$ (or equivalently, $Q_s/P_\perp$). 

This factorized expression involves three important ingredients.\\ (i) The WW gluon TMD which, in light cone gauge and light cone quantization, represents the number density of gluons per unit momentum in the target at small-$x$. This WW TMD obeys a kinematically constrained JIMWLK renormalization group equation in rapidity. \\
(ii) A Sudakov soft factor that resums to all orders double and single logarithms of $P_\perp/q_\perp$ appearing in the impact factor at NLO. \\
(iii) A NLO coefficient function that gathers all the $\mathcal{O}(\alpha_s)$ finite terms which are not enhanced by rapidity logarithms or logarithms in $P_\perp/q_\perp$. 

Following \cite{Caucal:2022ulg}, we confirmed from a detailed analysis of all virtual diagrams that the WW gluon TMD must satisfy a kinematically constrained JIMWLK evolution equation in rapidity.  
We were able to go further and unambiguously fix the transverse scale in the kinematic constraint that imposes lifetime ordering of successive gluon emissions. We leave for forthcoming work a detailed numerical study of Eq.\,\eqref{wwe} giving the kinematically constrained evolution of the WW gluon TMD.

Fixing the transverse scale of the kinematic constraint also enables us to eliminate ambiguities in the single Sudakov logarithms. In the end, both Sudakov double and single logarithms turn out to be independent of the small-$x$ resummation scheme. This extends to single logarithmic accuracy  the results in \cite{Mueller:2013wwa} in which the authors claim that Sudakov and small-$x$ resummation can be performed independently and simultaneously in the Regge limit.

We computed here for the first time remarkably compact analytical expressions for the  NLO coefficient function for longitudinally polarized virtual photons. Such compact results will be very beneficial for quantitative predictions of our NLO results for future data from the Electron-Ion Collider. Our NLO results however display a residual rapidity factorization scheme dependence that can be quantified further beyond NLO.

We performed a preliminary numerical study of our analytic results with the aim of providing the reader with a feel for the order of magnitude of individual contributions in our factorized formula at NLO coming from the soft and hard contributions to the impact factor. 
We demonstrated the significant impact of the Sudakov double logarithmic soft factor contributions relative to the LO impact factor and further showed the relative impact of the contributions from the Sudakov single logarithms to the soft factor. We observed that the running of the coupling in the soft factor has a sizable impact on the differential cross-sections as a function of the dijet imbalance $q_\perp$.

The effect of the NLO coefficient function can be summarized by distinguishing between the NLO hard factors coming from hard virtual and in-cone real gluons and from the rest of the contributions to the coefficient function that are associated with real soft out-of-cone gluon emissions. On the one hand, the NLO hard factors do not modify the shape of the cross-section as a function of $q_\perp$ (in sharp contrast to the Sudakov form factor), and contribute an overall shift which can be comparable in magnitude to the soft factors. However,  we observed that these contributions can qualitatively modify the shape of distributions as a function of $P_\perp$ since the NLO hard factor  depends on the ratio $P_\perp/Q$. On the other hand,  finite pieces  contributing to the NLO coefficient function that come from real soft gluons emissions outside the jet cone significantly alter the shape  (as a function of $q_\perp$) of the $v_2$ and $v_4$ azimuthal anisotropies. A fully quantitative analysis of the contribution of these NLO finite pieces requires small-$x$ evolution in order to mitigate the residual dependence of the hard factors on the rapidity factorization scale.

Our study can be developed further in several directions. Firstly, 
we can compute results for the hard factors in the scattering of transversely polarized virtual photons off the target to complement the results presented here for longitundinally polarized virtual photons. Secondly, it is crucial to compute the kinematically constrained small-$x$ evolution of the WW gluon TMD extending prior  work on the dipole distribution~\cite{Beuf:2020dxl,Iancu:2015joa}. 
It is also important to 
understand the contribution of residual logarithms of the ratio of the virtuality of the photon and the dijet invariant mass, which can in principle be large for a wide separation of the two scales. 
This comment also applies for the remaining $\ln(R)$ dependence in the impact factor; for such contributions, we note that resummation techniques are already available \cite{Dasgupta:2014yra,Kang:2016mcy}.
Another open question with regard to back-to-back dijets is the magnitude of the higher twist $q_\perp/P_\perp$ contributions that require one to go beyond the correlation limit and are not considered here\footnote{The study of power corrections has been recently an active area of research within SCET ~\cite{Ebert:2021jhy,Rodini:2022wic,Gamberg:2022lju} albeit not at small $x$.}. In principle, such contributions can be systematically computed following the method outlined in Appendix \ref{app:power-suppressed}, resulting in the appearance of distributions beyond the WW TMD.

Not least, a related and interesting follow-up calculation relevant for EIC phenomenology would be to replace our dijet computation with one for  dihadrons using fragmentation functions. Forthcoming papers will address the aforementioned important questions towards the goal of making quantitative predictions for the EIC with the aim of isolating the dynamics arising from the energy evolution of saturated gluons from the physics governing the suppression of radiation at the edges of phase space.

\section*{Acknowledgements}

We are grateful to Zhongbo Kang, Feng Yuan and Jian Zhou for valuable discussions.
F.S. is supported by the National Science Foundation under grant No. PHY-1945471, and partially supported by the UC Southern California Hub, with funding from the UC National Laboratories division of the University of California Office of the President.
T.S. kindly acknowledges support of the Polish National Science Center (NCN) Grants No.\,2019/32/C/ST2/00202 and 2021/43/D/ST2/03375.
B.P.S. and R.V. are supported by the U.S. Department of Energy, Office of Science, Office of Nuclear Physics, under DOE Contract No.~DE-SC0012704 and within the framework of the Saturated Glue (SURGE) Topical Theory Collaboration.R.V.'s work is also supported in part by an LDRD grant from Brookhaven Science Associates. 
\appendix

\section{Details of the calculation of virtual hard factors}
\label{app:virtual-details}
In this appendix, we shall provide mathematical details necessary to reproduce our results for the hard factors coming from virtual diagrams in the NLO back-to-back dijet cross-section.

\subsection{Computation of $\Hcal_{\rm NLO,1}$}
\label{app:NLO1}

In section~\ref{sub:V2-SE2}, the hard factor $\Hcal_{\rm NLO,1}$  was computed based on our knowledge of the $z_g\to0$ behaviour of the $z_g$ integrand appearing in $\Hcal_{\rm NLO,2}$. There is an alternative method of computing this hard factor that does not rely on the calculation of $\Hcal_{\rm NLO,2}$.

The building block of $\Hcal_{\rm NLO,1}$ is the integral 
\begin{align}
    I_0&=\int\frac{\der^2\ut}{(2\pi)}e^{-i\Pt\cdot\ut}K_0(\bar Qu_\perp)\ln(P_\perp u_\perp) \nonumber \\
    &=\frac{1}{\Pt^2}\int_0^\infty\der u \ u   \textrm{J}_0(u)K_0(\chi u)\ln(u) \,,
\end{align}
where, in the second line, we have performed the angular integration and the change of variable $u=P_\perp u_\perp$.

Using the standard ``replica" trick
\begin{equation}
\ln(u)=\lim\limits_{\alpha\to 0}\partial_\alpha u^\alpha\,,
\end{equation}
we can reexpress the $u$ integral in terms of the ordinary hypergeometric function ${}_2F_1(a,b,c;z)$:
\begin{align}
    \Pt^2I_0&=\lim\limits_{\alpha\to 0}\partial_\alpha  \int_0^\infty \der u \ u^{1+\alpha}\textrm{J}_0(u)K_0(\chi u) \nonumber \\
      &=\lim\limits_{\alpha\to 0}\partial_\alpha\left\{\frac{2^\alpha}{\chi^{2+\alpha}}\Gamma^2\left(1+\frac{\alpha}{2}\right){}_2F_1\left(1+\frac{\alpha}{2},1+\frac{\alpha}{2},1;-\chi^{-2}\right)\right\} \nonumber\\
      &=-\frac{\ln(\chi/c_0)}{1+\chi^2}+\frac{1}{2\chi^2}\left[\partial_a\,{}_2F_1(1,1,1;-\chi^{-2})+\partial_b\,{}_2F_1(1,1,1;-\chi^{-2})\right] \,.
\end{align}
To obtain the second line, we  used results from section 13.45 in \cite{watson1922treatise} for Weber-Schafheitlin type integrals.
Using the identity \cite{abramowitz1964handbook}
\begin{equation}
    {}_2F_1(a,b,c;z)=(1-z)^{-b}{}_2F_1\left(b,c-a,c;\frac{z}{z-1}\right)\,, \label{eq:ab-2F1}
\end{equation}
and ${}_2F_1(a,0,1;z)=1$,
the derivative with respect to $b$ gives
\begin{equation}
\partial_b\,{}_2F_1(1,1,1;z)=-\frac{\ln(1-z)}{1-z}\,.
\end{equation}
Then using ${}_2F_1(a,b,b;z)=(1-z)^{-a}$, the derivative with respect to $a$ can be easily computed:
\begin{equation}
\partial_a\,{}_2F_1(1,1,1;z)=-\frac{\ln(1-z)}{1-z}\,.
\end{equation}
The integral $I_0$ then reads
\begin{align}
    I_0=-\frac{1}{\Pt^2+\bar Q^2}\left[\ln\left(\frac{\chi}{c_0}\right)+\ln\left(1+\frac{1}{\chi^2}\right)\right] \,.
\end{align}
The first NLO hard factor can be related to the integral $I_0$
%$\Mcal$ 
since
\begin{align}
    \Hcal_{\rm NLO,1}^{\lambda=\rmL,ij}(\Pt)&=16z_1^3z_2^3Q^2\frac{i\Pt^j}{(\Pt^2+\bar Q^2)^2}\left[\int\frac{\der^2\ut}{(2\pi)}e^{-i\Pt\cdot\ut}\ut^i K_0(\bar Qu_\perp)\ln(\Pt^2\ut^2)\right] \nonumber \\
    &=16z_1^3z_2^3Q^2\frac{-\Pt^j}{(\Pt^2+\bar Q^2)^2}\left[2\partial^i_{\Pt}I_0-\frac{2\Pt^i}{\Pt^2(\Pt^2+\bar Q^2)}\right] \,.
\end{align}
After taking the derivative of $I_0$ with respect to $\Pt$, one ends up with the same result as Eq.\,\eqref{eq:NLO1-final} obtained 
using a different method.

\subsection{Computation of $\Hcal_{\rm NLO,2}$}
\label{app:NLO2}

This appendix deals with the calculation of the integral $I_1^i$ defined in Eq.\,\eqref{eq:Iint-def} by
\begin{equation}
I_1^i\equiv\int\frac{\der^2 \ut}{(2\pi)}e^{-i\Pt\cdot\ut}\ut^i K _0(\bar Q_{\rm V3}\ut')e^{i\frac{z_g}{z_1}\Pt\cdot\ut}\Jcal_{\odot}\left(\ut,\left(1-\frac{z_g}{z_1}\right)\Pt,\Delta_{\rm V3}\right)\,.
\end{equation}
The key manipulation consists of changing the variable  $\lt$ to $\lt-\Kt$, where
\begin{equation}
    \Kt =\left(1-\frac{z_g}{z_1}\right)\Pt \,,\label{eq:Kt-def}
\end{equation}
in the definition of $\Jcal_{\odot}$ in Eq.\,\eqref{eq:tough-integral} so that the phases within cancel except for a remaining $e^{i\lt\cdot\ut}$ factor. One can then switch the order of integration to perform the $\ut$ integral first:
\begin{align}
    I_1^i&=\int\frac{\der^2 \ut}{(2\pi)}\ut^iK_0(\bar Q_{\rm V3}\ut)\int\frac{\der^2\lt}{(2\pi)}\frac{2(\lt+\Kt) \cdot \Kt \ e^{i\lt \cdot \ut}}{(\lt+\Kt)^2\left[\lt^2-\Delta^2_{\rm V3} - i \epsilon\right]} \nonumber \\
    &=\int\frac{\der^2\lt}{(2\pi)}\frac{2(\lt+\Kt) \cdot \Kt}{(\lt+\Kt)^2\left[\lt^2-\Delta^2_{\rm V3} - i \epsilon\right]}\int\frac{\der^2 \ut}{(2\pi)}\ut^i\textrm{K}_0(\bar Q_{\rm V3}\ut)e^{i\lt \cdot \ut} \nonumber \\
    &=2i\int\frac{\der^2\lt}{(2\pi)}\frac{2(\lt+\Kt) \cdot \Kt \ \lt^i }{(\lt+\Kt)^2\left[\lt^2-\Delta^2_{\rm V3} - i \epsilon\right](\lt^2+\bar Q_{\rm V3}^2)^2} \,.
\end{align}
The remaining integral over $\lt$ can be performed in polar coordinates:
\begin{equation}
    I_1^i=4i\int_0^\infty \der\ell \frac{\ell^2\Kt^i}{(\ell^2+K_\perp^2)\left[\ell^2-\Delta^2_{\rm V3} - i \epsilon\right](\ell^2+\bar Q_{\rm V3}^2)^2}\int_0^{2\pi}\frac{\der\theta}{2\pi}\frac{\ell\cos^2(\theta)+K_\perp\cos(\theta)}{\left(1+\frac{2\ell K_\perp}{\ell^2+K_\perp^2}\cos(\theta)\right)} \,.
\end{equation}
The integral over $\theta$ is interesting. First one notices that for $\ell=K_\perp$, it gives 
\begin{equation}
    \int_0^{2\pi}\frac{\der\theta}{2\pi}\frac{\ell\cos^2(\theta)+K_\perp\cos(\theta)}{\left(1+\frac{2\ell K_\perp}{\ell^2+K_\perp^2}\cos(\theta)\right)}=\ell\int_0^{2\pi}\frac{\der\theta}{2\pi}\cos(\theta)=0 \,.
\end{equation}
For $\ell<K_\perp$, one finds
\begin{equation}
    \int_0^{2\pi}\frac{\der\theta}{2\pi}\frac{\ell\cos^2(\theta)+K_\perp\cos(\theta)}{\left(1+\frac{2\ell K_\perp}{\ell^2+K_\perp^2}\cos(\theta)\right)}=-\frac{\ell}{2K_\perp^2}(\ell^2+K_\perp^2) \,,
\end{equation}
whereas for $\ell>K_\perp$, one gets
\begin{equation}
    \int_0^{2\pi}\frac{\der\theta}{2\pi}\frac{\ell\cos^2(\theta)+K_\perp\cos(\theta)}{\left(1+\frac{2\ell K_\perp}{\ell^2+K_\perp^2}\cos(\theta)\right)}=\frac{1}{2\ell}(\ell^2+K_\perp^2) \,.
\end{equation}
Hence the $\theta$ integration leads to a discontinuity at  $\ell=K_\perp$ and the $\ell$ integrand vanishes over a set of zero measure. The condition $\ell=K_\perp$ corresponds to the condition $|\kgt-\Pt|=(1-z_g/z_1)|\Pt|$ in terms of the virtual gluon transverse momentum $\kgt$ \cite{Caucal:2021ent}. In particular, the kinematic configuration  $z_1\kgt-z_g\Pt=0$, namely, a virtual gluon collinear to the quark jet, does not contribute to the integral for this specific diagram.

Based on these results, one must separate the $\ell$ integral into two pieces
\begin{equation}
   I_1^i=I_{1,<}^i+I_{1,>}^i \,,
\end{equation}
with
\begin{align}
    I_{1,<}^i&=-\frac{2i\Kt^i}{\Kt^2}\int_0^{K_\perp} \der\ell\frac{\ell^3}{\left[\ell^2-\Delta^2_{\rm V3}\right](\ell^2+\bar Q_{\rm V3}^2)^2} \,, \\
    I_{1,>}^i&=2i\Kt^i\int_{K_\perp}^\infty \der\ell\frac{\ell}{\left[\ell^2-\Delta^2_{\rm V3} - i \epsilon\right](\ell^2+\bar Q_{\rm V3}^2)^2} \,.
\end{align}
In $I_{1,<}^i$ we have removed the $-i\epsilon$ prescription since $|\Kt|<\Delta_{\rm V3}$ and therefore the support of the integral does not encounter the pole. This integral reads
\begin{equation}
    I_{1,<}^i=\frac{-i\Kt^i}{(\Delta_{\rm V3}^2+\bar Q_{\rm V3}^2)^2}\left[\frac{\Delta_{\rm V3}^2+\bar Q_{\rm V3}^2}{\Kt^2+\bar Q_{\rm V3}^2}+\frac{\Delta^2_{\rm V3}}{\Kt^2}\ln\left(\frac{\bar Q_{\rm V3}^2(\Delta_{\rm V3}^2-\Kt^2)}{\Delta_{\rm V3}^2(\Kt^2+\bar Q_{\rm V3}^2)}\right)\right] \,. \label{eq:l3-int}
\end{equation}
For $I_{1,>}^i$, we use the formula
\begin{equation}
    \frac{1}{x-i\epsilon}=\textrm{PV}\frac{1}{x}+i\pi\delta(x) \,,
\end{equation}
to properly account for the $-i\epsilon$ prescription in the contour integration, and we find in the end
\begin{align}
    I_{1,>}^i&=\frac{-i\Kt^i}{(\Delta_{\rm V3}^2+\bar Q_{\rm V3}^2)^2}\left[\frac{\Delta_{\rm V3}^2+\bar Q_{\rm V3}^2}{\Kt^2+\bar Q_{\rm V3}^2}+\ln\left(\frac{\Delta_{\rm V3}^2-\Kt^2}{\Kt^2+\bar Q_{\rm V3}^2}\right)-i\pi\right] \,. \label{eq:l-int-inf}
\end{align}
Adding these two terms together, one finds Eq.\,\eqref{eq:IV3-final}.

\subsection{Computation of $\Hcal_{\rm NLO,3}$}
\label{app:NLO3}

In this section, we will provide details on the calculation of the integral $I_2$ defined by Eq.\,\eqref{eq:Lint-def}:
\begin{align}
I_2&\equiv\int\frac{\der^2\ut}{(2\pi)}e^{-i\Pt\cdot\ut}\int\frac{\der^2\rt}{(2\pi)}\frac{1}{\rt^2}\left[ K_0\left(\bar Q\sqrt{\ut^2+\omega\rt^2}\right)-e^{-\frac{\rt^2}{\ut^2e^{\gamma_E}}}K_0(\bar Qu_\perp)\right]\,.
\end{align}
We first consider the second (UV subtraction) term, whose dependence on $\ut$ through the exponential prefactor complicates the calculation. It is simpler to compute the integral with a UV subtraction term proportional to 
\begin{equation}
    \frac{1}{\rt^2}e^{-\frac{\rt^2}{2\xi}}K_0(\bar Q u_\perp) \,,
\end{equation}
with $\xi>0$ a fixed number independent of $\ut$ with the idea being that one adds and subtracts this term inside the $\rt$ integral. (See also \cite{Hanninen:2017ddy,Caucal:2021ent}.) Since~\cite{DBLP:journals/corr/abs-1912-05812}
\begin{equation}
    \int\frac{\der^2 \rt}{(2\pi)}\frac{1}{\rt^2}\left[e^{-\frac{\rt^2}{2\xi}}-e^{-\frac{\rt^2}{\ut^2e^{\gamma_E}}}\right]=\frac{1}{2}\ln\left(\frac{c_0\xi}{\ut^2}\right)\,,
\end{equation}
we can write the hard factor as
\begin{align}
   I_2&=\int\frac{\der^2\ut}{(2\pi)}e^{-i\Pt\cdot\ut}\int\frac{\der^2\rt}{(2\pi)}\frac{1}{\rt^2}\left[ K_0\left(\bar Q\sqrt{\ut^2+\omega\rt^2}\right)-e^{-\frac{\rt^2}{2\xi}}K_0(\bar Qu_\perp)\right]\nonumber\\
   &+\frac{1}{2}\int\frac{\der^2\ut}{(2\pi)}e^{-i\Pt\cdot\ut}\ln\left(\frac{c_0\xi}{\ut^2}\right)K_0(\bar Qu_\perp)\,.\label{eq:HNLO-3-undone}
\end{align}
The integral over $\ut$ in the second line is done using the identity in  Eq.\,\eqref{eq:app-LO-ln}.

To compute the integral over $\ut$ and $\rt$ of the first line in Eq.\,\eqref{eq:HNLO-3-undone}, the trick is to use the identity~\cite{Caucal:2022ulg}
\begin{equation}
    K_0\left(\bar Q\sqrt{\ut^2+\omega\rt^2}\right)=-i\rt^k\int\frac{\der^2\ltone}{(2\pi)}\int\frac{\der^2\lttwo}{(2\pi)}\frac{\lttwo^ke^{i\ltone \cdot\ut}e^{i\lttwo\cdot\rt}}{(\ltone^2+\bar Q^2)(\lttwo^2+\omega(\ltone^2+\bar Q^2))}\,,
\end{equation}
which introduces two supplemental transverse integrals with the benefit of putting the transverse coordinate dependence on $\ut$ and $\rt$ inside pure phases.
The integration over $\ut$ gives a Dirac delta function that enforces $\ltone=\Pt$ so that one easily performs the $\ut$ and $\ltone$ integrals, leading to
\begin{align}
     I_2&=\frac{1}{\Pt^2+\bar Q^2}\int\frac{\der^2\rt}{(2\pi)}\frac{(-i\rt^k)}{\rt^2}\int\frac{\der^2 \lttwo}{(2\pi)}\frac{\lttwo^k e^{i\lttwo\cdot\rt}}{\lttwo^2+\Delta^2}+\frac{1}{2(\Pt^2+\bar Q^2)}\ln\left(\frac{(\Pt^2+\bar Q^2)^2\xi}{c_0\bar Q^2}\right)\nonumber \\
     &=\frac{1}{\Pt^2+\bar Q^2}\int\frac{\der^2\rt}{(2\pi)}\left[\frac{\Delta}{r_\perp} K_1\left(\Delta r_\perp\right)-\frac{1}{\rt^2}e^{\frac{-\rt^2}{2\xi}}\right]+\frac{1}{2(\Pt^2+\bar Q^2)}\ln\left(\frac{(\Pt^2+\bar Q^2)^2\xi}{c_0\bar Q^2}\right) \,,
\end{align}
with $\Delta^2=\omega(\Pt^2+\bar Q^2)$. In the integral over $\rt$, the $\xi$ dependent UV subtraction term still regulates the UV $\rt\to0$ behaviour of the diagram, since at small $r_\perp$, $K_1(\Delta r_\perp)\sim (\Delta r_\perp)^{-1}$. In the IR regime, when $r_\perp$ is large, the modified Bessel function kills the power divergence beyond the scale $\sim 1/\Delta$. Since $\Delta^2\propto \omega\propto z_g$, one sees that in the slow gluon limit, the $\rt$ integral is more and more sensitive to this IR divergent domain. This explains the double logarithmic divergence of the resulting $z_g$ integral.

The last step of the transverse integration consists of computing the $\rt$ integral. Using the result Eq.\,\eqref{eq:app-K1-UV-int}, we end up with 
\begin{align}
    I_2&\equiv\frac{1}{2(\Pt^2+\bar Q^ 2)}\ln\left(\frac{c_0}{\omega(\Pt^2+\bar Q^2)\xi}\right)+\frac{1}{2(\Pt^2+\bar Q^2)}\ln\left(\frac{(\Pt^2+\bar Q^2)^2\xi}{c_0\bar Q^2}\right) \,.
\end{align}
As expected, the $\xi$ variable cancels between the two terms since the $\xi$ dependent term was added and subtracted in the first place.  The derivative w.r.t $\Pt$ is straightforward and leads to Eq.\,\eqref{eq:HNLO3-int}.

\subsection{Computation of $\Hcal_{\rm NLO,4}$}
\label{app:NLO4}
In this appendix, we shall provide details of the computation of the two integrals $I_3$ and $I_4^i$ in Eqs.\,\eqref{eq:I3-def} and \eqref{eq:I4-def} that appear in the calculation of the hard factor associated with diagram $\rm V_1\times LO$.

\paragraph{The integral $I_3$.} We begin by reminding the reader of the definition of the integral $I_3$ which gives the term proportional to $\ut^i\ut'^j$ in this hard factor, after taking the derivative w.r.t.\,$\Pt$:
\begin{align}
    I_3=\int\frac{\der^2\ut}{(2\pi)}\int\frac{\der^2\rt}{(2\pi)}e^{-i\Pt\cdot\ut}\frac{\rt\cdot(\rt+\ut)}{\rt^2(\rt+\ut)^2} K_0\left(\bar Q\sqrt{\ut^2+\omega \rt^2}\right)\,.
\end{align}
The structure of the vertex correction implies a dependence of the integral on the scalar product $\rt\cdot\ut$, which complicates the calculation. To address this issue, the trick is to write the kernel
\begin{equation}
    \frac{(\rt^i+\ut^i)}{(\rt+\ut)^2}=(-i)\int\frac{\der^2\ltthre}{(2\pi)}\frac{\ltthre^i}{\ltthre^2}e^{i\ltthre\cdot(\rt+\ut)}\,,
\end{equation}
and likewise, the Bessel function (as in the case for the hard factor for $\rm SE_1$), 
\begin{equation}
    K_0\left(\bar Q\sqrt{\ut^2+\omega\rt^2}\right)=-i\rt^k\int\frac{\der^2\ltone}{(2\pi)}\int\frac{\der^2\lttwo}{(2\pi)}\frac{\lttwo^ke^{i\ltone \cdot\ut}e^{i\lttwo\cdot\rt}}{(\ltone^2+\bar Q^2)(\lttwo^2+\omega(\ltone^2+\bar Q^2))}\,.
\end{equation}
As a result of these tricks, $\ut$ appears only inside phases; the $\ut$ integral gives a delta function that enforces $\ltone=\Pt-\ltthre$, resulting in 
\begin{align}
    I_3&=\int\frac{\der^2\rt}{(2\pi)}\int\frac{\der^2\ltthre}{(2\pi)}\int\frac{\der^2\lttwo}{(2\pi)}\frac{(-1) \rt^i\rt^k \ltthre^i \lttwo^ke^{i \ltthre\cdot\rt}e^{i\lttwo\cdot\rt}}{\rt^2 \ltthre^2 ((\ltthre-\Pt)^2+\bar Q^2)(\lttwo^2+\omega((\ltthre-\Pt)^2+\bar Q^2))} \nonumber \\
    &=(-i)\int\frac{\der^2\rt}{(2\pi)}\int\frac{\der^2\ltthre}{(2\pi)}\frac{\rt^i}{r_\perp}\frac{\ltthre^i}{\ltthre^2}\frac{e^{i\ltthre\cdot\rt}}{((\ltthre-\Pt)^2+\bar Q^2)}\Delta(\ltthre)K_1(\Delta(\ltthre)r_\perp)\,,
\end{align}
where we have defined
\begin{equation}
    \Delta(\ltthre)\equiv\omega((\ltthre-\Pt)^2+\bar Q^2)\,.
\end{equation}
It is more judicious to integrate first over $\rt$ and then do the auxiliary $\ltthre$ integral. Using
\begin{equation}
   \int\frac{\der^2\rt}{(2\pi)} \frac{\rt^i}{r_\perp}e^{i\ltthre\cdot\rt}K_1(\Delta r_\perp)=\frac{i\ltthre^i}{\Delta}\frac{1}{\Delta^2+\ltthre^2}\,,
\end{equation}
we find, after performing the shift $\ltthre-\Pt\to\ltthre$,
\begin{align}
    I_3&=\int\frac{\der^2\ltthre}{(2\pi)}\frac{1}{(\ltthre^2+\bar Q^2)\left[\omega(\ltthre^2+\bar Q^2)+(\ltthre+\Pt)^2\right]} \nonumber \\
    &=\int_0^\infty\der\ell\frac{\ell}{(\ell^2+\bar Q^2)}\frac{1}{\sqrt{\left[(\ell-P_\perp)^2+\omega(\ell^2+\bar Q^2)\right]\left[(\ell+P_\perp)^2+\omega(\ell^2+\bar Q^2)\right]}}\,.
\end{align}
The last integral can be computed explicitly with the result shown in Eq.\,\eqref{eq:Jcal-final}. 

\paragraph{The integral $I_4$.} We turn now to the term proportional to $\rt^i\ut'^j$ in $\Hcal_{\rm NLO,4}$, which involves the integral $I_4^l$:
\begin{align}
    I_{4}^l=\int\frac{\der^2\ut}{(2\pi)}\int\frac{\der^2\rt}{(2\pi)}e^{-i\Pt\cdot\ut}\rt^l\frac{\rt\cdot(\rt+\ut)}{\rt^2(\rt+\ut)^2} K_0\left(\bar Q\sqrt{\ut^2+\omega \rt^2}\right)\,.
\end{align}
The latter differs from $I_3$ by the additional $\rt^l$ factor in the integrand.
Following the same tricks and steps as in the calculation of $I_3$, one ends up with the $\rt$ integral
\begin{align}
   \int\frac{\der^2\rt}{(2\pi)} \frac{\rt^i\rt^l}{r_\perp}e^{i\ltthre\cdot\rt}K_1(\Delta r_\perp)&=\frac{\delta^{il}}{\Delta[\Delta^2+\ltthre^2]}-\frac{2\ltthre^i\ltthre^l}{\Delta} \frac{1}{(\ltthre^2+\Delta^2)^2}\,.
\end{align}
We have then
\begin{align}
    I_{4}^l&=(-i)\int\frac{\der^2\rt}{(2\pi)}\int\frac{\der^2\ltthre}{(2\pi)}\frac{\rt^i\rt^l}{r_\perp}\frac{\ltthre^i}{\ltthre^2}\frac{e^{i\ltthre\cdot\rt}}{((\ltthre-\Pt)^2+\bar Q^2)}\Delta(\ltthre)K_1(\Delta(\ltthre)r_\perp) \nonumber \\
    &=(-i)\int\frac{\der^2\ltthre}{(2\pi)}\frac{\lt^l+\Pt^l}{(\ltthre+\Pt)^2[\ltthre^2+\bar Q^2][(\ltthre+\Pt)^2+\omega(\ltthre^2+\bar Q^2)]}\nonumber\\
    &+(2i)\int\frac{\der^2\ltthre}{(2\pi)}\frac{\lt^l+\Pt^l}{[\ltthre^2+\bar Q^2][(\ltthre+\Pt)^2+\omega(\ltthre^2+\bar Q^2)]^2} \,.
\end{align}
The auxiliary $\ltthre$ integral is performed in polar coordinates
\begin{align}
    I_{4}^l&=\frac{-i\Pt^l}{2\omega\Pt^2}\int_0^\infty\der\ell  \ \frac{\ell[\ell^2-\Pt^2+\omega(\ell^2+\bar Q^2)]}{(\ell^2+\bar Q^2)^2}\frac{1}{\sqrt{D}}+\frac{i\Pt^l}{2\omega\Pt^2}\int_0^\infty\der\ell  \ \frac{\ell(\ell^2-\Pt^2)}{(\ell^2+\bar Q^2)^2}\frac{1}{|\ell^2-\Pt^2|}\nonumber\\
    &+2i\Pt^l\int_0^\infty\der \ell \ \frac{\ell[-\ell^2+\Pt^2+\omega(\ell^2+\bar Q^2)]}{\ell^2+\bar Q^2}\frac{1}{D\sqrt{D}}\,,
\end{align}
with the denominator
\begin{equation}
    D\equiv\left[(\ell-P_\perp)^2+\omega(\ell^2+\bar Q^2)\right]\left[(\ell+P_\perp)^2+\omega(\ell^2+\bar Q^2)\right]\,.
\end{equation}
The factor $1/\omega$ in the first two lines is potentially dangerous. Because $\omega\sim z_g$, it would lead to a power divergence in the slow gluon limit. (Recall that $\Hcal_{\rm NLO,4}$ has an external $z_g$ integral of the form $\int\der z_g/z_g$.) However one sees that this power divergence cancels between the first two terms since when $\omega\to0$,
\begin{equation}
    \frac{1}{\sqrt{D}}\sim\frac{1}{|\ell^2-\Pt^2|}\,.
\end{equation}
This is an important consistency check of our result since any power divergence encountered in the calculation would spoil NLO rapidity factorization. Again, the final integral can be computed explicitly as shown in Eq.\,\eqref{eq:Kcal-final}. This is a remarkably simple result.

\paragraph{The integrals $I_{3s}$ and $I_{4s}$.} We end this appendix with the calculation of the two integrals $I_{3s}$ and $I_{4s}$ defined by Eqs.\,\eqref{eq:I3s-def} and \eqref{eq:I4s-def}. These integrals are useful to obtain the kinematically constrained slow divergence of the hard factor coming from the dressed vertex corrections $\rm V_1$. 
The integral $I_{3s}$ is relatively easy, owing to the identity
\begin{equation}
    \frac{\rt\cdot(\rt+\ut)}{\rt^2(\rt+\ut)^2}=\frac{1}{2}\left[\frac{1}{(\rt+\ut)^2}+\frac{1}{\rt^2}-\frac{\ut^2}{\rt^2(\rt+\ut)^2}\right] \,,
\end{equation}
and going to polar coordinates,
\begin{align}
    I_{3s}&=\int \frac{\der^2 \ut }{2\pi} e^{-i\Pt\cdot \ut} K_0(\bar{Q} u_\perp) 
    \frac{1}{2}\int_0^{\sqrt{\frac{z_f}{z_g}}\frac{1}{Q_f}}\der r r\left[\frac{1}{|r^2-\ut^2|}+\frac{1}{r^2}-\frac{\ut^2}{r^2|r^2-\ut^2|}\right] \nonumber \\
    &=\int \frac{\der^2 \ut }{2\pi} e^{-i\Pt\cdot \ut} K_0(\bar{Q} u_\perp) 
    \frac{1}{2}\ln\left(\frac{z_f}{z_gQ_f^2\ut^2}\right)\Theta\left(\frac{z_f}{z_gQ_f^2}-\ut^2\right)\,.
\end{align}
In a similar fashion,
\begin{align}
       I_{4s}&=\int \frac{\der^2 \ut }{2\pi} e^{-i\Pt\cdot \ut} K_0(\bar{Q} u_\perp)  \frac{\ut^i}{4}\left[1-\ln\left(\frac{z_f}{z_gQ_f^2\ut^2}\right)\right]\Theta\left(\frac{z_f}{z_gQ_f^2}-\ut^2\right)\nonumber\,\\
       &+\frac{\ut^i}{4\ut^2}\frac{z_f}{z_gQ_f^2}\Theta\left(\ut^2-\frac{z_f}{z_gQ_f^2}\right)\,.
\end{align}
Note that since $Q_f\sim P_\perp\sim1/u_\perp$ and $z_g$ is small, one can neglect the case where $\ut^2\ge z_f/(z_gQ_f^2)$ when performing the $\ut$ integral. The latter is done using the identity Eq.\,\eqref{eq:app-LO-ln}:
\begin{align}
    I_{3s}&=\int \frac{\der^2 \ut }{2\pi} e^{-i\Pt\cdot \ut} K_0(\bar{Q} u_\perp) 
 \frac{1}{2}\ln\left(\frac{z_f}{z_gQ_f^2\ut^2}\right)+\mathcal{O}(z_g) \,,\nonumber \\
 & = \frac{1}{2(\Pt^2+\bar Q^2)} \left[\ln\left( \frac{z_f \Pt^2}{z_gQ_f^2} \right)+2\ln\left(\frac{\Pt^2+\bar Q^2}{c_0\bar Q P_\perp}\right) \right]\,,
\end{align}
which is identical to Eq.\,\eqref{eq:I3s}.
For $I_{4s}$, one ends up with Eq.\,\eqref{eq:I4s} using similar techniques.

\section{Power suppressed contributions}
\label{app:power-suppressed}

In this appendix, we will provide the mathematical justification for the power suppression of (i) the higher order terms in the NLO correlation expansion of the color correlator $\Xi_{\rm NLO,1}$ in section~\ref{app:power-suppressed-virtual}, and  (ii) the real diagrams in which the gluon crosses the shockwave in section~\ref{app:power-suppressed-real}.

We emphasize that the two calculations below are primarily illustrative. They focus on specific NLO Feynman diagrams computed in the next-to-leading order of the correlation expansion, and involve hard factors with an odd number of indices in the tensor structure. This means that, after including the quark-antiquark exchange diagrams, these terms would vanish anyway at the cross-section level. To get a nonzero contribution at cross-section level one must go to the next-to-next-to leading order in the correlation expansion. However based on the calculations detailed below, this order is even further power suppressed.

\subsection{Power suppressed terms in virtual diagrams}
\label{app:power-suppressed-virtual}

We  consider here the correlation expansion of the dressed quark self-energy ($\rm SE_1\times LO$) going beyond the leading terms. The  diagram depends on the CGC color correlator $\Xi_{\rm NLO,1}$ given by Eq.\,\eqref{eq:Xi-NLO1} whose Taylor expansion around $\bt$ and $\bt'$, up to third order, reads
\begin{align}
\widetilde\Xi_{\rm NLO,1}&=\left[C_F(\etz^i-\ety^i)\ut'^j-\frac{1}{2N_c}(\etx^i-\etz^i)\ut'^j\right]\frac{1}{N_c}\left\langle\textrm{Tr} \left(V_b[\partial^iV_b^\dagger][\partial^jV_{b'}]V_{b'}^\dagger\right)\right\rangle\nonumber\\
&+\left[\frac{N_c}{4}\etz^i\etz^j\ut'^k-\frac{1}{4N_c}\etx^i\etx^j\ut'^k\right]\frac{1}{N_c}\left\langle\textrm{Tr} \left([\partial^i\partial^jV_b]V_b^\dagger[\partial^kV_{b'}]V_{b'}^\dagger\right)\right\rangle\nonumber\\
&+\frac{C_F}{2}\ety^i\ety^j\ut'^k\frac{1}{N_c}\left\langle\textrm{Tr} \left(V_b[\partial^i\partial^j V_b^\dagger][\partial^kV_{b'}]V_{b'}^\dagger\right)\right\rangle\nonumber\\
&+\left[\frac{N_c}{2}\etz^i\ety^j\ut'^k-\frac{1}{2N_c}\etx^i\ety^j\ut'^k\right]\frac{1}{N_c}\left\langle\textrm{Tr} \left([\partial^iV_b][\partial^j V_b^\dagger][\partial^kV_{b'}]V_{b'}^\dagger\right)\right\rangle\nonumber\\
&+\frac{1}{2}z_1\ut'^i\ut'^j\left[C_F(\etz^k-\ety^k)-\frac{1}{2N_c}(\etx^k-\etz^k)\right]\frac{1}{N_c}\left\langle\textrm{Tr} \left(V_b[\partial^k V_b^\dagger][\partial^i\partial^j V_{b'}]V_{b'}^\dagger\right)\right\rangle\nonumber\\
&+\frac{1}{2}z_2\ut'^i\ut'^j\left[C_F(\etz^k-\ety^k)-\frac{1}{2N_c}(\etx^k-\etz^k)\right]\frac{1}{N_c}\left\langle\textrm{Tr} \left(V_b[\partial^k V_b^\dagger] V_{b'}[\partial^i\partial^j V_{b'}^\dagger]\right)\right\rangle\nonumber\\
&+\mathcal{O}\left(\ut^2\ut'^2,\rt^2\ut'^2,u_\perp r_\perp \ut'^2,...\right)\,,\label{eq:NLO1-b2b-2}
\end{align}
where we defined 
\begin{equation}
\etx=z_2\ut-\frac{z_g}{z_1-z_g}\rt\,,\quad \ety=-z_1\ut\,,\quad \etz=z_2\ut+\rt\,,
\end{equation}
and abbreviated $V_b\equiv V(\bt)$, $\partial^iV_b\equiv \partial^i V(\bt)$, etc. In the correlation expansion, we recall that one typically takes $|\etx|,|\ety|,|\etz|,|\ut|\ll b_\perp$. 
The first row in this Taylor expansion is the leading power term given in Eq.\,\eqref{eq:NLO1-b2b}. The other terms on the following rows involve new operators which differ from the WW gluon TMD by additional spatial derivatives. Hence in principle such terms break the factorization in terms of the WW gluon TMD although they can still be factorized into a hard factor and a nonperturbative operator.

Let us now infer the order of magnitude of the contributions associated with these new operators. The third term is clearly beyond leading power because of the $\ut^i\ut^j\ut'^k$ dependence of the hard factor which gives an additional power of $P_\perp$ suppression from the $\Pt$-derivative. In fact, the only nontrivial contribution is from the tensor structure $\rt^i\rt^j\ut'^k$ which appears in the second line. (The other terms can be obtained from $\Hcal_{\rm NLO,3}$ by taking the $\Pt$ derivative, resulting in a $1/P_\perp$ suppression.) Its contribution to  $\rm SE_1\times LO$ can be expressed as
\begin{align}
    \left.    \der\sigma^{\lambda=\rm L}_{\rm SE_1}\right|_{\textrm{not WW}}&=\alpha_{\rm em}\alpha_se_f^2\deltatwo\left[\frac{\alpha_s}{\pi}\mathcal{H}^{\lambda=\rmL,ijk}(\Pt)\right]\nonumber\\
    &\times \int\frac{\der^2\rbbpt}{(2\pi)^4}e^{-i\qt\cdot\rbbpt}\frac{2}{\alpha_s}\left\langle\textrm{Tr} \left([\partial^i\partial^jV_b]V_b^\dagger[\partial^kV_{b'}]V_{b'}^\dagger\right)\right\rangle_{Y_f}+\textrm{other operators} \,,
\end{align}
where  ``other operators"  refers to the non-WW operators in Eq.\,\eqref{eq:NLO1-b2b-2}.
The next-to-leading power hard factor $\mathcal{H}^{\lambda=\rmL,ijk}$ is defined as
\begin{align}
    &\mathcal{H}^{\lambda=\rmL,ijk}(\Pt)\equiv8z_1^2z_2^3Q^2\int\frac{\der^2\ut}{(2\pi)}\frac{\der^2\ut'}{(2\pi)}\ut'^ ke^{-i\Pt\cdot\ruupt}K_0\left(\bar Q u_\perp'\right)\int_0^{z_1}\frac{\der z_g}{z_g}\left(1-\frac{z_g}{z_1}+\frac{z_g^2}{2z_1^2}\right)\nonumber\\
    &\times\frac{\delta^{ij}}{2}\left[\frac{N_c}{4}-\frac{1}{4N_c}\left(\frac{z_g}{z_1-z_g}\right)^2\right]\int\frac{\der^2\rt}{(2\pi)}\left[ K_0\left(\bar Q\sqrt{\ut^2+\omega\rt^2}\right)-e^{-\frac{\rt^2}{\ut^2e^{\gamma_E}}}K_0(\bar Qu_\perp)\right]\nonumber\\
    &-    \mathcal{G}_{\rm slow}^{\lambda=\rmL,ijk}(\Pt)\,.
\end{align}
When compared to Eq.\,\eqref{eq:HNLO3-def}, we observe that the tensor structure $\rt^i\rt^j\ut'^k$ simplifies into $\delta^{ij}\rt^2\ut'^k/2$  which cancels the $1/\rt^2$ denominator in the integral over $\rt$ in Eq.\,\eqref{eq:HNLO3-def}. This hard factor can be computed analytically along the lines of appendix~\ref{app:NLO3}:
\begin{align}
\mathcal{H}^{\lambda=\rmL,ijk}(\Pt)&=8z_1^3z_2^3Q^2\frac{i\delta^{ij}\Pt^k}{(\Pt^2+\bar Q^2)^2}\int_0^{z_1}\frac{\der z_g}{z_g}\left(1-\frac{z_g}{z_1}+\frac{z_g^2}{2z_1^2}\right)\left[\frac{N_c}{4}-\frac{1}{4N_c}\left(\frac{z_g}{z_1-z_g}\right)^2\right]\nonumber\\
&\times\left\{\frac{2}{\omega(\Pt^2+\bar Q^2)^2}+\frac{e^{\gamma_E}}{2}\frac{6\Pt^2-2\bar Q^2}{(\Pt^2+\bar Q^2)^3}\right\}-    \mathcal{G}_{\rm slow}^{\lambda=\rmL,ijk}(\Pt)\,.
\label{eq:beyond-lp-Gfactor}
\end{align}
Parametrically, this hard factor falls like $1/P_\perp^7$ at large $P_\perp$ instead of $1/P_\perp^6$ for the leading power contribution computed previously. The $\mathcal{H}^{\lambda=\rmL,ijk}(\Pt)$ factor multiplies the $\qt$-distribution
\begin{equation}
    \int\frac{\der^2\rbbpt}{(2\pi)^4}e^{-i\qt\cdot\rbbpt}\frac{2}{\alpha_s}\left\langle\textrm{Tr} \left([\partial^i\partial^jV_b]V_b^\dagger[\partial^kV_{b'}]V_{b'}^\dagger\right)\right\rangle_{Y_f}= \mathcal{O}\left(\textrm{max}(q_\perp,Q_s)\times G^{jk}_{Y_f}\right)\,,
\end{equation}
which is parametrically of order $q_\perp$ or $Q_s$ relative to the WW gluon TMD because of the additional derivative. This demonstrates that the higher order terms in the expansion inside the $\rt$ integral of the CGC operator $\Xi_{\rm NLO,1}$ have the power suppressed form 
\begin{equation}
    \left.    \der\sigma^{\lambda=\rm L}_{\rm SE_1}\right|_{\textrm{first correction to WW}}=\mathcal{O}\left(\frac{\textrm{max}(q_\perp,Q_s)}{P_\perp}\right)\times \alpha_{\rm em}\alpha_se_f^2\deltatwo\left[\frac{\alpha_sC_F}{\pi}\Hcal_{\rm NLO,3}^{\lambda=\rmL,ij}(\Pt)\right]G^{ij}_{Y_f}(\qt)\,.
\end{equation}
A subtle point is that the next to leading power hard factor $\mathcal{H}^{\lambda=\rmL,ijk}(\Pt)$  displays a power divergence at  $z_g=0$ from the $1/\omega\propto 1/z_g$ factor in the curly bracket. This power divergence cancels against an identical term appearing in the next-to-leading power correction to the vertex diagram $\rm V_1\times LO$. We leave for future work the study of the NLO dijet cross-section beyond the correlation limit. The systematic expansion outlined here will be useful in estimating and possibly resumming the all order terms that are powers of $q_\perp/P_\perp$ . 

\subsection{Power suppressed real diagrams}
\label{app:power-suppressed-real}

We turn now to the real diagrams in which the gluon interacts with the shockwave of the target nucleus.
More specifically, we first look at the contribution from diagrams $\rm R_1\times \rm R_2$ and $\rm R_1'\times R_2$, that we call $\der\sigma_{\rm R_{1,1'}\times R_2}$. As a proof of concept, we integrate over the full gluon phase space; hence  the dijet system we are considering is always the quark-anti-quark pair\footnote{In principle, such a definition of the final state is not IRC safe since partons are not experimentally measurable. However it is sufficient to consider this parton definition of the measured dijet to prove that $\rm R_1\times R_2$ is power suppressed.}. One gets
\begin{align}
    \der\sigma_{\rm R_{1,1'}\times R_2}^{\lambda=\rm L}&=\frac{\alpha_{\rm em}e_f^2N_c}{(2\pi)^6}\,\int\der^8\Xt e^{-i\ktone\cdot\rxxtp-i\kttwo\cdot\ryytp}\frac{\alpha_s}{\pi}\int\frac{\der z_g}{z_g} \ 8z_1^2z_{2}^3(1-z_2)Q^2\nonumber\\
    &
    \times \int\frac{\der^2\zt}{\pi}e^{-i\frac{z_g}{z_1}\ktone\cdot\rzxpt}\left[-\kappa_1\frac{\rzxt\cdot\rzxpt}{\rzxt^2\rzxpt^2}+\kappa_2\frac{\rzyt\cdot\rzxpt}{\rzyt^2\rzxpt^2}\right]K_0(QX_{\rm R})K_0(\bar Q_{\mathrm{R}2}r_{x'y'})\nonumber\\
   &\times\Xi_{\rm NLO,1}(\xt,\yt,\zt;\xt',\yt')-\textrm{``slow divergence"}\label{eq:real-R1xR2}\,,
\end{align}
with $z_2=1-z_1-z_g$, $\bar Q_{\rm R2}^2=z_2(1-z_2)Q^2$,
\begin{equation}
    X_{\rm R}=z_1z_2\rxyt^2+z_1z_g\rzxt^2+z_2z_g\rzyt^2 \,,
\end{equation}
and $\kappa_1$ and $\kappa_2$ given by
\begin{align}
    \kappa_1&=1+\frac{z_g}{z_1}+\frac{z_g^2}{2z_1^2}\,,\\
    \kappa_2&=1+\frac{z_g}{z_1}+\frac{z_g}{2z_2}\,,
\end{align}
where only the leading term in the $z_g\to0$ limit turns out to be important for back-to-back kinematics.
The slow gluon subtraction term is implicit in Eq.\,\eqref{eq:real-R1xR2} as the logarithmic divergence when $z_g\to 0$ has to be regulated.
The following change of variables enables one to simplify the calculation of the back-to-back limit:
\begin{align}
    \xt &= \bt+z_2\ut-\frac{z_g(z_1+z_2)}{z_1}\rt \,, \nonumber \\
    \yt &= \bt-z_1\ut \,, \nonumber \\
    \zt &= \bt+z_2\ut+(z_1+z_2)\rt \,.
\end{align}
After this change of variables, we get
\begin{align}
       & \der\sigma_{\rm R_{1,1'}\times R_2}^{\lambda=\rm L}=\frac{\alpha_{\rm em}e_f^2N_c}{(2\pi)^6}\,\int\der^8\Xttilde e^{-i\Pt\cdot\ruupt-i\qt\cdot\rbbpt}\frac{\alpha_s}{\pi}\int\frac{\der z_g}{z_g} \ 8z_2^2(1-z_2)^2\bar Q^2 (z_1+z_2)^3\nonumber\\
    &\times\int\frac{\der^2\rt}{\pi} e^{-i\frac{z_g}{z_1(z_1+z_2)}(\Pt+z_1\qt)\cdot(\rbbpt+z_2\ruupt)}K_0\left((z_1+z_2)\bar Q_{\rm R2}\sqrt{\ut^2+\omega'\rt^2}\right)K_0(\bar Q_{\mathrm{R}2}u_\perp')\nonumber\\
    &\times\left[-\kappa_1\frac{\rt\cdot\left(\rbbpt+z_2\ruupt+(z_1+z_2)\rt\right)}{\rt^2\left(\rbbpt+z_2\ruupt+(z_1+z_2)\rt\right)^2}+\kappa_2\frac{(\rt+\ut)\cdot\left(\rbbpt+z_2\ruupt+(z_1+z_2)\rt\right)}{(\rt+\ut)^2\left(\rbbpt+z_2\ruupt+(z_1+z_2)\rt\right)^2}\right]\nonumber\\
   &\times\widetilde{\Xi}_{\rm NLO,1}(\ut,\bt,\rt;\ut',\bt')-\textrm{``slow divergence"}\,,
\end{align}
with
\begin{equation}
    \omega'=\frac{z_g}{z_1z_2}\,.
\end{equation}
At this stage, the expression remains exact for general dijet kinematics. One notes that the argument of the Bessel function  kills the $\rt$ integral when $\omega'\rt^2\gg \ut^2$, or parametrically $\kgt^2\ll z_g\Pt^2$, as claimed at the beginning of this discussion.

For back-to-back kinematics, one can further simplify the phase in the second line using $\Pt+z_1\qt\approx \Pt$. The phase then becomes
\begin{equation}
    e^{-\frac{z_g}{z_1(z_1+z_2)}\Pt\cdot\rbbpt}\,,
    \label{eq:R1xR2-phase}
\end{equation}
which suppresses the cross-section in the back-to-back limit $\Pt\cdot\rbbpt\to\infty$ unless $z_g$ is small. One can therefore set $z_g=0$ everywhere but in the singular $1/z_g$ term and in the variable $\omega'$ inside the $K_0$ modified Bessel function which effectively enforces the kinematic constraint.
One can also approximate the JIMWLK kernel by
\begin{align}
    \frac{\rt\cdot\left(\rbbpt+z_2\ruupt+(z_1+z_2)\rt\right)}{\rt^2\left(\rbbpt+z_2\ruupt+(z_1+z_2)\rt\right)^2}\approx\frac{\rt\cdot\rbbpt}{\rt^2\rbbpt^2}\,,
    \label{eq:kernel-real-b2b}
\end{align}
and expand the CGC color correlator to leading power as
\begin{align}
    \widetilde{\Xi}_{\rm NLO,1}&\approx(z_1+z_2)\left[C_F\left(\ut^i\ut'^j+\rt^i\ut'^j\right)+\frac{1}{2N_c}\left(1+\frac{z_g}{z_1}\right)\rt^i\ut'^j\right]\frac{\alpha_s}{2N_c}\hat G^{ij}(\rbbpt) \nonumber \\
    &= \left[C_F\ut^i\ut'^j+\frac{N_c}{2}\rt^i\ut'^j\right]\frac{\alpha_s}{2N_c}\hat G^{ij}(\rbbpt)+\mathcal{O}(z_g)\,.
\end{align}
Indeed, the correlation expansion of the CGC operator relies on the parametric orders of magnitude: $u_\perp\sim 1/P_\perp$, $r_\perp\sim1/P_\perp$ and $|\rbbpt|\sim 1/q_\perp$; hence $r_\perp, u_\perp \ll |\rbbpt|$. The presence of the transverse coordinate scale $\rbbpt$ in the JIMWLK kernel Eq.\,\eqref{eq:kernel-real-b2b} for real diagrams is the essential difference w.r.t.\ virtual corrections, since, as demonstrated below, it amounts to an extra power of $q_\perp/P_\perp$ in the back-to-back limit.

With these simplifications, diagrams $\rm R_1\times R_2$ and $\rm R_1'\times R_2$ factorize as
\begin{align}
 \der\sigma^{\lambda=\rmL}_{\rm R_{1,1'}\times R_2}&=\alpha_{\rm em}\alpha_se_f^2\int\frac{\der^2\rbbpt}{(2\pi)^4} e^{-i\qt\cdot\rbbpt}\frac{\rbbpt^k}{\rbbpt^2}\hat G^{ij}_{Y_f}(\rbbpt)\nonumber\\
 &\times\frac{\alpha_s}{\pi}\int_0^1\frac{\der\xi}{\xi}\left[e^{-i\xi\Pt\cdot\rbbpt}-1\right]\mathcal{H}^{\lambda=\rmL,ijk}_{\rm R_{1,1'}\times R_2}\left(\Pt\right)\,.\label{eq:R1R2-b2b} 
\end{align}
where the ``$-1$" inside the square bracket comes from the slow gluon subtraction term (including the kinematic constraint!) and with the hard factor given by
\begin{align}
   & \mathcal{H}^{\lambda=\rmL,ijk}_{\rm R_{1,1'}\times R_2}(\Pt)\equiv16z_1^2z_2^3\frac{iQ^2\Pt^j}{(\Pt^2+\bar Q^2)^2}\nonumber\\
    &\times \left\{-\frac{N_c}{2}\int\frac{\der^2\ut}{(2\pi)}e^{-i\Pt\cdot\ut}\int\frac{\der^2\rt}{(2\pi)}\frac{\rt^i\rt^k}{\rt^2} K_0\left(\bar Q\sqrt{\ut^2+\omega'\rt^2}\right)\right.\nonumber\\
    &\left.+\int\frac{\der^2\ut}{(2\pi)}e^{-i\Pt\cdot\ut}\int\frac{\der^2\rt}{(2\pi)}\left(C_F\ut^i+\frac{N_c}{2}\rt^i\right)\frac{(\rt+\ut)^k}{(\rt+\ut)^2} K_0\left(\bar Q\sqrt{\ut^2+\omega'\rt^2}\right)\right\}\,.
\end{align}
There is a slight abuse of notation in Eq.\,\eqref{eq:R1R2-b2b}. Indeed although the hard factor $\mathcal{H}^{\lambda=\rmL,ijk}_{\rm R_{1,1'}\times R_2}$ depends on $\xi\equiv z_g/z_1$ through the $\omega'$ variable, it is sufficient to consider the leading term in the $z_g\to0$ limit since the hard factor  factors out of the $\xi$ integral and does not depend on $\xi$.

The form of the kernel in Eq.\,\eqref{eq:kernel-real-b2b} brings in the additional $\rbbpt^k/\rbbpt^2$ dependence inside the Fourier transform of the WW gluon TMD in coordinate space in Eq.\,\eqref{eq:R1R2-b2b}. 
The two terms in the hard factor $\mathcal{H}^{\lambda=\rmL,ijk}_{\rm R_1\times R_2}$ (inside the curly bracket) come respectively from diagram $\rm R_1\times R_2$ (real emission from the quark before the shockwave with absorption in the c.c.\ by the quark after the shockwave) and diagram $\rm R_1'\times R_2$  (real emission from the anti-quark before the shockwave with absorption in the c.c.\ by the quark after the shockwave). It is very important to consider these two diagrams together as each individual diagram has a power divergence $\propto 1/\omega'$ in the limit $\omega'\propto z_g\to0$ which is potentially troublesome. This is because  such a power divergence combined with the phase $e^{-i\xi\Pt\cdot\rbbpt}$ would lead to a $P_\perp/q_\perp$ enhancement of the cross-section. Fortunately the power divergence cancels amongst the two diagrams with the net result
\begin{align}
    \mathcal{H}^{\lambda=\rmL,ijk}_{\rm R_1\times R_2}(\Pt)\equiv16z_1^2z_2^3\frac{iQ^2\Pt^j}{(\Pt^2+\bar Q^2)^4}&\left\{\delta^{ik}\left[\frac{N_c}{2}\frac{1}{1+\chi^2}-\frac{1}{2N_c}\right]\right.\nonumber\\
    &\left.-\frac{\Pt^i\Pt^k}{\Pt^2}\left(N_c-\frac{2}{N_c}\right)\frac{1}{1+\chi^2}\right\}\,,
\end{align}
up to powers of $z_g$ corrections which are sub-leading in the back-to-back limit. Using $\bar Q\sim P_\perp$ and $\rbbpt^k/\rbbpt^2\sim \nabla_{\qt}^k/\nabla^2_{\qt}$ in Fourier space, we find that the leading contribution of diagrams $\rm R_1\times R_2$ plus $\rm R_1'\times R_2$ behave parametrically like
\begin{align}
    \der\sigma^{\lambda=\rmL}_{\rm R_1\times R_2}&=\alpha_{\rm em}\alpha_s^2\times \mathcal{O}\left(\frac{1}{P_\perp^5}\times\nabla_{\qt}^{-1}G^{ij}_{Y_f}(\qt)\right)\\
    &=\alpha_s\der\sigma^{\lambda=\rmL}_{\rm LO}\times\mathcal{O}\left(\frac{\textrm{max}(q_\perp,Q_s)}{P_\perp}\right)\,.
\end{align}
This concludes our proof of the power suppression of this diagram relative to the leading power terms. 

As highlighted in the introduction of this appendix, the odd tensor structure of the hard factor $\mathcal{H}^{\lambda=\rmL,ijk}_{\rm R_1\times R_2}(\Pt)$ implies that the contribution computed in this section would vanish at cross-section level, after including the quark-antiquark exchanged diagrams. To get a non-vanishing (but power-suppressed like $\textrm{max}(q_\perp^2,Q_s^2)/P_\perp^2$) contribution at cross-section level, one would need the next term in the small $\rt$, $\ut$ or $\ut'$ expansion. There are several ways to get an even contribution under $q\leftrightarrow\bar q$ exchange: one of them is to include $\mathcal{O}(\ut'^2)$ corrections from the LO diagram in the complex conjugate.

The physical and mathematical arguments presented in this section to demonstrate that diagrams $\rm R_1\times R_2$ and $\rm R_1'\times R_2$ 
contribute beyond leading power in $q_\perp/P_\perp$ extend to all real diagrams where the gluon interacts at least once with the shockwave -- in particular, the diagram $\rm R_1\times \rm R_1$. Thus in the back-to-back limit, such diagrams do not contribute to the hard factor but only to the rapidity evolution of the WW gluon TMD.

\section{Sudakov form factor with two loop running coupling}
\label{app:sudakov}
The Sudakov form factor computed in this paper reads (cf. Eq.\,\eqref{eq:Sudakov-final})
\begin{equation}
   \mathcal{\tilde{S}}(\mu_h^2,\rbbpt^2)=\exp\left\{-\int_{c_0^2/\rbbpt^2}^{\mu_h^2}\frac{\der\mu^2}{\mu^2}\frac{\alpha_s(\mu^2)N_c}{\pi}\left[\frac{1}{2}\ln\left(\frac{\mu_h^2}{\mu^2}\right)+\frac{C_F}{N_c}s_0-\tilde{s}_f-\frac{\pi\beta_0}{N_c}\right]\right\} \,,
\end{equation}
with the hard scale
\begin{equation}
    \mu_h^2=\Pt^2+\bar Q^2 \,,
\end{equation}
chosen to avoid large negative double logarithms of $Q^2/M_{q\bar q}^2$ in the impact factor. The single log coefficients $s_0$ and $\tilde{s}_f$ read
\begin{equation}
    s_0 = \ln\left(\frac{2(1+\cosh(\Delta \eta_{12}))}{R^2}\right)+\mathcal{O}(R^2)\,,\quad \tilde{s}_f=1- \frac{1}{2}\ln\left(\frac{\Pt^2+\bar Q^2}{\Pt^2}\right)\,.
\end{equation}
\subsection{Fixed coupling}
\label{fixed_coup_sud_sect}
For fixed coupling $\alpha_s(\mu^2)\simeq\alpha_s(\mu_h^2)$, we find
\begin{equation}
   \mathcal{\tilde{S}}(\mu_h^2,\rbbpt^2)=\exp\left\{-\frac{\alpha_s(\mu_h^2)N_c}{\pi}\left[\frac{1}{4}L^2 +\left(\frac{C_F}{N_c}s_0-\tilde{s}_f-\frac{\pi\beta_0}{N_c}\right)L  \right]\right\} \,,
\end{equation}
with the large logarithm in the resummation
\begin{align}
    L \equiv \ln\left( \frac{\mu_h^2 \rbbpt^2}{c_0^2} \right).
\end{align}

\subsection{Running coupling at two loop}
The running of the coupling at two loops reads (see for instance Appendix A in \cite{Marzani:2019hun})
\begin{align}
    \alpha_s(\mu^2) = \frac{\alpha_s(\mu_h^2)}{1+\alpha_s(\mu_h^2) \beta_0 \ln\left( \frac{\mu^2}{\mu_h^2} \right)} \left[1- \alpha_s(\mu_h^2) \frac{\beta_1}{\beta_0} \frac{\ln \left(1+ \alpha_s(\mu_h^2) \beta_0 \ln\left( \frac{\mu^2}{\mu_h^2} \right) \right)}{1+\alpha_s(\mu_h^2) \beta_0 \ln\left( \frac{\mu^2}{\mu_h^2} \right)} \right] \,,
\end{align}
where $\beta_0 = \frac{(11 C_A - 2 n_f)}{12\pi}$ and $\beta_1 = \frac{17 C_A^2 - 5 C_A n_f - 3 C_F n_f }{24\pi^2}$.
Performing the $\mu^2$ integral, the Sudakov factor simplifies into
\begin{align}
    \mathcal{\tilde{S}}(\mu_h^2,\rbbpt^2)=  \exp\left\{ -\frac{N_c}{\pi}\left[ \frac{1}{2} J_1  + \left( \frac{C_F}{N_c} s_0 - \tilde{s}_f -\frac{\pi\beta_0}{N_c}\right) J_2 \right] \right\} \,,
\end{align}
where
\begin{align}
    J_1 &= \frac{1}{\alpha_s(\mu_h^2) \beta_0^2} \left[\ln\left(\frac{1}{1-\alpha_s(\mu_h^2) \beta_0 L }\right) -\alpha_s(\mu_h^2) \beta_0 L \right]  \nonumber \\
    & + \frac{\beta_1}{\beta_0^3} \left[ 1-\frac{1 - \ln\left(\frac{1}{1-\alpha_s(\mu_h^2) \beta_0 L}\right) }{1-\alpha_s(\mu_h^2) \beta_0 L} - \frac{1}{2}\ln^2\left(\frac{1}{1-\alpha_s(\mu_h^2) \beta_0 L} \right)   \right] \,, \nonumber \\
    J_2 &= \frac{1}{\beta_0} \ln\left( \frac{1}{1- \alpha_s(\mu_h^2) \beta_0 L}\right) + \alpha_s(\mu_h^2) \frac{\beta_1}{\beta_0^2} \left[ 1-\frac{1-\ln\left(\frac{1}{1-\alpha_s(\mu_h^2) \beta_0 L}\right)}{1-\alpha_s(\mu_h^2) \beta_0 L} \right] \,.
    \label{eq:2loop-resummed}
\end{align} 

\section{Useful integrals}
\label{app:useful-int}
The following two integrals are useful in computing the LO hard factor for longitudinally polarized photons. They also appear in the LO complex conjugate amplitude in the calculation of the NLO hard factors associated with virtual corrections:
\begin{equation}
    \int\frac{\der^2\ut}{(2\pi)}e^{i\Pt\cdot\ut}K_0(\bar Qu_\perp)=\frac{1}{\Pt^2+\bar Q^2}\label{eq:app-LO} \,.
\end{equation}
Differentiating w.r.t.\ $\Pt$, one gets
\begin{equation}
        \int\frac{\der^2\ut}{(2\pi)}\ut^ie^{i\Pt\cdot\ut}K_0(\bar Qu_\perp)=\frac{2i\Pt^i}{(\Pt^2+\bar Q^2)^2}\label{eq:app-LO-vector} \,.
\end{equation}
Additional useful integrals in appendix~\ref{app:NLO1} are as follows:
\begin{equation}
    \int\frac{\der^2\ut}{(2\pi)}e^{-i\Pt\cdot\ut}K_0(\bar Qu_\perp)\ln(P_\perp u_\perp)=-\frac{1}{\Pt^2+\bar Q^2}\ln\left(\frac{\Pt^2+\bar Q^2}{c_0P_\perp \bar Q}\right)\label{eq:app-LO-ln} \,.
\end{equation}
Differentiating w.r.t.\ $\Pt$, we find
\begin{align}
        \int\frac{\der^2\ut}{(2\pi)}\ut^i e^{-i\Pt\cdot\ut}K_0(\bar Qu_\perp)\ln(P_\perp u_\perp)=\frac{2i\Pt^i}{(\Pt^2+\bar Q^2)^2}\left[-1+\ln\left(\frac{\Pt^2+\bar Q^2}{c_0P_\perp \bar Q}\right)\right]\label{eq:app-LO-vector-ln} \,.
\end{align}
This integral is useful in the calculation of the hard factor for the dressed self-energy:
\begin{equation}
    \int_0^\infty\der x\left[\textrm{K}_1(x)-\frac{1}{x}\exp\left(-\frac{x^2}{2\Delta^2}\right)\right]=\frac{1}{2}\ln\left(\frac{c_0}{\Delta^2}\right)\label{eq:app-K1-UV-int} \,.
\end{equation}

\bibliographystyle{utcaps}
\bibliography{dijet-factorization}

\end{document}